\documentclass[bibauthoryear]{PHEA}

\usepackage{graphicx}
\usepackage{txfonts}
\usepackage{amsmath}
\usepackage{url}
\usepackage{placeins}          

\def\apjl{Astrophys. J. Lett.}
\def\apjs{Astrophys. J. Suppl. Ser.}
\def\ao{Appl. Opt.}

\newcommand{\EQ}[1]{Eq.~(\ref{#1})}
\newcommand{\FIG}[1]{Fig.~\ref{#1}}
\newcommand{\TAB}[1]{Table~\ref{#1}}
\newcommand{\SUPTAB}[1]{Table~\ref{#1}}
\newcommand{\SUPFIG}[1]{Fig.~\ref{#1}}

\newcommand{\tar}{\mbox{GRB 221009A}}

\makeatletter
\renewcommand*\@fnsymbol[1]{%
  \ensuremath{%
    \ifcase#1
    \or \star
    \or \dagger
    \or \ddagger
    \or \mathsection
    \or \mathparagraph
    \or \|
    \or \star\star
    \or \dagger\dagger
    \or \ddagger\ddagger
    \else \@ctrerr
    \fi
  }%
}
\makeatother

\begin{document}

\pheaNum{XXXXXX}

\titlerunning{Extreme superluminal expansion of GRB 221009A}
\authorrunning{Geng et al.}

\title{Extreme superluminal expansion traces an ultra-narrow GRB jet across
relativistic regimes}

\author{J.-J.~Geng\inst{1,2}\thanks{These authors contributed equally to this work.}
   \and Y.-K.~Zhang\inst{3,4}\footnotemark[1]
   \and H.-X.~Gao\inst{1}
   \and J.~Hu\inst{1,5}
   \and F.~Xu\inst{6}
   \and B.~Li\inst{7}
   \and T.-R.~Sun\inst{1}
   \and A.-L.~Wang\inst{8,9}
   \and Z.-J.~Xu\inst{3}
   \and Y.-Q.~Liu\inst{4}
   \and J.~Yang\inst{10}
   \and C.-R.~Hu\inst{11}
   \and L.~Rhodes\inst{12,13}
   \and L.~Li\inst{14,15}
   \and Y.~Wang\inst{16}
   \and Y.~Li\inst{1,2}
   \and D.~Xiao\inst{1,2}
   \and J.~Ren\inst{1}
   \and B.~Zhang\inst{17,18}
   \and T.~An\inst{5}\corrauth{antao2008@ustc.edu.cn}
   \and X.-F.~Wu\inst{1,5,2}\corrauth{xfwu@pmo.ac.cn}
   \and Y.-F.~Huang\inst{11,19}\corrauth{hyf@nju.edu.cn}
   \and Z.-G.~Dai\inst{5}\corrauth{daizg@ustc.edu.cn}
}

\institute{Purple Mountain Observatory, Chinese Academy of Sciences, Nanjing 210023, P.~R.~China
   \and State Key Laboratory of Radio Astronomy and Technology, Purple Mountain Observatory, Chinese Academy of Sciences, Nanjing 210023, P.~R.~China
   \and Shanghai Astronomical Observatory, Chinese Academy of Sciences, Shanghai 200030, P.~R.~China
   \and State Key Laboratory of Radio Astronomy and Technology, Shanghai Astronomical Observatory, Chinese Academy of Sciences, Shanghai 200030, P.~R.~China
   \and School of Astronomy and Space Sciences, University of Science and Technology of China, Hefei 230026, P.~R.~China
   \and Institute of Space Weather, School of Atmospheric Physics, Nanjing University of Information Science \& Technology, Nanjing 210044, P.~R.~China
   \and Department of Astronomy, Tsinghua University, Beijing 100084, P.~R.~China
   \and Yunnan Observatories, Chinese Academy of Sciences, Kunming 650216, P.~R.~China
   \and Key Laboratory of Particle Astrophysics, Institute of High Energy Physics, Chinese Academy of Sciences, Beijing 100049, P.~R.~China
   \and Department of Space, Earth and Environment, Chalmers University of Technology, Onsala Space Observatory, SE-439 92 Onsala, Sweden
   \and School of Astronomy and Space Science, Nanjing University, Nanjing 210023, P.~R.~China
   \and Trottier Space Institute at McGill, 3550 Rue University, Montreal, Quebec H3A 2A7, Canada
   \and Department of Physics, McGill University, 3600 Rue University, Montreal, Quebec H3A 2T8, Canada
   \and Institute of Fundamental Physics and Quantum Technology, Ningbo University, Ningbo, Zhejiang 315211, P.~R.~China
   \and Department of Physics, School of Physical Science and Technology, Ningbo University, Ningbo, Zhejiang 315211, P.~R.~China
   \and ICRANet, Piazza della Repubblica 10, I-65122 Pescara, Italy
   \and The Hong Kong Institute for Astronomy and Astrophysics, The University of Hong Kong, Pokfulam Road, Hong Kong, P.~R.~China
   \and Department of Physics, The University of Hong Kong, Pokfulam Road, Hong Kong, P.~R.~China
   \and Key Laboratory of Modern Astronomy and Astrophysics (Nanjing University), Ministry of Education, Nanjing 210023, P.~R.~China}

\date{}

\abstract
  {The long-term evolution of gamma-ray burst (GRB) jets from 
   ultra-relativistic launch to a slowly expanding radio source remains a 
   fundamental problem in high-energy astrophysics, rarely resolved directly by imaging.} 
  {We use very long baseline interferometry (VLBI) to trace the radio angular
   scale of GRB 221009A, the brightest GRB ever detected, from 5 to 299 days
   after the burst, and constrain the jet's geometry and dynamics.}
  {The Very Long Baseline Array data at 8.4 and 15.2 GHz are calibrated and imaged
   with AIPS and Difmap, and the source size is characterized by a circular-Gaussian
   model fitted to the complex visibilities. 
   The jet dynamics, multi-wavelength emission, and source size are modeled within a two-shell collision scenario. 
   A hydrodynamic simulation confirms that such an ultra-narrow jet undergoes lateral spreading.}
  {The observations reveal apparent superluminal expansion, with the expansion speed
   declining from about $90c$ at 5 days to about $6c$ at 300 days, tracing the jet's
   deceleration from the ultra-relativistic to the mildly relativistic regime and
   placing GRB 221009A far beyond the luminosity--velocity envelope of active
   galactic nuclei and microquasar jets. Combined with the multi-wavelength afterglow,
   they establish an ultra-narrow relativistic jet with a half-opening angle of about
   0.01--0.03 radians, propagating through a wind-like medium at small radii before
   entering an interstellar-medium-like profile at larger radii. Joint modeling
   favors a two-phase size evolution with tentative evidence for lateral spreading.}
  {These results provide a direct geometric view of a GRB outflow and establish
   GRB 221009A as a benchmark for relativistic jet physics.}


\maketitle

\nolinenumbers

\section{Introduction}

Gamma-ray bursts (GRBs) are flashes of gamma rays arising from cataclysmic cosmic events.
A fundamental aspect of GRB physics is the relativistic jet that powers these explosions \citep{KumarZhang15}.
The GRB outflow has traditionally been modeled as a conical jet with a half-opening angle of the order of $0.1$ radians, supported by observations of a steepening break in their afterglow lightcurves (``jet break''). While early models assumed a simple uniform structure, an angle-dependent structured jet \citep{Meszaros98,Dai01,Rossi02,ZhangB02,Kumar03} may better represent the true nature of these outflows, as demonstrated in some GRBs including the gravitational-wave-associated short GRB 170817A \citep{Lamb17,Troja17,Granot17,Lazzati18,Kathirgamaraju18,Gottlieb18,Geng19,Troja19,Ryan20}.
Very long baseline interferometry (VLBI) is the most powerful tool to directly image nearby GRB jets and study their complex geometry and dynamics. The successful size measurements and proper motion detection in GRB 030329 \citep{Taylor04,Taylor05,Pihlstrom07} and GRB 170817A \citep{Ghirlanda19,Mooley18,Mooley22} have provided crucial clues to unveil the nature of those jets.

GRB 221009A, discovered at 13:16:59.99 UTC on 2022 October 9 \citep{Lesage23} (which we define as $T_0$) by the Fermi Gamma-ray Burst Monitor (GBM), is the record holder as the brightest GRB ever detected, with isotropic equivalent energy exceeding $10^{55}$ erg \citep{Lesage23,AnZH23,LHAASO23,Burns23,Frederiks23}. Its energy level and relative proximity (redshift of $z = 0.151$, \citealt{deUgarte22,Castro-Tirado22}) make GRB 221009A an ideal natural laboratory for detailed multi-wavelength investigations.
The extensive follow-up campaigns have revealed its temporal and spectral evolution across the electromagnetic spectrum. Starting as early as the prompt stage, the tera-electron volt (TeV) emission was detected by the Large High Altitude Air Shower Observatory (LHAASO) Collaboration \citep{LHAASO23}.
Sustained emission is recorded in the TeV, GeV, X-ray, optical, and radio bands \citep{LHAASO23,HAWC22,LST25,Ravasio24,Tavani23,Axelsson25,Kann23,OConnor23,Levan23,Williams23,Laskar23,Bright23,Rhodes24}, providing a rich dataset to probe the jet properties.
Recently acquired high-resolution imaging through VLBI has added a crucial new dimension to our understanding of this exceptional event \citep{Giarratana24}.
Several models have been proposed to interpret the multi-wavelength emission of GRB 221009A, invoking non-uniform jet structures \citep{Gill23,Sato23,OConnor23,Ren24,ZhangB23,Zheng24}, specific burst environments \citep{Ren23,Zheng24}, multiple emission components \citep{Isravel23,ZhangBT25}, and evolving microphysics shock parameters \citep{Fraija24,Foffano24}.
Despite the wealth of observational data gathered on GRB 221009A, a coherent physical picture of GRB 221009A remains elusive, particularly when considering the complex details revealed by VLBI observations.
Existing VLBI measurements with the European VLBI Network (EVN) and the Very Long Baseline Array (VLBA) resolve the radio source from 40 to 262 days post-burst \citep{Giarratana24}, but do not sample the first month, when the angular ,size most strongly constrains the early outflow dynamics, jet opening angle, and circumburst environment.

\section{Results}

\subsection{VLBI observation}

Here, we combine new and archival VLBI observations at 8 and 15 GHz to trace the radio angular size from 5 to 299 days after the burst (\TAB{tab:ObsSummary}).
The VLBA observations and data reduction are described in Appendix~\ref{sec:vlba_obs} and \ref{sec:vlba_reduction}, respectively.
Our analyses uncover important clues regarding the geometry, motion, and evolving nature of the outflow.
Throughout our VLBA monitoring campaign, GRB 221009A maintained its appearance as a compact radio source (see \FIG{fig:Image}).
The absence of a substantial centroid shift suggests an on-axis jet aligned with our line of sight.
We characterize the source size by the circular-Gaussian full width at half maximum $\theta_{\rm G}$, obtained by fitting the complex visibilities directly in the Fourier domain (see Appendix~\ref{sec:vlbi_error_structure}).
At 9 days post-burst, $\theta_{\rm G}$ of GRB 221009A is about five times larger than that of GRB 030329 at a similar distance (\FIG{fig:size_evol}).
Given that the expansion rate of the jet head radius scales with the burst energy, this discrepancy directly reflects the significant difference in energy between the two bursts, approximately $10^{55}$ erg for GRB 221009A versus about $10^{52}$ erg for GRB 030329 (see Appendix~\ref{sec:superluminal}).
The combined VLBI measurements at 15 and 8 GHz show that the source size evolution can be described by a single power law, $\propto t^{0.18}_{\rm obs}$, shallower than the $\propto t^{0.43}_{\rm obs}$ found for GRB 030329. A broken power law with a late transition to steeper growth yields a modest reduction in chi-squared (see Appendix~\ref{sec:size_evolution}), which alone is not statistically decisive.
However, the inferred transition time coincides with the onset of a prominent bump in the 15-GHz lightcurve \citep{Rhodes24}, suggesting a change in the jet's dynamical state. This temporal coincidence provides independent, albeit indirect, support for a two-phase description.

The apparent expansion speed $\beta_{\rm app} = (1+z)D_{\rm A}\theta_{\rm G}/(2t_{\rm obs})$ provides a characteristic expansion rate for the radio-emitting region, where $D_{\rm A}$ is the angular distance, $c$ is the speed of light, and $t_{\rm obs}$ is the observational time since $T_0$.
At approximately 9 days post-burst, we measure $\beta_{\rm app} \sim 60 c$, larger than the component speeds measured in most active galactic nuclei (AGN) and Galactic X-ray binaries (\FIG{fig:beta_Lnu}) (see Appendix~\ref{sec:superluminal}).
Multi-epoch VLBI monitoring further reveals that $\beta_{\rm app}$ declines from $\sim 90 c$ to $\sim 6 c$ between $\sim 5$ and $\sim 300$ days, tracing the pronounced deceleration of the outflow.
This high apparent speed and its rapid decline place GRB 221009A in a distinct region of the $\beta_{\rm app}$ versus spectral luminosity diagram, where VLBI has historically mapped the apparent transverse velocities and spectral luminosities of relativistic jets in AGN and microquasars (\FIG{fig:beta_Lnu}).
In this plane, varying the viewing angle for a fixed intrinsic Lorentz factor and comoving luminosity produces a family of aspect curves \citep{Cohen07}.
The AGN population lies within the envelope defined by an intrinsic Lorentz factor of 50 and a comoving-frame spectral luminosity of $\simeq 3 \times 10^{31}$ erg~s$^{-1}$~Hz$^{-1}$.
GRB 221009A, by contrast, extends far beyond this envelope and evolves from exceptionally high to moderate apparent speeds. This trajectory is naturally explained as the on-axis deceleration of a jet with an initial Lorentz factor exceeding 100.

\subsection{Analyses of size evolution}
We consider a relativistic top-hat jet with an initial half-opening angle of $\theta_{\rm j,0}$ propagating into a circumburst medium with a power-law profile of $\propto R^{-s}$, where $R$ is the distance from the explosion center, and $s=0$ and $s=2$ correspond to a uniform interstellar medium (ISM) and a free stellar wind, respectively. A jet with an initial bulk Lorentz factor $\Gamma_{0}$ (typically $> 100$)
decelerates once it has swept up enough mass from the medium, typically within a few hundred seconds.
After the deceleration time, the jet dynamics follow the Blandford--McKee (BM) solution \citep{Blandford76}, i.e., the radius $R$ grows as $t^{1/(4-s)}_{\rm obs}$ and the Lorentz factor $\Gamma$ decays as $t^{(s-3)/(8-2s)}_{\rm obs}$. A simplified version: Interpreting the early size scales on 5 and 9 days with the classical afterglow theory gives a lower limit of $\theta_{\rm j,0} \ge 0.01$~rad. Their large values are more readily accommodated by a wind-like density profile than by an ISM with a typical density value (see Appendix~\ref{sec:tophat}). The optical and X-ray afterglows independently support a wind-like medium during this early phase \citep{OConnor23,Ren23}.

After the jet break at $\sim 0.8$~day~\citep{OConnor23}, a non-spreading top-hat jet in an ISM has a projected scale proportional to $t_{\rm obs}^{1/4}$, while a wind profile gives $t_{\rm obs}^{1/2}$. The measured index of 0.18 by VLBI is close to the ISM scaling. This dynamical comparison supports that the jet has expanded into an ISM-like environment at larger radii. The late-time ISM environment is consistent with the location of the burst in a dense star-forming region inferred from James Webb Space Telescope (JWST) spectroscopy \citep{Blanchard24}. A detailed temporal analysis of a structured jet also disfavors a wind profile at large radii (see Appendix~\ref{sec:structured}).

After approximately 100 days, we find tentative evidence for a transition to steeper size growth, with $\theta_{\rm G} \propto t^{0.91}_{\rm obs}$.
Concurrently, the 15-GHz lightcurve exhibits a bump followed by a steep decline, a feature that can be produced by lateral spreading in the adopted dynamical model.
Such behavior is theoretically expected for very narrow jets with $\theta_{\rm j,0} \ll 10^{-1.5} \approx 0.03$ \citep{Granot12}, which begin to spread appreciably when $\Gamma \approx \zeta/\theta_{\rm j,0}$.
In this context, our observations constrain the dimensionless spreading parameter to $\zeta < 0.5$ (see Appendix~\ref{sec:spreading}).
An independent upper limit of $\theta_{\rm j,0} \le 0.03$~rad follows from the temporal break in the X-ray and optical afterglows \citep{OConnor23}.
Together, the VLBI size constraints, afterglow breaks, and JWST environmental diagnostics \citep{Blanchard24} converge on a relativistic jet with an initial half-opening angle $\theta_{\rm j,0} \simeq 0.01-0.03$~rad. The combined data favor a jet break followed by a transition from a wind-like medium into an ISM-like environment, and likely lateral spreading (see Appendix~\ref{sec:size_analysis}).

\subsection{Multi-wavelength emission}
\begin{figure*}
  \centering
  \includegraphics[width=\textwidth]{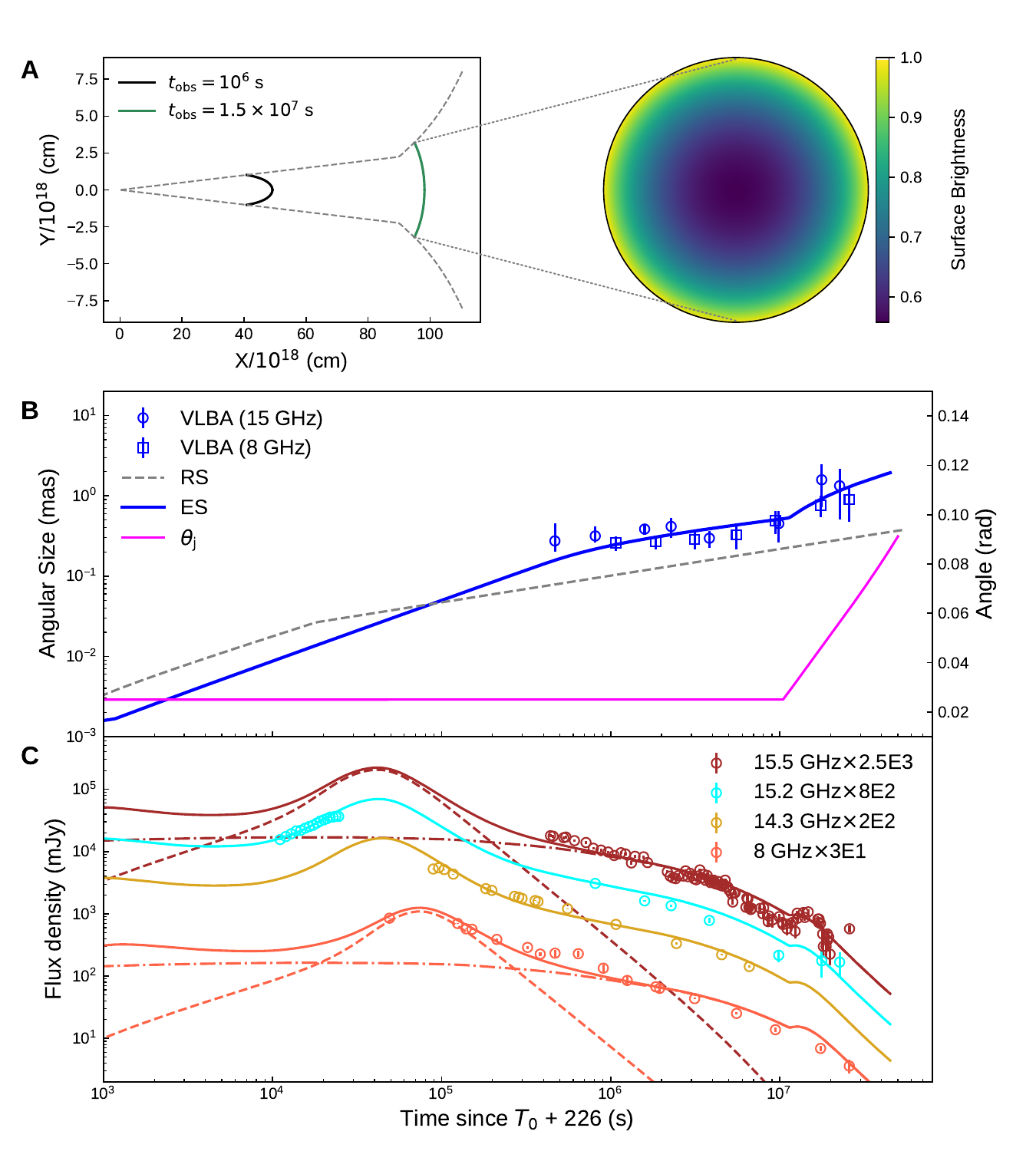}
  \caption{
  {\bf Modeling the source size evolution.}
  (\textbf{A}) Equal-arrival-time surfaces (arc-shaped curves) used to compute the numerical image size, shown on a Cartesian grid. Grey dashed lines trace the evolution of the jet's physical boundaries. The inset contour map illustrates the normalized limb-brightened surface brightness profile at radio frequencies, obtained via line-of-sight projection.
  (\textbf{B}) Numerical image size of the expanding external shock (ES; blue solid line) and the dynamical size of the reverse shock (RS; grey dashed line) lagging behind the ES. The magenta line shows the evolution of the ES half-opening angle in the model.
  (\textbf{C}) Lightcurves around 15 GHz and 8 GHz.
  The RS emission (dashed lines) becomes negligible after $3 \times 10^5$~s (also see Fig.~\ref{fig:Fitting}); thereafter, the observed source size is determined by the ES emission (dash-dotted lines).
  }
 \label{fig:SizeFit}
\end{figure*}

These dynamical insights motivate us to delve into understanding the complex multi-wavelength properties of GRB 221009A. The multi-pulsed prompt emission and the joint afterglow analysis lead us to propose a two-shell collision scenario (see Appendix~\ref{sec:twoshell}), in which the entire emission can be decomposed into three phases (see Appendix~\ref{sec:twoshell}).
In this scenario, the early TeV emission arises from the forward shock generated as a fast rear shell catches up and interacts with a slow preceding shell.
Radio brightening bumps originate from the reverse shock that propagates back into the fast shell, while the forward shock, after breaking through the preceding shell, transitions into the external shock and powers the long-lived multi-wavelength afterglow.
Our detailed numerical calculations consistently interpret both the broadband lightcurves and the observed size evolution in GRB 221009A (see \FIG{fig:SizeFit}).
While our hydrodynamic simulation confirms that such an ultra-narrow jet undergoes lateral spreading, more accurate size and lightcurve predictions will require fully coupled radiation--hydrodynamics modeling (see Appendix~\ref{sec:twoshell}).
Early high-cadence radio observations of GRB 221009A additionally constrain the magnetization of the fast shell to the order of $10^{-3}$, implying that the original outflow was at least moderately magnetized, consistent with a magnetic reconnection origin for the prompt emission \citep{Zhang11}.

\section{Discussion}

Our VLBI monitoring resolves the angular scale of GRB 221009A from 5 to 299 days after the burst, spanning the transition from the ultra-relativistic to the mildly relativistic regime.
The size evolution alone is consistent with a single power law $\theta_{\rm G}\propto t_{\rm obs}^{0.18}$. Joint modeling with the multi-wavelength afterglow, however, favors a more complex picture in which a narrow jet with an initial half-opening angle of $\simeq 0.01$--$0.03$~rad decelerates from a wind-like medium into an ISM-like environment.
Although the VLBI size data alone do not require a break in the power-law evolution, the coincidence of the inferred transition time with a bump in the 15-GHz lightcurve, together with theoretical expectations for narrow-jet dynamics, provides tentative evidence for lateral spreading contributing to accelerated angular growth after approximately 100 days.
The VLBI measurements provide direct geometric constraints on the jet opening angle, while the conversion to physical radius and speed depends on the adopted afterglow image model. These results demonstrate how resolved VLBI imaging, combined with multi-wavelength modeling, can probe the geometry, dynamics, and circumburst environment of GRB jets across their full relativistic evolution.

\begin{acknowledgements}
We appreciate the Fermi-GBM and Konus-Wind teams for sharing the data, and Hendrik Van Eerten for useful discussions.\\
\noindent \textbf{Funding.} This work is partially supported by the National Natural Science Foundation of China (12321003, 12393812, 12393813, 12622309, 12273113, 12233002, 12103055, 12041301, 12133007, 12103089), the National Key Research and Development Program of China (2021YFA0718500), the Strategic Priority Research Program of the Chinese Academy of Sciences (XDB0550400), and the International Partnership Program of Chinese Academy of Sciences for Grand Challenges (114332KYSB20210018). JJG acknowledges support from the Youth Innovation Promotion Association (2023331). HXG acknowledges support from Jiangsu Funding Program for Excellent Postdoctoral Talent. BL acknowledges support from the National Astronomical Science Data Center Young Data Scientist Program (NADC2023YDS-04). YL acknowledges support from the Natural Science Foundation of Jiangsu Province (BK20211000). YFH and TA also acknowledge support from the Xinjiang Tianchi Program. TA acknowledges support from the Shanghai Oriental Talent Project.\\
\noindent \textbf{Author contributions.} JJG, XFW, TA, BZ, YFH and ZGD wrote the majority of the manuscript; TA, YKZ, ALW and ZJX led the VLBI observations, processed the data, and performed data analyses; YQL and JY participated in VLBI observations and data processing discussions; JJG, HXG and JH performed numerical calculations and simulations; FX, TRS, and LR collected the multi-wavelength data; JJG, XFW, ZGD, YFH and BZ analyzed the results and provided theoretical models; HXG, JR, DX, YL, LL, and YW discussed the results; BL drew the schematic picture of the model; YKZ, JJG and CRH drew the rest of the plots. All authors contributed to the analysis and interpretation of the data and to the final version of the manuscript.\\
\noindent \textbf{Data availability.} Raw VLBA data can be found in the NRAO Data Archive under project codes TG015, BA161 (PI: Tao An), BA160 (PI: Pikky Atri), and BL307 (PI: Tanmoy Laskar). Calibrated VLBI visibility data and the reduction pipeline have been deposited in the Dryad database (\url{https://doi.org/10.5061/dryad.1vhhmgr9z}).\\
\noindent \textbf{Code availability.} The data calibration and imaging were performed using the VLBI pipeline (\url{https://github.com/ykzhang112233/VLBI-pipeline-for-GRB221009A.git}) at the China Square Kilometre Array (SKA) Regional Center. Other softwares used are \textsc{AIPS} (\url{https://www.aips.nrao.edu/index.shtml}), \textsc{ParselTongue} (\url{https://github.com/jive-vlbi/ParselTongue}), \textsc{GAMMA} (\url{https://github.com/eliotayache/GAMMA}), and \textsc{Emcee} (\url{https://github.com/dfm/emcee}).\\
\noindent \textbf{Competing interests.} The authors declare that they have no competing financial interests.\\
\noindent \textbf{Declaration of generative AI and AI-assisted technologies.} During the preparation of this work the authors used DeepSeek to check and debug the code used to generate the figures. After using this tool, the authors reviewed and edited the content as needed and take full responsibility for the content of the published article.
\end{acknowledgements}

\begin{table*}
\small
\caption{Summary of the VLBI observation campaign and fitting results
of \tar.}
\begin{center}
\setlength{\tabcolsep}{3.0pt}
\begin{tabular}{lcccccccccc}
\hline\hline
Code  & Date & $\Delta T$ & Freq.  & Beam  &  Noise  & $S_{\rm peak}$ & $S_{\nu}$ & Size & $\beta_{\rm app}$  \\
      & &  (days) & (GHz) & (mas$\times$mas, $\degr$) & (mJy/b) & (mJy/b) & (mJy) & (mas) & ($c$) \\
(1) & (2) & (3) & (4) & (5) & (6) & (7) & (8)  & (9) & (10) \\ \hline
TG015   & 2022-10-14 &  5 & 15.2 & 1.9$\times$0.5, $-$16.7                    & 0.048   & 0.842   & 0.92$\pm$0.06 & 0.274$\pm$0.072 & 93.3$\pm$24.5 \\
BA161a  & 2022-10-18 &  9 & 15.2 & 2.3$\times$0.5, $-$12.9
   & 0.073  & 3.230 & 3.84$\pm$0.09 & 0.316$\pm$0.054 & 62.6$\pm$10.7 \\
BL307a  & 2022-10-22 &  12 &  8.4 & 2.0$\times$0.9, 2.0                       & 0.024  & 1.570   & 1.67$\pm$0.03 & 0.260$\pm$0.053 & 39.1$\pm$8.0 \\
BA161b  & 2022-10-28 &  18 & 15.2 & 1.3$\times$0.5, $-$6.2
   & 0.042  & 1.550 & 2.02$\pm$0.06 & 0.385$\pm$0.057 & 39.0$\pm$5.8 \\
BL307b  & 2022-10-30 &  21 &  8.4 & 3.1$\times$1.0, $-$15.1
   & 0.026  & 2.165 & 2.25$\pm$0.03 & 0.268$\pm$0.054 & 23.4$\pm$4.7 \\
BA161c  & 2022-11-04 &  26 & 15.2 & 1.3$\times$0.5, $-$4.9
   & 0.043  & 1.250 & 1.69$\pm$0.06 & 0.415$\pm$0.116 & 29.4$\pm$8.2 \\
BL307c  & 2022-11-14 &  36 &  8.4 & 2.6$\times$0.9, $-$1.4
   & 0.036  & 1.368 & 1.45$\pm$0.04 & 0.285$\pm$0.070 & 14.7$\pm$3.6 \\
BA160b  & 2022-11-22 &  44 & 15.2 & 1.3$\times$0.4, $-$8.6                    & 0.054  & 0.768 & 0.98$\pm$0.09 & 0.296$\pm$0.071 & 12.5$\pm$3.0 \\
BL307d  & 2022-12-12 &  64 &  8.4 & 2.2$\times$1.2, 8.2
   & 0.026  & 0.780 & 0.84$\pm$0.03 & 0.326$\pm$0.111 & 9.5$\pm$3.2 \\
BL307e  & 2023-01-26 & 109 &  8.4 & 2.9$\times$1.1, $-$15.8
   & 0.028  & 0.404 & 0.46$\pm$0.04 & 0.491$\pm$0.155 & 8.4$\pm$2.7 \\
BA160c  & 2023-01-31 & 114 & 15.2 & 1.3$\times$0.5, $-$0.6                    & 0.039  & 0.100 & 0.27$^{+0.06}_{-0.05}$ & 0.450$\pm$0.190 & 7.4$\pm$3.1 \\
BL307f  & 2023-04-29 & 202 &  8.4 & 2.6$\times$1.3, $-$7.9
   & 0.018  & 0.191 & 0.23$\pm$0.02 & 0.764$\pm$0.228 & 7.1$\pm$2.1 \\
BA160c1 & 2023-05-02 & 205 & 15.2 & 2.2$\times$0.5, $-$18.7                   & 0.035  & 0.084 & 0.22$\pm$0.10 & 1.584$\pm$0.865 & 14.5$\pm$7.9 \\
BA160d  & 2023-06-28 & 262 & 15.2 & 1.7$\times$0.5, $-$3.2
   & 0.038  & 0.104 & 0.21$\pm$0.09 & 1.332$\pm$0.825 & 9.5$\pm$5.9 \\
BL307g  & 2023-08-04 & 299 &  8.4 & 2.2$\times$1.1, 1.3
   & 0.020  & 0.082 & 0.12$\pm$0.03 & 0.899$\pm$0.425 & 5.6$\pm$2.7 \\
\hline
\end{tabular}
\end{center}
Notes: Col.~1 -- Project code and epoch label. Col.~2 -- Observation date. Col.~3 --  Time interval since the trigger $T_0$ = 2022 October 9 13:16:59.99 UTC. Col.~4 -- Observing frequency averaged over intermediate frequencies. Col.~5 -- CLEAN beam major axis, minor axis and position angle (from north to east). Col.~6 -- Image r.m.s.\ noise level. Col.~7 -- Peak flux density of the image. Col.~8 -- Flux density from the complex-visibility circular-Gaussian fit; BL307g uses the position-constrained local fit anchored to the BL307b astrometric position, and BA160c is the independent published value of Giarratana et al. \citep{Giarratana24}. Col.~9 -- Primary circular-Gaussian angular scale $\theta_{\rm G}$ and its continuous $1\sigma$ interval, which combines the statistical, station-amplitude and residual-coherence terms (see Appendix~\ref{sec:vlbi_error_structure}). Col.~10 -- The average apparent expansion speed calculated from the source size.
\label{tab:ObsSummary}
\end{table*}

\begin{figure*}
    \centering
    \includegraphics[width=0.32\textwidth]{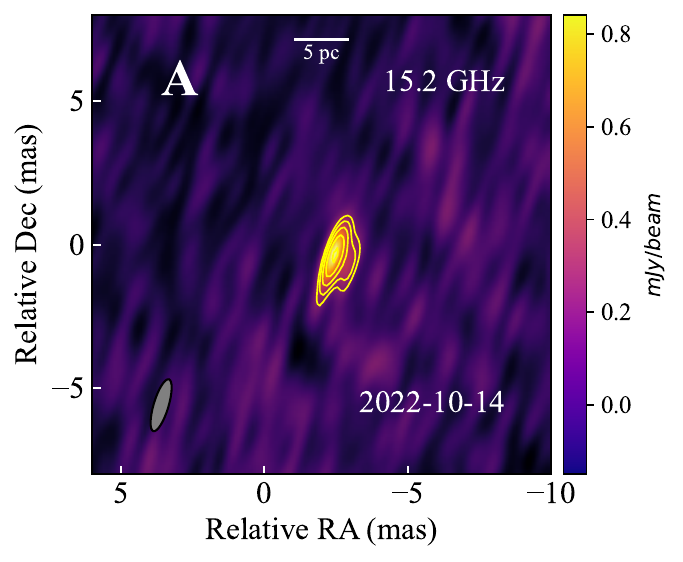}
    \includegraphics[width=0.32\textwidth]{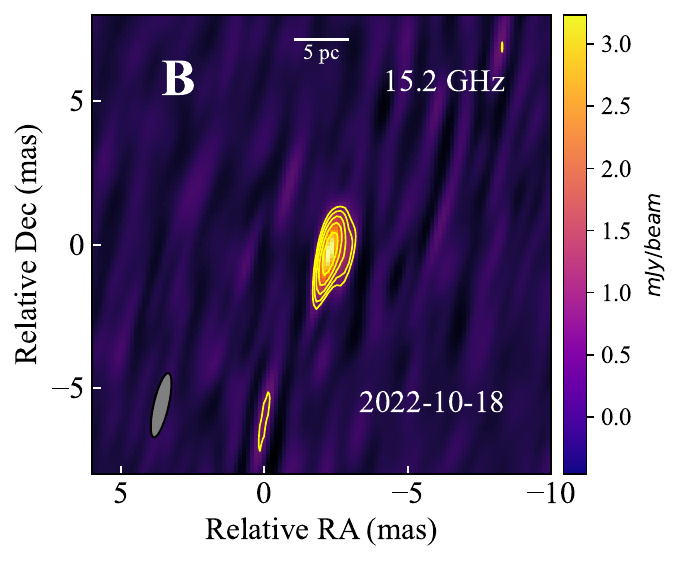}
    \includegraphics[width=0.32\textwidth]{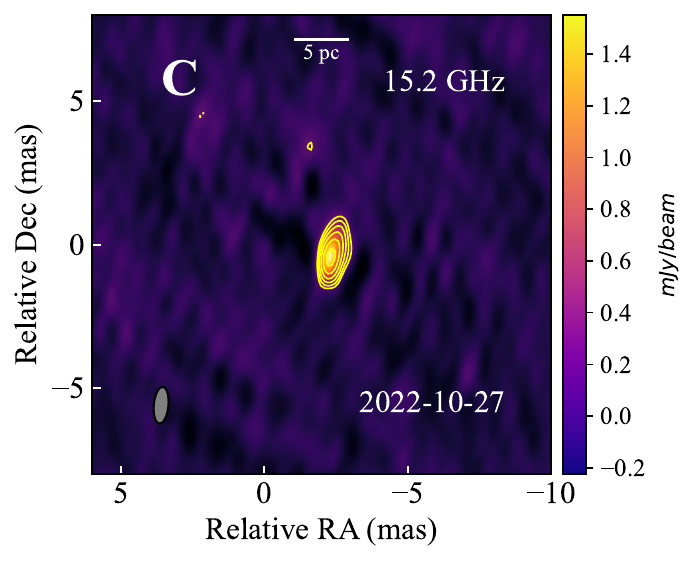} \\
    \includegraphics[width=0.32\textwidth]{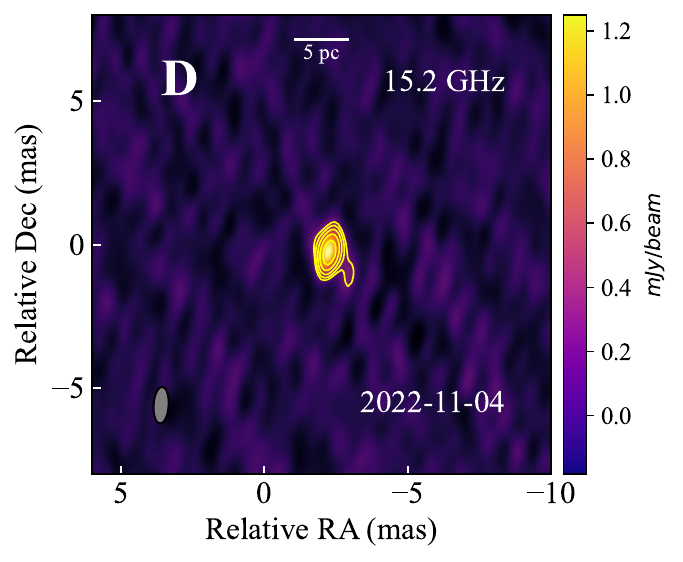}
    \includegraphics[width=0.32\textwidth]{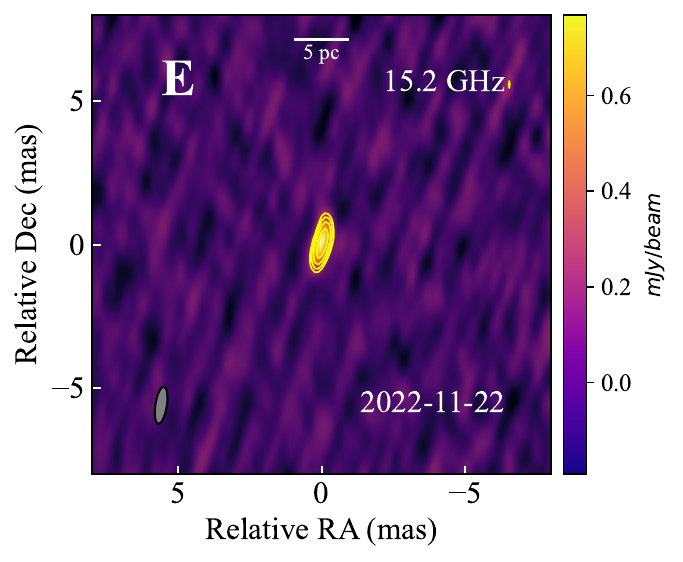}
    \includegraphics[width=0.32\textwidth]{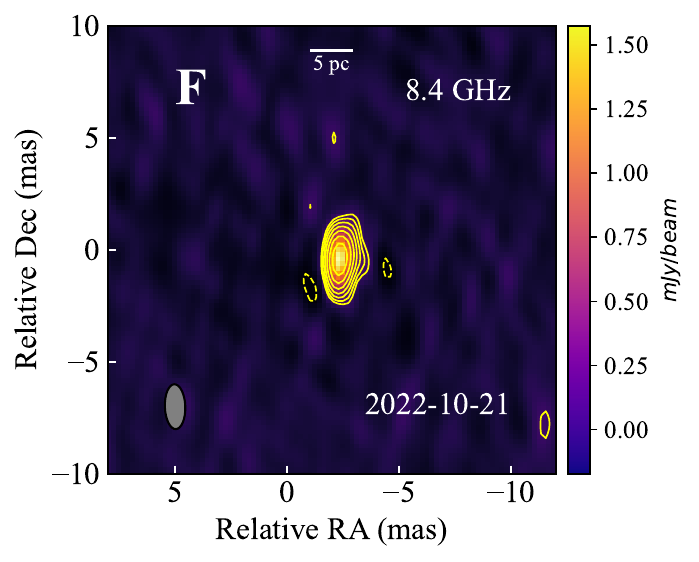}
    \\
    \includegraphics[width=0.32\textwidth]{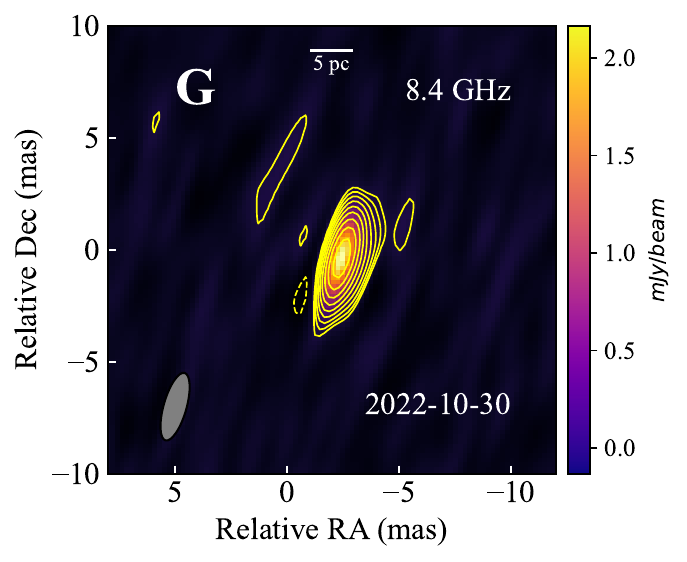}
    \includegraphics[width=0.32\textwidth]{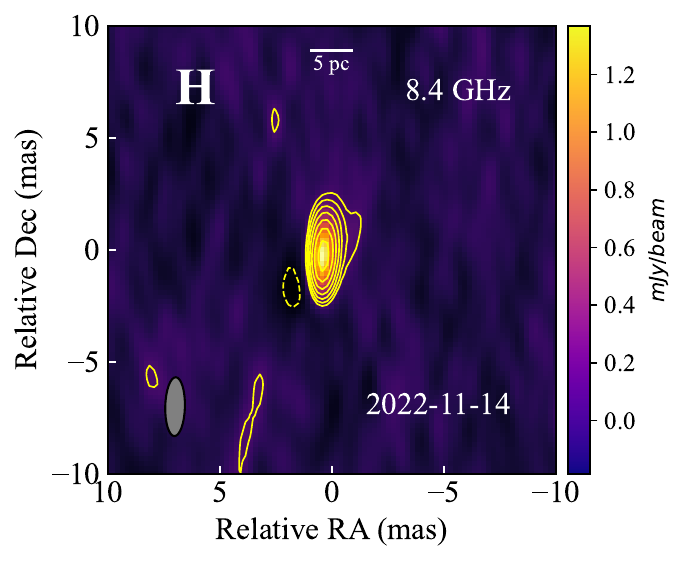}
    \includegraphics[width=0.32\textwidth]{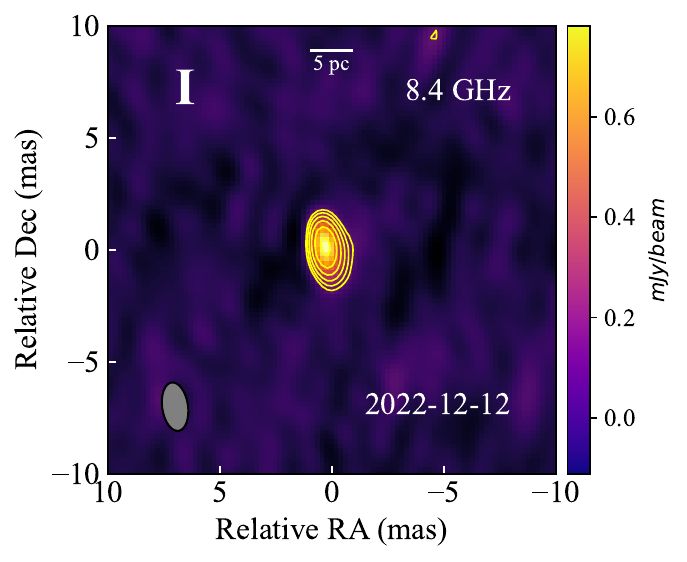}
    \\
    \includegraphics[width=0.32\textwidth]{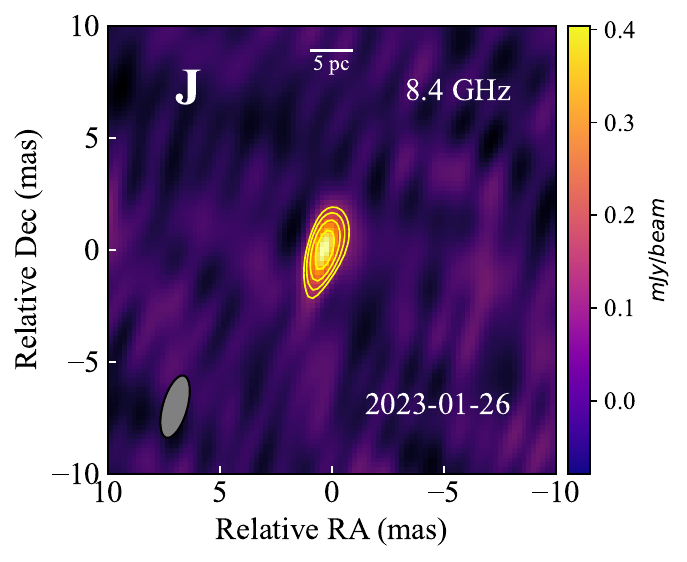}
    \includegraphics[width=0.32\textwidth]{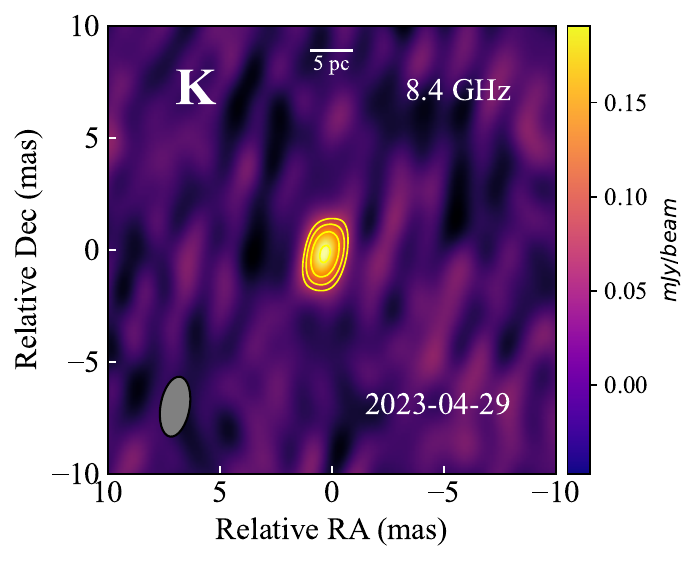}
    \includegraphics[width=0.32\textwidth]{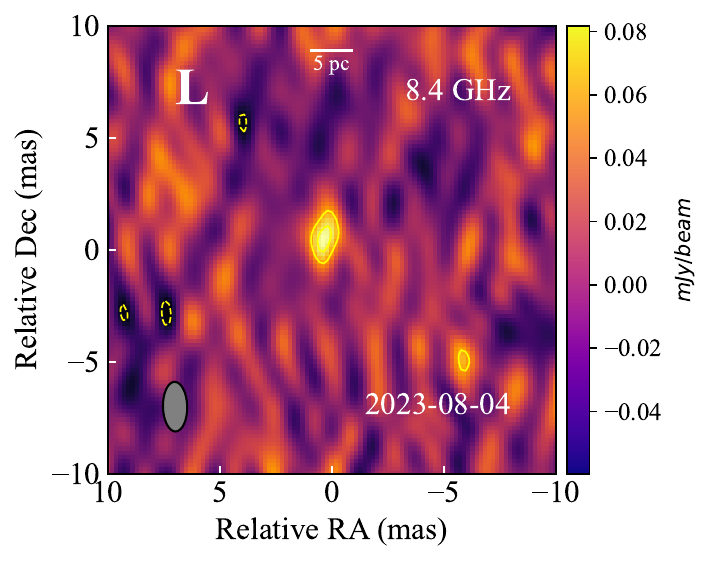}
    \caption{
    {\bf Naturally weighted VLBI images of GRB 221009A at 15.2 GHz and 8.4 GHz.}
    Epochs shown correspond to (\textbf{A}) 5 d, (\textbf{B}) 9 d, (\textbf{C}) 18 d, and (\textbf{D}) 26 d and (\textbf{E}) 44 d post-burst at 15 GHz, and (\textbf{F}) 12 d, (\textbf{G}) 21 d, (\textbf{H}) 36 d, (\textbf{I}) 64 d, (\textbf{J}) 109 d, (\textbf{K}) 202 d, and (\textbf{L}) 299 d at 8.4 GHz.
    The observation results are summarized in Table~\ref{tab:ObsSummary}, and further observational details are provided in Table~\ref{tab:obslog}. The beam shape is shown at the bottom left as a grey ellipse in each image. The image center has been shifted to the brightness peak. In each map image, the lowest contours represent 3 times the post-CLEAN r.m.s. noise, with successive contours increasing by a factor of $\sqrt{2}$.}
    \label{fig:Image}
\end{figure*}

\begin{figure*}
  \centering
  \includegraphics[width=0.83\textwidth]{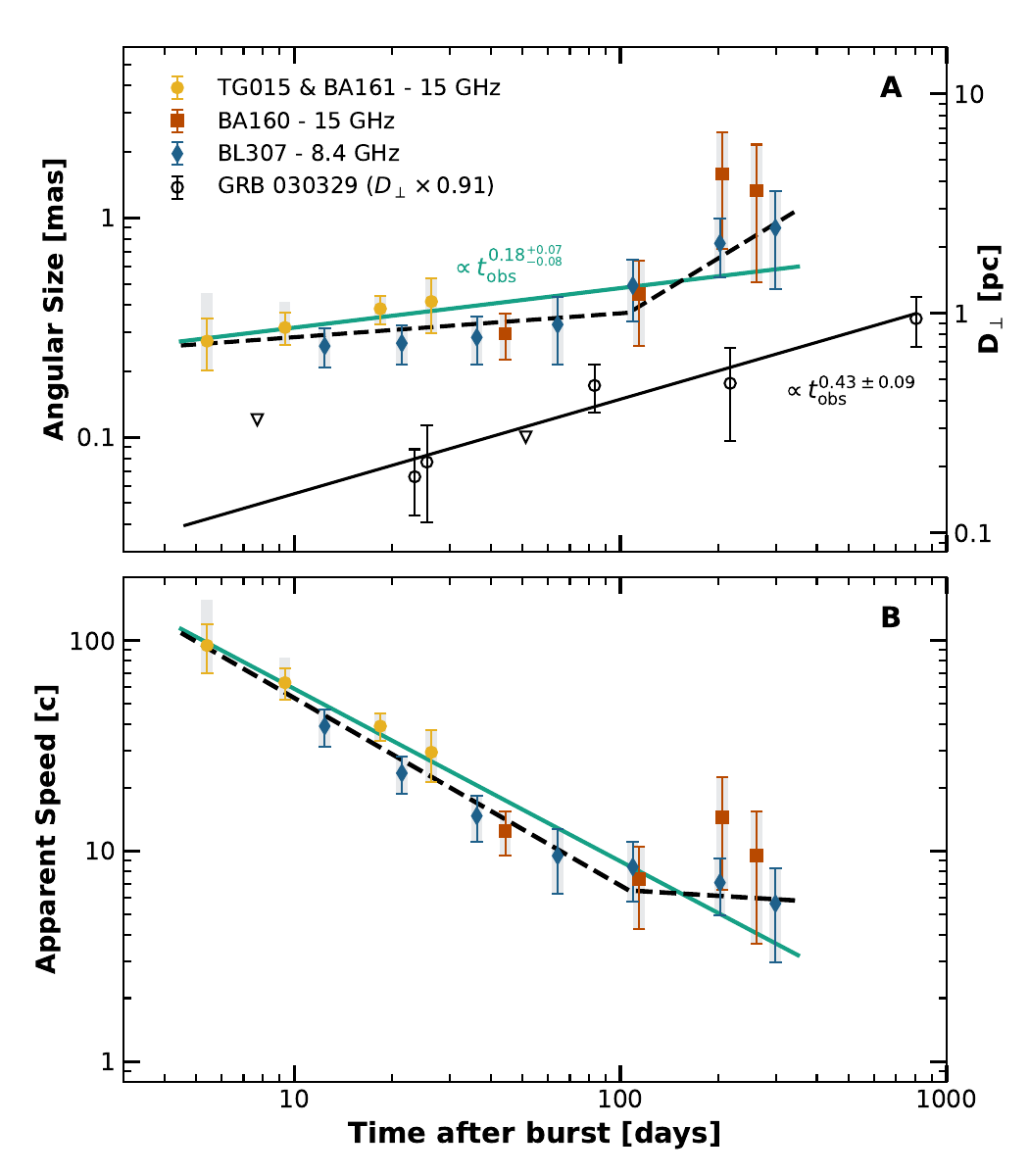}
  \caption{
  {\bf The dynamical evolution of GRB 221009A.}
  (\textbf{A}) Source size (angular diameter $D_{\perp}$) as a function of observer time (since $T_0$). The combined VLBA measurements are described by a single power law of
  $t_{\rm obs}^{0.18^{+0.07}_{-0.08}}$ (colored solid line) or by a broken power law (dashed line) with a transition at $t_{\rm tr} \approx 106$~day, and indices $0.11^{+0.09}_{-0.07}$
  and $0.91^{+0.82}_{-0.38}$ before and after the break, respectively. All uncertainties are quoted at the $1\sigma$ confidence level.
  As a comparison, the size evolution of GRB 030329 \citep{Taylor04,Taylor05,Pihlstrom07} at a comparable redshift ($z = 0.169$) is shifted to the same distance and fitted with a single power-law (black solid line), with a temporal index of $0.43 \pm 0.09$ ($1\sigma$).
  The shaded bars indicate the conservative error envelope of the measured sizes (see Appendix~\ref{sec:vlbi_error_structure}).
  (\textbf{B}) Apparent expansion speed (in units of the speed of light $c$) as a function of observer time.
  }
  \label{fig:size_evol}
\end{figure*}

\begin{figure*}
  \centering
  \includegraphics[width=\textwidth]{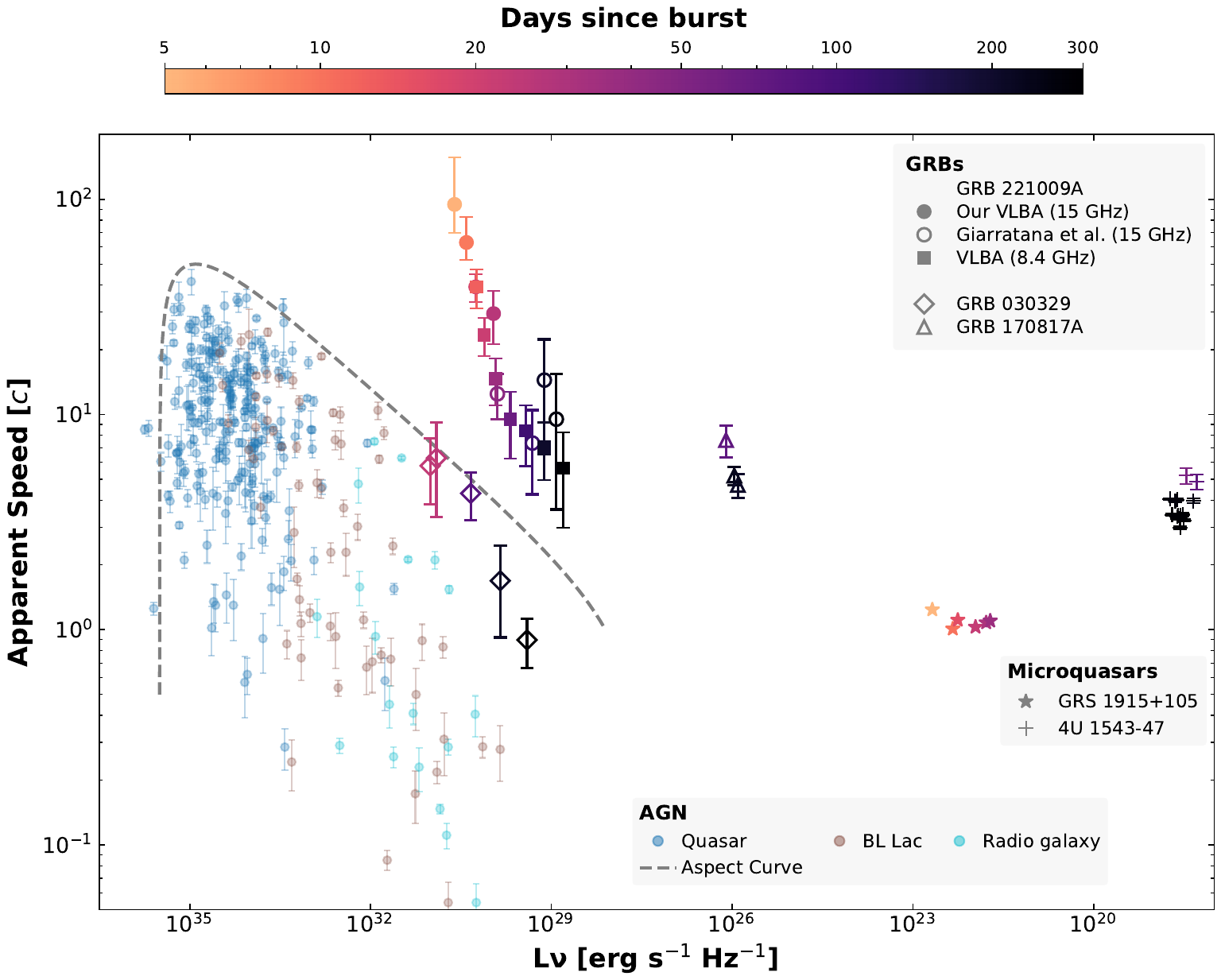}
  \caption{
  {\bf Apparent velocity versus spectral luminosity diagram for transient relativistic jets.}
  Data are shown for three GRBs: GRB 030329 (the first case with multi-epoch VLBI monitoring, \citealt{Taylor04,Taylor05,Pihlstrom07}), the off-axis-viewed jet GRB 170817A \citep{Mooley18,Ghirlanda19,Mooley22}, and GRB 221009A (the brightest burst, reported here).
  The AGN sample consists of 318 jets from the MOJAVE program \citep{Homan21}; for each source, the maximum apparent velocity across epochs and the average spectral luminosity are adopted.
  Microquasar data include two representative sources: GRS 1915+105, the first galactic superluminal source \citep{Mirabel94}, and the record holder 4U 1543--47 \citep{Fender25,ZhangX25}.
  The dashed curve marks the aspect line for an intrinsic Lorentz factor of 50 and a comoving-frame spectral luminosity of $\simeq 3 \times 10^{31}$ erg~s$^{-1}$~Hz$^{-1}$.
  }
  \label{fig:beta_Lnu}
\end{figure*}

\clearpage  

\bibliographystyle{PHEA}
\bibliography{GRB}

@ARTICLE{Asaki07,
       author = {{Asaki}, Yoshiharu and {Sudou}, Hiroshi and {Kono}, Yusuke and {Doi}, Akihiro and {Dodson}, Richard and {Pradel}, Nicolas and {Murata}, Yasuhiro and {Mochizuki}, Nanako and {Edwards}, Philip G. and {Sasao}, Tetsuo and {Fomalont}, Edward B.},
        title = "{Verification of the Effectiveness of VSOP-2 Phase Referencing with a Newly Developed Simulation Tool, ARIS}",
      journal = {\pasj},
         year = 2007,
        month = apr,
       volume = {59},
        pages = {397-418},
          doi = {10.1093/pasj/59.2.397},
archivePrefix = {arXiv},
       eprint = {0707.0558},
 primaryClass = {astro-ph},
       adsurl = {https://ui.adsabs.harvard.edu/abs/2007PASJ...59..397A}
}

@ARTICLE{Ayache22,
       author = {{Ayache}, Eliot H. and {van Eerten}, Hendrik J. and {Eardley}, Rupert W.},
        title = "{GAMMA: a new method for modelling relativistic hydrodynamics and non-thermal emission on a moving mesh}",
      journal = {\mnras},
         year = 2022,
        month = feb,
       volume = {510},
       number = {1},
        pages = {1315-1330},
          doi = {10.1093/mnras/stab3509},
archivePrefix = {arXiv},
       eprint = {2104.09397},
 primaryClass = {astro-ph.HE},
       adsurl = {https://ui.adsabs.harvard.edu/abs/2022MNRAS.510.1315A}
}

@ARTICLE{LST25,
       author = {{Abe}, K. and {Abe}, S. and {Abhishek}, A. and {Acero}, F. and {Aguasca-Cabot}, A. and {Agudo}, I. and {Alispach}, C. and {Ambrosino}, D. and {Ambrosino}, F. and {Antonelli}, L.~A. and {Aramo}, C. and {Arbet-Engels}, A. and {Arcaro}, C. and {Arnesen}, T.~T.~H. and {Asano}, K. and {Aubert}, P. and {Baktash}, A. and {Balbo}, M. and {Bamba}, A. and {Baquero Larriva}, A. and {Barres de Almeida}, U. and {Barrio}, J.~A. and {Barrios Jim{\'e}nez}, L. and {Batkovic}, I. and {Baxter}, J. and {Becerra Gonz{\'a}lez}, J. and {Bernardini}, E. and {Bernete}, J. and {Berti}, A. and {Bezshyiko}, I. and {Bigongiari}, C. and {Bissaldi}, E. and {Blanch}, O. and {Bonnoli}, G. and {Bordas}, P. and {Borkowski}, G. and {Brunelli}, G. and {Bulgarelli}, A. and {Bunse}, M. and {Burelli}, I. and {Burmistrov}, L. and {Cardillo}, M. and {Caroff}, S. and {Carosi}, A. and {Carraro}, R. and {Carrasco}, M.~S. and {Cassol}, F. and {Cerasole}, D. and {Ceribella}, G. and {Cervi{\~n}o Cort{\'\i}nez}, A. and {Chai}, Y. and {Cheng}, K. and {Chiavassa}, A. and {Chikawa}, M. and {Chon}, G. and {Chytka}, L. and {Cicciari}, G.~M. and {Cifuentes}, A. and {Contreras}, J.~L. and {Cortina}, J. and {Costantini}, H. and {Dalchenko}, M. and {Da Vela}, P. and {Dazzi}, F. and {De Angelis}, A. and {de Bony de Lavergne}, M. and {Del Burgo}, R. and {Delgado}, C. and {Delgado Mengual}, J. and {Dellaiera}, M. and {della Volpe}, D. and {De Lotto}, B. and {Del Peral}, L. and {de Menezes}, R. and {De Palma}, G. and {D{\'\i}az}, C. and {Di Piano}, A. and {Di Pierro}, F. and {Di Tria}, R. and {Di Venere}, L. and {Dominik}, R.~M. and {Dominis Prester}, D. and {Donini}, A. and {Dorner}, D. and {Doro}, M. and {Eisenberger}, L. and {Els{\"a}sser}, D. and {Emery}, G. and {Escudero}, J. and {Fallah Ramazani}, V. and {Ferrarotto}, F. and {Fiasson}, A. and {Foffano}, L. and {Fr{\'\i}as Garc{\'\i}a-Lago}, F. and {Fr{\"o}se}, S. and {Fukazawa}, Y. and {Gallozzi}, S. and {Garcia L{\'o}pez}, R. and {Garcia Soto}, S. and {Gasbarra}, C. and {Gasparrini}, D. and {Geyer}, D. and {Giesbrecht Paiva}, J. and {Giglietto}, N. and {Giordano}, F. and {Godinovic}, N. and {Gradetzke}, T. and {Grau}, R. and {Green}, D. and {Green}, J. and {Gunji}, S. and {G{\"u}nther}, P. and {Hackfeld}, J. and {Hadasch}, D. and {Hahn}, A. and {Hashizume}, M. and {Hassan}, T. and {Hayashi}, K. and {Heckmann}, L. and {Heller}, M. and {Herrera Llorente}, J. and {Hirotani}, K. and {Hoffmann}, D. and {Horns}, D. and {Houles}, J. and {Hrabovsky}, M. and {Hrupec}, D. and {Hui}, D. and {Iarlori}, M. and {Imazawa}, R. and {Inada}, T. and {Inome}, Y. and {Inoue}, S. and {Ioka}, K. and {Iori}, M. and {Itokawa}, T. and {Iuliano}, A. and {Jahanvi}, J. and {Jimenez Martinez}, I. and {Jimenez Quiles}, J. and {Jorge Rodrigo}, I. and {Jurysek}, J. and {Kagaya}, M. and {Kalashev}, O. and {Karas}, V. and {Katagiri}, H. and {Kerszberg}, D. and {Kiyomot}, T. and {Kobayashi}, Y. and {Kohri}, K. and {Kong}, A. and {Kornecki}, P. and {Kubo}, H. and {Kushida}, J. and {Lacave}, B. and {Lainez}, M. and {Lamanna}, G. and {Lamastra}, A. and {Lemoigne}, L. and {Linhoff}, M. and {Lombardi}, S. and {Longo}, F. and {L{\'o}pez-Coto}, R. and {L{\'o}pez-Moya}, M. and {L{\'o}pez-Oramas}, A. and {Loporchio}, S. and {Lorini}, A. and {Lozano Bahilo}, J. and {Lucarelli}, F. and {Luciani}, H. and {Luque-Escamilla}, P.~L. and {Majumdar}, P. and {Makariev}, M. and {Mallamaci}, M. and {Mandat}, D. and {Manganaro}, M. and {Maniadakis}, D.~K. and {Manic{\`o}}, G. and {Mannheim}, K. and {Marchesi}, S. and {Marini}, F. and {Mariotti}, M. and {Marquez}, P. and {Marsella}, G. and {Mart{\'\i}}, J. and {Martinez}, O. and {Mart{\'\i}nez}, G. and {Mart{\'\i}nez}, M. and {Mas-Aguilar}, A. and {Massa}, M. and {Maurin}, G. and {Mazin}, D. and {M{\'e}ndez-Gallego}, J. and {Menon}, S. and {Mestre Guillen}, E. and {Miceli}, D. and {Miener}, T. and {Miranda}, J.~M. and {Mirzoyan}, R. and {Mizote}, M.},
        title = "{GRB 221009A: Observations with LST-1 of CTAO and Implications for Structured Jets in Long Gamma-Ray Bursts}",
      journal = {\apjl},
         year = 2025,
        month = jul,
       volume = {988},
       number = {2},
          eid = {L42},
        pages = {L42},
          doi = {10.3847/2041-8213/ade4cf},
archivePrefix = {arXiv},
       eprint = {2507.03077},
 primaryClass = {astro-ph.HE},
       adsurl = {https://ui.adsabs.harvard.edu/abs/2025ApJ...988L..42A}
}

@ARTICLE{Tavani23,
       author = {{Tavani}, Marco and {Piano}, Giovanni and {Bulgarelli}, Andrea and {Foffano}, Luca and {Ursi}, Alessandro and {Verrecchia}, Francesco and {Pittori}, Carlotta and {Casentini}, Claudio and {Giuliani}, Andrea and {Longo}, Francesco and {Panebianco}, Gabriele and {Di Piano}, Ambra and {Baroncelli}, Leonardo and {Fioretti}, Valentina and {Parmiggiani}, Nicol{\`o} and {Argan}, Andrea and {Trois}, Alessio and {Vercellone}, Stefano and {Cardillo}, Martina and {Antonelli}, Lucio Angelo and {Barbiellini}, Guido and {Caraveo}, Patrizia and {Cattaneo}, Paolo W. and {Chen}, Andrew W. and {Costa}, Enrico and {Del Monte}, Ettore and {Di Cocco}, Guido and {Donnarumma}, Immacolata and {Evangelista}, Yuri and {Feroci}, Marco and {Gianotti}, Fulvio and {Labanti}, Claudio and {Lazzarotto}, Francesco and {Lipari}, Paolo and {Lucarelli}, Fabrizio and {Marisaldi}, Martino and {Mereghetti}, Sandro and {Morselli}, Aldo and {Pacciani}, Luigi and {Pellizzoni}, Alberto and {Perotti}, Francesco and {Picozza}, Piergiorgio and {Pilia}, Maura and {Rapisarda}, Massimo and {Rappoldi}, Andrea and {Rubini}, Alda and {Soffitta}, Paolo and {Trifoglio}, Massimo and {Vittorini}, Valerio and {D'Amico}, Fabio},
        title = "{AGILE Gamma-Ray Detection of the Exceptional GRB 221009A}",
      journal = {\apjl},
         year = 2023,
        month = oct,
       volume = {956},
       number = {1},
          eid = {L23},
        pages = {L23},
          doi = {10.3847/2041-8213/acfaff},
archivePrefix = {arXiv},
       eprint = {2309.10515},
 primaryClass = {astro-ph.HE},
       adsurl = {https://ui.adsabs.harvard.edu/abs/2023ApJ...956L..23T}
}

@ARTICLE{Kann23,
       author = {{Kann}, D.~A. and {Agayeva}, S. and {Aivazyan}, V. and {Alishov}, S. and {Andrade}, C.~M. and {Antier}, S. and {Baransky}, A. and {Bendjoya}, P. and {Benkhaldoun}, Z. and {Beradze}, S. and {Berezin}, D. and {Bo{\"e}r}, M. and {Broens}, E. and {Brunier}, S. and {Bulla}, M. and {Burkhonov}, O. and {Burns}, E. and {Chen}, Y. and {Chen}, Y.~P. and {Conti}, M. and {Coughlin}, M.~W. and {Cui}, W.~W. and {Daigne}, F. and {Delaveau}, B. and {Devillepoix}, H.~A.~R. and {Dietrich}, T. and {Dornic}, D. and {Dubois}, F. and {Ducoin}, J. -G. and {Durand}, E. and {Duverne}, P. -A. and {Eggenstein}, H. -B. and {Ehgamberdiev}, S. and {Fouad}, A. and {Freeberg}, M. and {Froebrich}, D. and {Ge}, M.~Y. and {Gervasoni}, S. and {Godunova}, V. and {Gokuldass}, P. and {Gurbanov}, E. and {Han}, D.~W. and {Hasanov}, E. and {Hello}, P. and {Hussenot-Desenonges}, T. and {Inasaridze}, R. and {Iskandar}, A. and {Ismailov}, N. and {Janati}, A. and {du Laz}, T. Jegou and {Jia}, S.~M. and {Karpov}, S. and {Kaeouach}, A. and {Kiendrebeogo}, R.~W. and {Klotz}, A. and {Kneip}, R. and {Kochiashvili}, N. and {Kunert}, N. and {Lekic}, A. and {Leonini}, S. and {Li}, C.~K. and {Li}, W. and {Li}, X.~B. and {Liao}, J.~Y. and {Logie}, L. and {Lu}, F.~J. and {Mao}, J. and {Marchais}, D. and {M{\'e}nard}, R. and {Morris}, D. and {Natsvlishvili}, R. and {Nedora}, V. and {Noonan}, K. and {Noysena}, K. and {Orange}, N.~B. and {Pang}, P.~T.~H. and {Peng}, H.~W. and {Pellouin}, C. and {Peloton}, J. and {Pradier}, T. and {Pyshna}, O. and {Rajabov}, Y. and {Rau}, S. and {Rinner}, C. and {Rivet}, J. -P. and {Romanov}, F.~D. and {Rosi}, P. and {Rupchandani}, V.~A. and {Serrau}, M. and {Shokry}, A. and {Simon}, A. and {Smith}, K. and {Sokoliuk}, O. and {Soliman}, M. and {Song}, L.~M. and {Takey}, A. and {Tillayev}, Y. and {Ramirez}, L.~M. Tinjaca and {e Melo}, I. Tosta and {Turpin}, D. and {de Ugarte Postigo}, A. and {Vanaverbeke}, S. and {Vasylenko}, V. and {Vernet}, D. and {Vidadi}, Z. and {Wang}, C. and {Wang}, J. and {Wang}, L.~T. and {Wang}, X.~F. and {Xiong}, S.~L. and {Xu}, Y.~P. and {Xue}, W.~C. and {Zeng}, X. and {Zhang}, S.~N. and {Zhao}, H.~S. and {Zhao}, X.~F.},
        title = "{GRANDMA and HXMT Observations of GRB 221009A: The Standard Luminosity Afterglow of a Hyperluminous Gamma-Ray Burst-In Gedenken an David Alexander Kann}",
      journal = {\apjl},
         year = 2023,
        month = may,
       volume = {948},
       number = {2},
          eid = {L12},
        pages = {L12},
          doi = {10.3847/2041-8213/acc8d0},
archivePrefix = {arXiv},
       eprint = {2302.06225},
 primaryClass = {astro-ph.HE},
       adsurl = {https://ui.adsabs.harvard.edu/abs/2023ApJ...948L..12K}
}

@ARTICLE{Ai21,
       author = {{Ai}, Shunke and {Zhang}, Bing},
        title = "{A mechanical model for magnetized relativistic blastwaves}",
      journal = {\mnras},
         year = 2021,
        month = oct,
       volume = {507},
       number = {2},
        pages = {1788-1794},
          doi = {10.1093/mnras/stab2000},
archivePrefix = {arXiv},
       eprint = {2104.06450},
 primaryClass = {astro-ph.HE},
       adsurl = {https://ui.adsabs.harvard.edu/abs/2021MNRAS.507.1788A}
}

@ARTICLE{Axelsson25,
       author = {{Axelsson}, M. and {Ajello}, M. and {Arimoto}, M. and {Baldini}, L. and {Ballet}, J. and {Baring}, M.~G. and {Bartolini}, C. and {Bastieri}, D. and {Becerra Gonzalez}, J. and {Bellazzini}, R. and {Berenji}, B. and {Bissaldi}, E. and {Blandford}, R.~D. and {Bonino}, R. and {Bruel}, P. and {Buson}, S. and {Cameron}, R.~A. and {Caputo}, R. and {Caraveo}, P.~A. and {Cavazzuti}, E. and {Cheung}, C.~C. and {Chiaro}, G. and {Cibrario}, N. and {Ciprini}, S. and {Cozzolongo}, G. and {Cristarella Orestano}, P. and {Crnogorcevic}, M. and {Cuoco}, A. and {Cutini}, S. and {D'Ammando}, F. and {De Gaetano}, S. and {Di Lalla}, N. and {Dinesh}, A. and {Di Tria}, R. and {Di Venere}, L. and {Dom{\'\i}nguez}, A. and {Fegan}, S.~J. and {Ferrara}, E.~C. and {Fiori}, A. and {Franckowiak}, A. and {Fukazawa}, Y. and {Funk}, S. and {Fusco}, P. and {Galanti}, G. and {Gargano}, F. and {Gasbarra}, C. and {Germani}, S. and {Giacchino}, F. and {Giglietto}, N. and {Giliberti}, M. and {Gill}, R. and {Giordano}, F. and {Giroletti}, M. and {Granot}, J. and {Green}, D. and {Grenier}, I.~A. and {Guiriec}, S. and {Gustafsson}, M. and {Hashizume}, M. and {Hays}, E. and {Hewitt}, J.~W. and {Horan}, D. and {Kayanoki}, T. and {Kuss}, M. and {Laviron}, A. and {Li}, J. and {Liodakis}, I. and {Longo}, F. and {Loparco}, F. and {Lorusso}, L. and {Lott}, B. and {Lovellette}, M.~N. and {Lubrano}, P. and {Maldera}, S. and {Malyshev}, D. and {Manfreda}, A. and {Mart{\'\i}-Devesa}, G. and {Martinelli}, R. and {Martinez Castellanos}, I. and {Mazziotta}, M.~N. and {McEnery}, J.~E. and {Mereu}, I. and {Meyer}, M. and {Michelson}, P.~F. and {Mirabal}, N. and {Mitthumsiri}, W. and {Mizuno}, T. and {Monti-Guarnieri}, P. and {Monzani}, M.~E. and {Morishita}, T. and {Morselli}, A. and {Moskalenko}, I.~V. and {Negro}, M. and {Niwa}, R. and {Omodei}, N. and {Orienti}, M. and {Orlando}, E. and {Paneque}, D. and {Panzarini}, G. and {Persic}, M. and {Pesce-Rollins}, M. and {Petrosian}, V. and {Pillera}, R. and {Piron}, F. and {Porter}, T.~A. and {Principe}, G. and {Racusin}, J.~L. and {Rain{\`o}}, S. and {Rando}, R. and {Rani}, B. and {Razzano}, M. and {Razzaque}, S. and {Reimer}, A. and {Reimer}, O. and {Ryde}, F. and {S{\'a}nchez-Conde}, M. and {Saz Parkinson}, P.~M. and {Serini}, D. and {Sgr{\`o}}, C. and {Sharma}, V. and {Siskind}, E.~J. and {Spandre}, G. and {Spinelli}, P. and {Suson}, D.~J. and {Tajima}, H. and {Tak}, D. and {Thayer}, J.~B. and {Torres}, D.~F. and {Valverde}, J. and {Zaharijas}, G. and {Fermi LAT Collaboration} and {Lesage}, S. and {Briggs}, M.~S. and {Burns}, E. and {Bala}, S. and {Bhat}, P.~N. and {Cleveland}, W.~H. and {Dalessi}, S. and {de Barra}, C. and {Gibby}, M. and {Giles}, M.~M. and {Hamburg}, R. and {Hristov}, B.~A. and {Kocevski}, D. and {Hui}, C.~M. and {Mailyan}, B. and {Malacaria}, C. and {McBreen}, S. and {Poolakkil}, S. and {Roberts}, O.~J. and {Scotton}, L. and {Veres}, P. and {von Kienlin}, A. and {Wilson-Hodge}, C.~A. and {Wood}, J. and {Fermi GBM Collaboration}},
        title = "{GRB 221009A: The B.O.A.T. Burst that Shines in Gamma Rays}",
      journal = {\apjs},
         year = 2025,
        month = mar,
       volume = {277},
       number = {1},
          eid = {24},
        pages = {24},
          doi = {10.3847/1538-4365/ada272},
archivePrefix = {arXiv},
       eprint = {2409.04580},
 primaryClass = {astro-ph.HE},
       adsurl = {https://ui.adsabs.harvard.edu/abs/2025ApJS..277...24A}
}

@ARTICLE{AnZH23,
       author = {{An}, Zheng-Hua and {Antier}, S. and {Bi}, Xing-Zi and {Bu}, Qing-Cui and {Cai}, Ce and {Cao}, Xue-Lei and {Camisasca}, Anna-Elisa and {Chang}, Zhi and {Chen}, Gang and {Chen}, Li and {Chen}, Tian-Xiang and {Chen}, Wen and {Chen}, Yi-Bao and {Chen}, Yong and {Chen}, Yu-Peng and {Coughlin}, Michael W. and {Cui}, Wei-Wei and {Dai}, Zi-Gao and {Hussenot-Desenonges}, T. and {Du}, Yan-Qi and {Du}, Yuan-Yuan and {Du}, Yun-Fei and {Fan}, Cheng-Cheng and {Frontera}, Filippo and {Gao}, He and {Gao}, Min and {Ge}, Ming-Yu and {Gong}, Ke and {Gu}, Yu-Dong and {Guan}, Ju and {Guo}, Dong-Ya and {Guo}, Zhi-Wei and {Guidorzi}, Cristiano and {Han}, Da-Wei and {He}, Jian-Jian and {He}, Jun-Wang and {Hou}, Dong-Jie and {Huang}, Yue and {Huo}, Jia and {Ji}, Zhen and {Jia}, Shu-Mei and {Jiang}, Wei-Chun and {Kann}, David Alexander and {Klotz}, A. and {Kong}, Ling-Da and {Lan}, Lin and {Li}, An and {Li}, Bing and {Li}, Chao-Yang and {Li}, Cheng-Kui and {Li}, Gang and {Li}, Mao-Shun and {Li}, Ti-Pei and {Li}, Wei and {Li}, Xiao-Bo and {Li}, Xin-Qiao and {Li}, Xu-Fang and {Li}, Yan-Guo and {Li}, Zheng-Wei and {Liang}, Jing and {Liang}, Xiao-Hua and {Liao}, Jin-Yuan and {Lin}, Lin and {Liu}, Cong-Zhan and {Liu}, He-Xin and {Liu}, Hong-Wei and {Liu}, Jia-Cong and {Liu}, Xiao-Jing and {Liu}, Ya-Qing and {Liu}, Yu-Rong and {Lu}, Fang-Jun and {Lu}, Hong and {Lu}, Xue-Feng and {Luo}, Qi and {Luo}, Tao and {Ma}, Bin-Yuan and {Ma}, Fu-Li and {Ma}, Rui-Can and {Ma}, Xiang and {Maccary}, Romain and {Mao}, Ji-Rong and {Meng}, Bin and {Nie}, Jian-Yin and {Orlandini}, Mauro and {Ou}, Ge and {Peng}, Jing-Qiang and {Peng}, Wen-Xi and {Qiao}, Rui and {Qu}, Jin-Lu and {Ren}, Xiao-Qin and {Shi}, Jing-Yan and {Shi}, Qi and {Song}, Li-Ming and {Song}, Xin-Ying and {Su}, Ju and {Sun}, Gong-Xing and {Sun}, Liang and {Sun}, Xi-Lei and {Tan}, Wen-Jun and {Tan}, Ying and {Tao}, Lian and {Tuo}, You-Li and {Turpin}, Damien and {Wang}, Jin-Zhou and {Wang}, Chen and {Wang}, Chen-Wei and {Wang}, Hong-Jun and {Wang}, Hui and {Wang}, Jin and {Wang}, Ling-Jun and {Wang}, Peng-Ju and {Wang}, Ping and {Wang}, Wen-Shuai and {Wang}, Xiang-Yu and {Wang}, Xi-Lu and {Wang}, Yu-Sa and {Wang}, Yue and {Wen}, Xiang-Yang and {Wu}, Bo-Bing and {Wu}, Bai-Yang and {Wu}, Hong and {Xiao}, Sheng-Hui and {Xiao}, Shuo and {Xiao}, Yun-Xiang and {Xie}, Sheng-Lun and {Xiong}, Shao-Lin and {Xiong}, Sen-Lin and {Xu}, Dong and {Xu}, He and {Xu}, Yan-Jun and {Xu}, Yan-Bing and {Xu}, Ying-Chen and {Xu}, Yu-Peng and {Xue}, Wang-Chen and {Yang}, Sheng and {Yang}, Yan-Ji and {Yang}, Zi-Xu and {Ye}, Wen-Tao and {Yi}, Qi-Bin and {Yi}, Shu-Xu and {Yin}, Qian-Qing and {You}, Yuan and {Yu}, Yun-Wei and {Yu}, Wei and {Yu}, Wen-Hui and {Zeng}, Ming and {Zhang}, Bing and {Zhang}, Bin-Bin and {Zhang}, Da-Li and {Zhang}, Fan and {Zhang}, Hong-Mei and {Zhang}, Juan and {Zhang}, Liang and {Zhang}, Peng and {Zhang}, Peng and {Zhang}, Shu and {Zhang}, Shuang-Nan and {Zhang}, Wan-Chang and {Zhang}, Xiao-Feng and {Zhang}, Xiao-Lu and {Zhang}, Yan-Qiu and {Zhang}, Yan-Ting and {Zhang}, Yi-Fei and {Zhang}, Yuan-Hang and {Zhang}, Zhen and {Zhao}, Guo-Ying and {Zhao}, Hai-Sheng and {Zhao}, Hong-Yu and {Zhao}, Qing-Xia and {Zhao}, Shu-Jie and {Zhao}, Xiao-Yun and {Zhao}, Xiao-Fan and {Zhao}, Yi and {Zheng}, Chao and {Zheng}, Shi-Jie and {Zhou}, Deng-Ke and {Zhou}, Xing and {Zhu}, Xiao-Cheng},
        title = "{Insight-HXMT and GECAM-C observations of the brightest-of-all-time GRB 221009A}",
      journal = {arXiv e-prints},
         year = 2023,
        month = mar,
          eid = {arXiv:2303.01203},
        pages = {arXiv:2303.01203},
          doi = {10.48550/arXiv.2303.01203},
archivePrefix = {arXiv},
       eprint = {2303.01203},
 primaryClass = {astro-ph.HE},
       adsurl = {https://ui.adsabs.harvard.edu/abs/2023arXiv230301203A}
}

@ARTICLE{An22,
       author = {{An}, Tao and {Wu}, Xiaocong and {Lao}, Baoqiang and {Guo}, Shaoguang and {Xu}, Zhijun and {Lv}, Weijia and {Zhang}, Yingkang and {Zhang}, Zhongli},
        title = "{Status and progress of China SKA Regional Centre prototype}",
      journal = {Sci. China. Phys. Mech. Astron.},
         year = 2022,
        month = dec,
       volume = {65},
       number = {12},
          eid = {129501},
        pages = {129501},
          doi = {10.1007/s11433-022-1981-8},
archivePrefix = {arXiv},
       eprint = {2206.13022},
 primaryClass = {astro-ph.IM},
       adsurl = {https://ui.adsabs.harvard.edu/abs/2022SCPMA..6529501A}
}

@ARTICLE{An19,
       author = {{An}, Tao and {Wu}, Xiang-Ping and {Hong}, Xiaoyu},
        title = "{SKA data take centre stage in China}",
      journal = {Nature Astronomy},
         year = 2019,
        month = nov,
       volume = {3},
        pages = {1030-1030},
          doi = {10.1038/s41550-019-0943-4},
       adsurl = {https://ui.adsabs.harvard.edu/abs/2019NatAs...3.1030A}
}

@ARTICLE{HAWC22,
       author = {{Ayala}, Hugo and {HAWC Collaboration}},
        title = "{GRB 221009A: Upper limits from HAWC 8 hours after trigger}",
      journal = {GRB Coordinates Network},
         year = 2022,
        month = oct,
       volume = {32683},
        pages = {1},
       adsurl = {https://ui.adsabs.harvard.edu/abs/2022GCN.32683....1A}
}

@ARTICLE{Barlow26,
       author = {{Barlow}, Roger and {Brazzale}, Alessandra Rosalba and {Volobouev}, Igor},
        title = "{Asymmetric errors}",
      journal = {Nuclear Instruments and Methods in Physics Research A},
         year = 2026,
        month = feb,
       volume = {1082},
          eid = {170857},
        pages = {170857},
          doi = {10.1016/j.nima.2025.170857},
archivePrefix = {arXiv},
       eprint = {2411.15499},
 primaryClass = {stat.ME},
       adsurl = {https://ui.adsabs.harvard.edu/abs/2026NIMPA108270857B}
}

@ARTICLE{Beloborodov06,
       author = {{Beloborodov}, Andrei M. and {Uhm}, Zuhngwhi Lucas},
        title = "{Mechanical Model for Relativistic Blast Waves}",
      journal = {\apjl},
         year = 2006,
        month = nov,
       volume = {651},
       number = {1},
        pages = {L1-L4},
          doi = {10.1086/508807},
archivePrefix = {arXiv},
       eprint = {astro-ph/0607641},
 primaryClass = {astro-ph},
       adsurl = {https://ui.adsabs.harvard.edu/abs/2006ApJ...651L...1B}
}

@ARTICLE{Blanchard24,
       author = {{Blanchard}, Peter K. and {Villar}, V. Ashley and {Chornock}, Ryan and {Laskar}, Tanmoy and {Li}, Yijia and {Leja}, Joel and {Pierel}, Justin and {Berger}, Edo and {Margutti}, Raffaella and {Alexander}, Kate D. and {Barnes}, Jennifer and {Cendes}, Yvette and {Eftekhari}, Tarraneh and {Kasen}, Daniel and {LeBaron}, Natalie and {Metzger}, Brian D. and {Muzerolle Page}, James and {Rest}, Armin and {Sears}, Huei and {Siegel}, Daniel M. and {Karthik Yadavalli}, S.},
        title = "{JWST Observations of the Extraordinary GRB 221009A Reveal an Ordinary Supernova Without Signs of $r$-Process Enrichment in a Low-Metallicity Galaxy}",
      journal = {Nature Astronomy},
        year = 2024,
        month = apr,
        volume = {},
        pages = {},
        eid = {arXiv:2308.14197},
        doi = {10.1038/s41550-024-02237-4},
archivePrefix = {arXiv},
       eprint = {2308.14197},
 primaryClass = {astro-ph.HE},
       adsurl = {https://ui.adsabs.harvard.edu/abs/2023arXiv230814197B}
}

@ARTICLE{Blandford76,
       author = {{Blandford}, R.~D. and {McKee}, C.~F.},
        title = "{Fluid dynamics of relativistic blast waves}",
      journal = {Physics of Fluids},
         year = 1976,
        month = aug,
       volume = {19},
        pages = {1130-1138},
          doi = {10.1063/1.861619},
       adsurl = {https://ui.adsabs.harvard.edu/abs/1976PhFl...19.1130B}
}

@ARTICLE{Blumenthal70,
       author = {{Blumenthal}, George R. and {Gould}, Robert J.},
        title = "{Bremsstrahlung, Synchrotron Radiation, and Compton Scattering of High-Energy Electrons Traversing Dilute Gases}",
      journal = {Rev. Mod. Phys.},
         year = 1970,
        month = jan,
       volume = {42},
       number = {2},
        pages = {237-271},
          doi = {10.1103/RevModPhys.42.237},
       adsurl = {https://ui.adsabs.harvard.edu/abs/1970RvMP...42..237B}
}

@ARTICLE{Bright23,
       author = {{Bright}, Joe S. and {Rhodes}, Lauren and {Farah}, Wael and {Fender}, Rob and {van der Horst}, Alexander J. and {Leung}, James K. and {Williams}, David R.~A. and {Anderson}, Gemma E. and {Atri}, Pikky and {DeBoer}, David R. and {Giarratana}, Stefano and {Green}, David A. and {Heywood}, Ian and {Lenc}, Emil and {Murphy}, Tara and {Pollak}, Alexander W. and {Premnath}, Pranav H. and {Scott}, Paul F. and {Sheikh}, Sofia Z. and {Siemion}, Andrew and {Titterington}, David J.},
        title = "{Precise measurements of self-absorbed rising reverse shock emission from gamma-ray burst 221009A}",
      journal = {Nature Astronomy},
         year = 2023,
        month = aug,
       volume = {7},
        pages = {986-995},
          doi = {10.1038/s41550-023-01997-9},
archivePrefix = {arXiv},
       eprint = {2303.13583},
 primaryClass = {astro-ph.HE},
       adsurl = {https://ui.adsabs.harvard.edu/abs/2023NatAs...7..986B}
}

@ARTICLE{Burns23,
       author = {{Burns}, Eric and {Svinkin}, Dmitry and {Fenimore}, Edward and {Kann}, D. Alexander and {Ag{\"u}{\'\i} Fern{\'a}ndez}, Jos{\'e} Feliciano and {Frederiks}, Dmitry and {Hamburg}, Rachel and {Lesage}, Stephen and {Temiraev}, Yuri and {Tsvetkova}, Anastasia and {Bissaldi}, Elisabetta and {Briggs}, Michael S. and {Dalessi}, Sarah and {Dunwoody}, Rachel and {Fletcher}, Cori and {Goldstein}, Adam and {Hui}, C. Michelle and {Hristov}, Boyan A. and {Kocevski}, Daniel and {Lysenko}, Alexandra L. and {Mailyan}, Bagrat and {Mangan}, Joseph and {McBreen}, Sheila and {Racusin}, Judith and {Ridnaia}, Anna and {Roberts}, Oliver J. and {Ulanov}, Mikhail and {Veres}, Peter and {Wilson-Hodge}, Colleen A. and {Wood}, Joshua},
        title = "{GRB 221009A: The Boat}",
      journal = {\apjl},
         year = 2023,
        month = mar,
       volume = {946},
       number = {1},
          eid = {L31},
        pages = {L31},
          doi = {10.3847/2041-8213/acc39c},
archivePrefix = {arXiv},
       eprint = {2302.14037},
 primaryClass = {astro-ph.HE},
       adsurl = {https://ui.adsabs.harvard.edu/abs/2023ApJ...946L..31B}
}

@ARTICLE{Castro-Tirado22,
       author = {{Castro-Tirado}, A.~J. and {Sanchez-Ramirez}, R. and {Hu}, Y. -D. and {Caballero-Garcia}, M.~D. and {Castro Tirado}, M.~A. and {Fernandez-Garcia}, E. and {Perez-Garcia}, I. and {Lombardi}, G. and {Pandey}, S.~B. and {Yang}, J. and {Zhang}, B. -B.},
        title = "{GRB 221009A: 10.4m GTC spectroscopic redshift confirmation}",
      journal = {GRB Coordinates Network},
         year = 2022,
        month = oct,
       volume = {32686},
        pages = {1},
       adsurl = {https://ui.adsabs.harvard.edu/abs/2022GCN.32686....1C}
}

@ARTICLE{Chrimes22,
       author = {{Chrimes}, A.~A. and {Gompertz}, B.~P. and {Kann}, D.~A. and {van Marle}, A.~J. and {Eldridge}, J.~J. and {Groot}, P.~J. and {Laskar}, T. and {Levan}, A.~J. and {Nicholl}, M. and {Stanway}, E.~R. and {Wiersema}, K.},
        title = "{Towards an understanding of long gamma-ray burst environments through circumstellar medium population synthesis predictions}",
      journal = {\mnras},
         year = 2022,
        month = sep,
       volume = {515},
       number = {2},
        pages = {2591-2611},
          doi = {10.1093/mnras/stac1796},
archivePrefix = {arXiv},
       eprint = {2206.13595},
 primaryClass = {astro-ph.HE},
       adsurl = {https://ui.adsabs.harvard.edu/abs/2022MNRAS.515.2591C}
}

@ARTICLE{Dai01,
       author = {{Dai}, Z.~G. and {Gou}, L.~J.},
        title = "{Gamma-Ray Burst Afterglows from Anisotropic Jets}",
      journal = {\apj},
         year = 2001,
        month = may,
       volume = {552},
       number = {1},
        pages = {72-80},
          doi = {10.1086/320463},
archivePrefix = {arXiv},
       eprint = {astro-ph/0010261},
 primaryClass = {astro-ph},
       adsurl = {https://ui.adsabs.harvard.edu/abs/2001ApJ...552...72D}
}

@ARTICLE{Cohen07,
       author = {{Cohen}, M.~H. and {Lister}, M.~L. and {Homan}, D.~C. and {Kadler}, M. and {Kellermann}, K.~I. and {Kovalev}, Y.~Y. and {Vermeulen}, R.~C.},
        title = "{Relativistic Beaming and the Intrinsic Properties of Extragalactic Radio Jets}",
      journal = {\apj},
         year = 2007,
        month = mar,
       volume = {658},
       number = {1},
        pages = {232-244},
          doi = {10.1086/511063},
archivePrefix = {arXiv},
       eprint = {astro-ph/0611642},
 primaryClass = {astro-ph},
       adsurl = {https://ui.adsabs.harvard.edu/abs/2007ApJ...658..232C}
}

@ARTICLE{deUgarte22,
       author = {{de Ugarte Postigo}, A. and {Izzo}, L. and {Pugliese}, G. and {Xu}, D. and {Schneider}, B. and {Fynbo}, J.~P.~U. and {Tanvir}, N.~R. and {Malesani}, D.~B. and {Saccardi}, A. and {Kann}, D.~A. and {Wiersema}, K. and {Gompertz}, B.~P. and {Thoene}, C.~C. and {Levan}, A.~J. and {Stargate Collaboration}},
        title = "{GRB 221009A: Redshift from X-shooter/VLT}",
      journal = {GRB Coordinates Network},
         year = 2022,
        month = oct,
       volume = {32648},
        pages = {1},
       adsurl = {https://ui.adsabs.harvard.edu/abs/2022GCN.32648....1D}
}

@ARTICLE{Eldridge06,
       author = {{Eldridge}, J.~J. and {Genet}, F. and {Daigne}, F. and {Mochkovitch}, R.},
        title = "{The circumstellar environment of Wolf-Rayet stars and gamma-ray burst afterglows}",
      journal = {\mnras},
         year = 2006,
        month = mar,
       volume = {367},
       number = {1},
        pages = {186-200},
          doi = {10.1111/j.1365-2966.2005.09938.x},
archivePrefix = {arXiv},
       eprint = {astro-ph/0509749},
 primaryClass = {astro-ph},
       adsurl = {https://ui.adsabs.harvard.edu/abs/2006MNRAS.367..186E}
}

@ARTICLE{Fender25,
       author = {{Fender}, R.~P. and {Motta}, S.~E.},
        title = "{The connection between the fastest astrophysical jets and the spin axis of their black hole}",
      journal = {Nature Astronomy},
         year = 2025,
        month = dec,
       volume = {9},
        pages = {1854-1859},
          doi = {10.1038/s41550-025-02665-w},
       adsurl = {https://ui.adsabs.harvard.edu/abs/2025NatAs...9.1854F}
}

@ARTICLE{Foffano24,
       author = {{Foffano}, Luca and {Tavani}, Marco and {Piano}, Giovanni},
        title = "{Theoretical Modeling of the Exceptional GRB 221009A Afterglow}",
      journal = {\apjl},
         year = 2024,
        month = oct,
       volume = {973},
       number = {2},
          eid = {L44},
        pages = {L44},
          doi = {10.3847/2041-8213/ad76a3},
archivePrefix = {arXiv},
       eprint = {2409.02859},
 primaryClass = {astro-ph.HE},
       adsurl = {https://ui.adsabs.harvard.edu/abs/2024ApJ...973L..44F}
}

@ARTICLE{Fraija24,
       author = {{Fraija}, Nissim and {Dainotti}, Maria G. and {Betancourt Kamenetskaia}, Boris and {Galv{\'a}n-G{\'a}mez}, Antonio and {Aguilar-Ruiz}, Edilberto},
        title = "{Microphysical parameter variation in gamma-ray burst stratified afterglows and closure relations: from sub-GeV to TeV observations}",
      journal = {\mnras},
         year = 2024,
        month = jan,
       volume = {527},
       number = {2},
        pages = {1884-1909},
          doi = {10.1093/mnras/stad3272},
archivePrefix = {arXiv},
       eprint = {2311.02859},
 primaryClass = {astro-ph.HE},
       adsurl = {https://ui.adsabs.harvard.edu/abs/2024MNRAS.527.1884F}
}

@ARTICLE{Frederiks23,
       author = {{Frederiks}, D. and {Svinkin}, D. and {Lysenko}, A.~L. and {Molkov}, S. and {Tsvetkova}, A. and {Ulanov}, M. and {Ridnaia}, A. and {Lutovinov}, A.~A. and {Lapshov}, I. and {Tkachenko}, A. and {Levin}, V.},
        title = "{Properties of the Extremely Energetic GRB 221009A from Konus-WIND and SRG/ART-XC Observations}",
      journal = {\apjl},
         year = 2023,
        month = may,
       volume = {949},
       number = {1},
          eid = {L7},
        pages = {L7},
          doi = {10.3847/2041-8213/acd1eb},
archivePrefix = {arXiv},
       eprint = {2302.13383},
 primaryClass = {astro-ph.HE},
       adsurl = {https://ui.adsabs.harvard.edu/abs/2023ApJ...949L...7F}
}

@ARTICLE{Fulton23,
       author = {{Fulton}, M.~D. and {Smartt}, S.~J. and {Rhodes}, L. and {Huber}, M.~E. and {Villar}, V.~A. and {Moore}, T. and {Srivastav}, S. and {Schultz}, A.~S.~B. and {Chambers}, K.~C. and {Izzo}, L. and {Hjorth}, J. and {Chen}, T. -W. and {Nicholl}, M. and {Foley}, R.~J. and {Rest}, A. and {Smith}, K.~W. and {Young}, D.~R. and {Sim}, S.~A. and {Bright}, J. and {Zenati}, Y. and {de Boer}, T. and {Bulger}, J. and {Fairlamb}, J. and {Gao}, H. and {Lin}, C. -C. and {Lowe}, T. and {Magnier}, E.~A. and {Smith}, I.~A. and {Wainscoat}, R. and {Coulter}, D.~A. and {Jones}, D.~O. and {Kilpatrick}, C.~D. and {McGill}, P. and {Ramirez-Ruiz}, E. and {Lee}, K. -S. and {Narayan}, G. and {Ramakrishnan}, V. and {Ridden-Harper}, R. and {Singh}, A. and {Wang}, Q. and {Kong}, A.~K.~H. and {Ngeow}, C. -C. and {Pan}, Y. -C. and {Yang}, S. and {Davis}, K.~W. and {Piro}, A.~L. and {Rojas-Bravo}, C. and {Sommer}, J. and {Yadavalli}, S.~K.},
        title = "{The Optical Light Curve of GRB 221009A: The Afterglow and the Emerging Supernova}",
      journal = {\apjl},
         year = 2023,
        month = mar,
       volume = {946},
       number = {1},
          eid = {L22},
        pages = {L22},
          doi = {10.3847/2041-8213/acc101},
archivePrefix = {arXiv},
       eprint = {2301.11170},
 primaryClass = {astro-ph.HE},
       adsurl = {https://ui.adsabs.harvard.edu/abs/2023ApJ...946L..22F}
}

@ARTICLE{Geng16,
       author = {{Geng}, J.~J. and {Wu}, X.~F. and {Huang}, Y.~F. and {Li}, L. and {Dai}, Z.~G.},
        title = "{Imprints of Electron-Positron Winds on the Multiwavelength Afterglows of Gamma-ray Bursts}",
      journal = {\apj},
         year = 2016,
        month = jul,
       volume = {825},
       number = {2},
          eid = {107},
        pages = {107},
          doi = {10.3847/0004-637X/825/2/107},
archivePrefix = {arXiv},
       eprint = {1605.01334},
 primaryClass = {astro-ph.HE},
       adsurl = {https://ui.adsabs.harvard.edu/abs/2016ApJ...825..107G}
}

@ARTICLE{Geng16b,
       author = {{Geng}, Jin-Jun and {Zhang}, Bing and {Kuiper}, Rolf},
        title = "{Propagation of Relativistic, Hydrodynamic, Intermittent Jets in a Rotating, Collapsing GRB Progenitor Star}",
      journal = {\apj},
         year = 2016,
        month = dec,
       volume = {833},
       number = {1},
          eid = {116},
        pages = {116},
          doi = {10.3847/1538-4357/833/1/116},
archivePrefix = {arXiv},
       eprint = {1608.01715},
 primaryClass = {astro-ph.HE},
       adsurl = {https://ui.adsabs.harvard.edu/abs/2016ApJ...833..116G}
}

@ARTICLE{Geng18,
       author = {{Geng}, Jin-Jun and {Huang}, Yong-Feng and {Wu}, Xue-Feng and {Zhang}, Bing and {Zong}, Hong-Shi},
        title = "{Low-energy Spectra of Gamma-Ray Bursts from Cooling Electrons}",
      journal = {\apjs},
         year = 2018,
        month = jan,
       volume = {234},
       number = {1},
          eid = {3},
        pages = {3},
          doi = {10.3847/1538-4365/aa9e84},
archivePrefix = {arXiv},
       eprint = {1709.05899},
 primaryClass = {astro-ph.HE},
       adsurl = {https://ui.adsabs.harvard.edu/abs/2018ApJS..234....3G}
}

@ARTICLE{Geng19,
       author = {{Geng}, Jin-Jun and {Zhang}, Bing and {K{\"o}lligan}, Anders and {Kuiper}, Rolf and {Huang}, Yong-Feng},
        title = "{Propagation of a Short GRB Jet in the Ejecta: Jet Launching Delay Time, Jet Structure, and GW170817/GRB 170817A}",
      journal = {\apjl},
         year = 2019,
        month = jun,
       volume = {877},
       number = {2},
          eid = {L40},
        pages = {L40},
          doi = {10.3847/2041-8213/ab224b},
archivePrefix = {arXiv},
       eprint = {1904.02326},
 primaryClass = {astro-ph.HE},
       adsurl = {https://ui.adsabs.harvard.edu/abs/2019ApJ...877L..40G}
}

@ARTICLE{Ghirlanda19,
       author = {{Ghirlanda}, G. and {Salafia}, O.~S. and {Paragi}, Z. and {Giroletti}, M. and {Yang}, J. and {Marcote}, B. and {Blanchard}, J. and {Agudo}, I. and {An}, T. and {Bernardini}, M.~G. and {Beswick}, R. and {Branchesi}, M. and {Campana}, S. and {Casadio}, C. and {Chassande-Mottin}, E. and {Colpi}, M. and {Covino}, S. and {D'Avanzo}, P. and {D'Elia}, V. and {Frey}, S. and {Gawronski}, M. and {Ghisellini}, G. and {Gurvits}, L.~I. and {Jonker}, P.~G. and {van Langevelde}, H.~J. and {Melandri}, A. and {Moldon}, J. and {Nava}, L. and {Perego}, A. and {Perez-Torres}, M.~A. and {Reynolds}, C. and {Salvaterra}, R. and {Tagliaferri}, G. and {Venturi}, T. and {Vergani}, S.~D. and {Zhang}, M.},
        title = "{Compact radio emission indicates a structured jet was produced by a binary neutron star merger}",
      journal = {Science},
         year = 2019,
        month = mar,
       volume = {363},
       number = {6430},
        pages = {968-971},
          doi = {10.1126/science.aau8815},
archivePrefix = {arXiv},
       eprint = {1808.00469},
 primaryClass = {astro-ph.HE},
       adsurl = {https://ui.adsabs.harvard.edu/abs/2019Sci...363..968G}
}

@ARTICLE{Giarratana24,
       author = {{Giarratana}, S. and {Salafia}, O.~S. and {Giroletti}, M. and {Ghirlanda}, G. and {Rhodes}, L. and {Atri}, P. and {Marcote}, B. and {Yang}, J. and {An}, T. and {Anderson}, G. and {Bright}, J.~S. and {Farah}, W. and {Fender}, R. and {Leung}, J.~K. and {Motta}, S.~E. and {P{\'e}rez-Torres}, M. and {van der Horst}, A.~J.},
        title = "{The expansion of the GRB 221009A afterglow}",
      journal = {\aap},
         year = 2024,
        month = oct,
       volume = {690},
          eid = {A74},
        pages = {A74},
          doi = {10.1051/0004-6361/202348524},
archivePrefix = {arXiv},
       eprint = {2311.05527},
 primaryClass = {astro-ph.HE},
       adsurl = {https://ui.adsabs.harvard.edu/abs/2024A&A...690A..74G}
}

@ARTICLE{Giannios09,
       author = {{Giannios}, Dimitrios and {Spitkovsky}, Anatoly},
        title = "{Signatures of a Maxwellian component in shock-accelerated electrons in GRBs}",
      journal = {\mnras},
         year = 2009,
        month = nov,
       volume = {400},
       number = {1},
        pages = {330-336},
          doi = {10.1111/j.1365-2966.2009.15454.x},
archivePrefix = {arXiv},
       eprint = {0905.1970},
 primaryClass = {astro-ph.HE},
       adsurl = {https://ui.adsabs.harvard.edu/abs/2009MNRAS.400..330G}
}

@ARTICLE{Gill23,
       author = {{Gill}, Ramandeep and {Granot}, Jonathan},
        title = "{GRB 221009A afterglow from a shallow angular structured jet}",
      journal = {\mnras},
         year = 2023,
        month = sep,
       volume = {524},
       number = {1},
        pages = {L78-L83},
          doi = {10.1093/mnrasl/slad075},
archivePrefix = {arXiv},
       eprint = {2304.14331},
 primaryClass = {astro-ph.HE},
       adsurl = {https://ui.adsabs.harvard.edu/abs/2023MNRAS.524L..78G}
}

@INPROCEEDINGS{Greisen03,
       author = {{Greisen}, E.~W.},
        title = "{AIPS, the VLA, and the VLBA}",
    booktitle = {Information Handling in Astronomy - Historical Vistas},
         year = 2003,
       editor = {{Heck}, Andr{\'e}},
       series = {Astrophysics and Space Science Library},
       volume = {285},
        month = mar,
        pages = {109},
          doi = {10.1007/0-306-48080-8_7},
       adsurl = {https://ui.adsabs.harvard.edu/abs/2003ASSL..285..109G}
}

@ARTICLE{Gottlieb18,
       author = {{Gottlieb}, Ore and {Nakar}, Ehud and {Piran}, Tsvi and {Hotokezaka}, Kenta},
        title = "{A cocoon shock breakout as the origin of the {\ensuremath{\gamma}}-ray emission in GW170817}",
      journal = {\mnras},
         year = 2018,
        month = sep,
       volume = {479},
       number = {1},
        pages = {588-600},
          doi = {10.1093/mnras/sty1462},
archivePrefix = {arXiv},
       eprint = {1710.05896},
 primaryClass = {astro-ph.HE},
       adsurl = {https://ui.adsabs.harvard.edu/abs/2018MNRAS.479..588G}
}

@ARTICLE{Gottlieb20,
       author = {{Gottlieb}, Ore and {Bromberg}, Omer and {Singh}, Chandra B. and {Nakar}, Ehud},
        title = "{The structure of weakly magnetized {\ensuremath{\gamma}}-ray burst jets}",
      journal = {\mnras},
         year = 2020,
        month = nov,
       volume = {498},
       number = {3},
        pages = {3320-3333},
          doi = {10.1093/mnras/staa2567},
archivePrefix = {arXiv},
       eprint = {2007.11590},
 primaryClass = {astro-ph.HE},
       adsurl = {https://ui.adsabs.harvard.edu/abs/2020MNRAS.498.3320G}
}

@ARTICLE{Gould67,
       author = {{Gould}, Robert J. and {Schr{\'e}der}, G{\'e}rard P.},
        title = "{Pair Production in Photon-Photon Collisions}",
      journal = {Phys. Rev.},
         year = 1967,
        month = mar,
       volume = {155},
       number = {5},
        pages = {1404-1407},
          doi = {10.1103/PhysRev.155.1404},
       adsurl = {https://ui.adsabs.harvard.edu/abs/1967PhRv..155.1404G}
}

@ARTICLE{Govreen24,
       author = {{Govreen-Segal}, Taya and {Nakar}, Ehud},
        title = "{The structure and evolution of relativistic jetted blast waves}",
      journal = {\mnras},
         year = 2024,
        month = jun,
       volume = {531},
       number = {1},
        pages = {1704-1720},
          doi = {10.1093/mnras/stae1224},
archivePrefix = {arXiv},
       eprint = {2311.09297},
 primaryClass = {astro-ph.HE},
       adsurl = {https://ui.adsabs.harvard.edu/abs/2024MNRAS.531.1704G}
}

@ARTICLE{Granot99,
       author = {{Granot}, Jonathan and {Piran}, Tsvi and {Sari}, Re'em},
        title = "{Images and Spectra from the Interior of a Relativistic Fireball}",
      journal = {\apj},
         year = 1999,
        month = mar,
       volume = {513},
       number = {2},
        pages = {679-689},
          doi = {10.1086/306884},
archivePrefix = {arXiv},
       eprint = {astro-ph/9806192},
 primaryClass = {astro-ph},
       adsurl = {https://ui.adsabs.harvard.edu/abs/1999ApJ...513..679G}
}

@ARTICLE{Granot05,
       author = {{Granot}, Jonathan and {Ramirez-Ruiz}, Enrico and {Loeb}, Abraham},
        title = "{Implications of the Measured Image Size for the Radio Afterglow of GRB 030329}",
      journal = {\apj},
         year = 2005,
        month = jan,
       volume = {618},
       number = {1},
        pages = {413-425},
          doi = {10.1086/425899},
archivePrefix = {arXiv},
       eprint = {astro-ph/0407182},
 primaryClass = {astro-ph},
       adsurl = {https://ui.adsabs.harvard.edu/abs/2005ApJ...618..413G}
}

@ARTICLE{Granot12,
       author = {{Granot}, Jonathan and {Piran}, Tsvi},
        title = "{On the lateral expansion of gamma-ray burst jets}",
      journal = {\mnras},
         year = 2012,
        month = mar,
       volume = {421},
       number = {1},
        pages = {570-587},
          doi = {10.1111/j.1365-2966.2011.20335.x},
archivePrefix = {arXiv},
       eprint = {1109.6468},
 primaryClass = {astro-ph.HE},
       adsurl = {https://ui.adsabs.harvard.edu/abs/2012MNRAS.421..570G}
}

@ARTICLE{Granot17,
       author = {{Granot}, Jonathan and {Guetta}, Dafne and {Gill}, Ramandeep},
        title = "{Lessons from the Short GRB 170817A: The First Gravitational-wave Detection of a Binary Neutron Star Merger}",
      journal = {\apjl},
         year = 2017,
        month = dec,
       volume = {850},
       number = {2},
          eid = {L24},
        pages = {L24},
          doi = {10.3847/2041-8213/aa991d},
archivePrefix = {arXiv},
       eprint = {1710.06407},
 primaryClass = {astro-ph.HE},
       adsurl = {https://ui.adsabs.harvard.edu/abs/2017ApJ...850L..24G}
}

@ARTICLE{Homan21,
       author = {{Homan}, D.~C. and {Cohen}, M.~H. and {Hovatta}, T. and {Kellermann}, K.~I. and {Kovalev}, Y.~Y. and {Lister}, M.~L. and {Popkov}, A.~V. and {Pushkarev}, A.~B. and {Ros}, E. and {Savolainen}, T.},
        title = "{MOJAVE. XIX. Brightness Temperatures and Intrinsic Properties of Blazar Jets}",
      journal = {\apj},
         year = 2021,
        month = dec,
       volume = {923},
       number = {1},
          eid = {67},
        pages = {67},
          doi = {10.3847/1538-4357/ac27af},
archivePrefix = {arXiv},
       eprint = {2109.04977},
 primaryClass = {astro-ph.HE},
       adsurl = {https://ui.adsabs.harvard.edu/abs/2021ApJ...923...67H}
}

@ARTICLE{Huang00,
       author = {{Huang}, Y.~F. and {Gou}, L.~J. and {Dai}, Z.~G. and {Lu}, T.},
        title = "{Overall Evolution of Jetted Gamma-Ray Burst Ejecta}",
      journal = {\apj},
         year = 2000,
        month = nov,
       volume = {543},
       number = {1},
        pages = {90-96},
          doi = {10.1086/317076},
archivePrefix = {arXiv},
       eprint = {astro-ph/9910493},
 primaryClass = {astro-ph},
       adsurl = {https://ui.adsabs.harvard.edu/abs/2000ApJ...543...90H}
}

@ARTICLE{Kobayashi00,
       author = {{Kobayashi}, Shiho and {Sari}, Re'em},
        title = "{Optical Flashes and Radio Flares in Gamma-Ray Burst Afterglow: Numerical Study}",
      journal = {\apj},
         year = 2000,
        month = oct,
       volume = {542},
       number = {2},
        pages = {819-828},
          doi = {10.1086/317021},
archivePrefix = {arXiv},
       eprint = {astro-ph/9910241},
 primaryClass = {astro-ph},
       adsurl = {https://ui.adsabs.harvard.edu/abs/2000ApJ...542..819K}
}

@ARTICLE{Kumar03,
       author = {{Kumar}, Pawan and {Granot}, Jonathan},
        title = "{The Evolution of a Structured Relativistic Jet and Gamma-Ray Burst Afterglow Light Curves}",
      journal = {\apj},
         year = 2003,
        month = jul,
       volume = {591},
       number = {2},
        pages = {1075-1085},
          doi = {10.1086/375186},
archivePrefix = {arXiv},
       eprint = {astro-ph/0303174},
 primaryClass = {astro-ph},
       adsurl = {https://ui.adsabs.harvard.edu/abs/2003ApJ...591.1075K}
}

@ARTICLE{Isravel23,
       author = {{Isravel}, Hebzibha and {B{\'e}gu{\'e}}, Damien and {Pe'er}, Asaf},
        title = "{Hybrid Emission Modeling of GRB 221009A: Shedding Light on TeV Emission Origins in Long GRBs}",
      journal = {\apj},
         year = 2023,
        month = oct,
       volume = {956},
       number = {1},
          eid = {12},
        pages = {12},
          doi = {10.3847/1538-4357/acefcd},
archivePrefix = {arXiv},
       eprint = {2308.06994},
 primaryClass = {astro-ph.HE},
       adsurl = {https://ui.adsabs.harvard.edu/abs/2023ApJ...956...12I}
}

@ARTICLE{Gao24,
       author = {{Gao}, Hao-Xuan and {Geng}, Jin-Jun and {Sun}, Tian-Rui and {Li}, Liang and {Huang}, Yong-Feng and {Wu}, Xue-Feng},
        title = "{Probing Thermal Electrons in Gamma-Ray Burst Afterglows}",
      journal = {\apj},
         year = 2024,
        month = aug,
       volume = {971},
       number = {1},
          eid = {81},
        pages = {81},
          doi = {10.3847/1538-4357/ad5443},
archivePrefix = {arXiv},
       eprint = {2311.06451},
 primaryClass = {astro-ph.HE},
       adsurl = {https://ui.adsabs.harvard.edu/abs/2024ApJ...971...81G}
}

@ARTICLE{Kathirgamaraju18,
       author = {{Kathirgamaraju}, Adithan and {Barniol Duran}, Rodolfo and {Giannios}, Dimitrios},
        title = "{Off-axis short GRBs from structured jets as counterparts to GW events}",
      journal = {\mnras},
         year = 2018,
        month = jan,
       volume = {473},
       number = {1},
        pages = {L121-L125},
          doi = {10.1093/mnrasl/slx175},
archivePrefix = {arXiv},
       eprint = {1708.07488},
 primaryClass = {astro-ph.HE},
       adsurl = {https://ui.adsabs.harvard.edu/abs/2018MNRAS.473L.121K}
}

@ARTICLE{Kong24,
       author = {{Kong}, De-Feng and {Wang}, Xiang-Gao and {Zheng}, WeiKang and {L{\"u}}, Hou-Jun and {Xin}, L.~P. and {Lin}, Da-Bin and {Cao}, Jia-Xin and {Lu}, Ming-Xuan and {Ren}, B. and {Vidal}, Edgar P. and {Wei}, J.~Y. and {Liang}, En-Wei and {Filippenko}, Alexei V.},
        title = "{GRB 221009A/SN 2022xiw: A Supernova Obscured by a Gamma-Ray Burst Afterglow?}",
      journal = {\apj},
         year = 2024,
        month = aug,
       volume = {971},
       number = {1},
          eid = {56},
        pages = {56},
          doi = {10.3847/1538-4357/ad5ce1},
archivePrefix = {arXiv},
       eprint = {2407.00639},
 primaryClass = {astro-ph.HE},
       adsurl = {https://ui.adsabs.harvard.edu/abs/2024ApJ...971...56K}
}

@ARTICLE{Kumar00,
       author = {{Kumar}, Pawan and {Piran}, Tsvi},
        title = "{Some Observational Consequences of Gamma-Ray Burst Shock Models}",
      journal = {\apj},
         year = 2000,
        month = mar,
       volume = {532},
       number = {1},
        pages = {286-293},
          doi = {10.1086/308537},
archivePrefix = {arXiv},
       eprint = {astro-ph/9906002},
 primaryClass = {astro-ph},
       adsurl = {https://ui.adsabs.harvard.edu/abs/2000ApJ...532..286K}
}

@ARTICLE{KumarZhang15,
       author = {{Kumar}, Pawan and {Zhang}, Bing},
        title = "{The physics of gamma-ray bursts \& relativistic jets}",
      journal = {\physrep},
         year = 2015,
        month = feb,
       volume = {561},
        pages = {1-109},
          doi = {10.1016/j.physrep.2014.09.008},
archivePrefix = {arXiv},
       eprint = {1410.0679},
 primaryClass = {astro-ph.HE},
       adsurl = {https://ui.adsabs.harvard.edu/abs/2015PhR...561....1K}
}

@ARTICLE{Lamb17,
       author = {{Lamb}, Gavin P. and {Kobayashi}, Shiho},
        title = "{Electromagnetic counterparts to structured jets from gravitational wave detected mergers}",
      journal = {\mnras},
         year = 2017,
        month = dec,
       volume = {472},
       number = {4},
        pages = {4953-4964},
          doi = {10.1093/mnras/stx2345},
archivePrefix = {arXiv},
       eprint = {1706.03000},
 primaryClass = {astro-ph.HE},
       adsurl = {https://ui.adsabs.harvard.edu/abs/2017MNRAS.472.4953L}
}

@ARTICLE{Laskar23,
       author = {{Laskar}, Tanmoy and {Alexander}, Kate D. and {Margutti}, Raffaella and {Eftekhari}, Tarraneh and {Chornock}, Ryan and {Berger}, Edo and {Cendes}, Yvette and {Duerr}, Anne and {Perley}, Daniel A. and {Ravasio}, Maria Edvige and {Yamazaki}, Ryo and {Ayache}, Eliot H. and {Barclay}, Thomas and {Duran}, Rodolfo Barniol and {Bhandari}, Shivani and {Brethauer}, Daniel and {Christy}, Collin T. and {Coppejans}, Deanne L. and {Duffell}, Paul and {Fong}, Wen-fai and {Gomboc}, Andreja and {Guidorzi}, Cristiano and {Kennea}, Jamie A. and {Kobayashi}, Shiho and {Levan}, Andrew and {Lobanov}, Andrei P. and {Metzger}, Brian D. and {Ros}, Eduardo and {Schroeder}, Genevieve and {Williams}, P.~K.~G.},
        title = "{The Radio to GeV Afterglow of GRB 221009A}",
      journal = {\apjl},
         year = 2023,
        month = mar,
       volume = {946},
       number = {1},
          eid = {L23},
        pages = {L23},
          doi = {10.3847/2041-8213/acbfad},
archivePrefix = {arXiv},
       eprint = {2302.04388},
 primaryClass = {astro-ph.HE},
       adsurl = {https://ui.adsabs.harvard.edu/abs/2023ApJ...946L..23L}
}

@ARTICLE{Lazzati12,
       author = {{Lazzati}, Davide and {Morsony}, Brian J. and {Blackwell}, Christopher H. and {Begelman}, Mitchell C.},
        title = "{Unifying the Zoo of Jet-driven Stellar Explosions}",
      journal = {\apj},
         year = 2012,
        month = may,
       volume = {750},
       number = {1},
          eid = {68},
        pages = {68},
          doi = {10.1088/0004-637X/750/1/68},
archivePrefix = {arXiv},
       eprint = {1111.0970},
 primaryClass = {astro-ph.HE},
       adsurl = {https://ui.adsabs.harvard.edu/abs/2012ApJ...750...68L}
}

@ARTICLE{Lazzati18,
       author = {{Lazzati}, Davide and {Perna}, Rosalba and {Morsony}, Brian J. and {Lopez-Camara}, Diego and {Cantiello}, Matteo and {Ciolfi}, Riccardo and {Giacomazzo}, Bruno and {Workman}, Jared C.},
        title = "{Late Time Afterglow Observations Reveal a Collimated Relativistic Jet in the Ejecta of the Binary Neutron Star Merger GW170817}",
      journal = {\prl},
         year = 2018,
        month = jun,
       volume = {120},
       number = {24},
          eid = {241103},
        pages = {241103},
          doi = {10.1103/PhysRevLett.120.241103},
archivePrefix = {arXiv},
       eprint = {1712.03237},
 primaryClass = {astro-ph.HE},
       adsurl = {https://ui.adsabs.harvard.edu/abs/2018PhRvL.120x1103L}
}

@ARTICLE{Lesage23,
       author = {{Lesage}, S. and {Veres}, P. and {Briggs}, M.~S. and {Goldstein}, A. and {Kocevski}, D. and {Burns}, E. and {Wilson-Hodge}, C.~A. and {Bhat}, P.~N. and {Huppenkothen}, D. and {Fryer}, C.~L. and {Hamburg}, R. and {Racusin}, J. and {Bissaldi}, E. and {Cleveland}, W.~H. and {Dalessi}, S. and {Fletcher}, C. and {Giles}, M.~M. and {Hristov}, B.~A. and {Hui}, C.~M. and {Mailyan}, B. and {Malacaria}, C. and {Poolakkil}, S. and {Roberts}, O.~J. and {von Kienlin}, A. and {Wood}, J. and {Ajello}, M. and {Arimoto}, M. and {Baldini}, L. and {Ballet}, J. and {Baring}, M.~G. and {Bastieri}, D. and {Gonzalez}, J. Becerra and {Bellazzini}, R. and {Bissaldi}, E. and {Blandford}, R.~D. and {Bonino}, R. and {Bruel}, P. and {Buson}, S. and {Cameron}, R.~A. and {Caputo}, R. and {Caraveo}, P.~A. and {Cavazzuti}, E. and {Chiaro}, G. and {Cibrario}, N. and {Ciprini}, S. and {Orestano}, P. Cristarella and {Crnogorcevic}, M. and {Cuoco}, A. and {Cutini}, S. and {D'Ammando}, F. and {De Gaetano}, S. and {Di Lalla}, N. and {Di Venere}, L. and {Dom{\'\i}nguez}, A. and {Fegan}, S.~J. and {Ferrara}, E.~C. and {Fleischhack}, H. and {Fukazawa}, Y. and {Funk}, S. and {Fusco}, P. and {Galanti}, G. and {Gammaldi}, V. and {Gargano}, F. and {Gasbarra}, C. and {Gasparrini}, D. and {Germani}, S. and {Giacchino}, F. and {Giglietto}, N. and {Gill}, R. and {Giroletti}, M. and {Granot}, J. and {Green}, D. and {Grenier}, I.~A. and {Guiriec}, S. and {Gustafsson}, M. and {Hays}, E. and {Hewitt}, J.~W. and {Horan}, D. and {Hou}, X. and {Kuss}, M. and {Latronico}, L. and {Laviron}, A. and {Lemoine-Goumard}, M. and {Li}, J. and {Liodakis}, I. and {Longo}, F. and {Loparco}, F. and {Lorusso}, L. and {Lovellette}, M.~N. and {Lubrano}, P. and {Maldera}, S. and {Manfreda}, A. and {Mart{\'\i}-Devesa}, G. and {Mazziotta}, M.~N. and {McEnery}, J.~E. and {Mereu}, I. and {Meyer}, M. and {Michelson}, P.~F. and {Mizuno}, T. and {Monzani}, M.~E. and {Morselli}, A. and {Moskalenko}, I.~V. and {Negro}, M. and {Nuss}, E. and {Omodei}, N. and {Orlando}, E. and {Ormes}, J.~F. and {Paneque}, D. and {Panzarini}, G. and {Persic}, M. and {Pesce-Rollins}, M. and {Pillera}, R. and {Piron}, F. and {Poon}, H. and {Porter}, T.~A. and {Principe}, G. and {Rain{\`o}}, S. and {Rando}, R. and {Rani}, B. and {Razzano}, M. and {Razzaque}, S. and {Reimer}, A. and {Reimer}, O. and {Ryde}, F. and {S{\'a}nchez-Conde}, M. and {Parkinson}, P.~M. Saz and {Scotton}, L. and {Serini}, D. and {Sgr{\`o}}, C. and {Sharma}, V. and {Siskind}, E.~J. and {Spandre}, G. and {Spinelli}, P. and {Tajima}, H. and {Torres}, D.~F. and {Valverde}, J. and {Venters}, T. and {Wadiasingh}, Z. and {Wood}, K. and {Zaharijas}, G.},
        title = "{Fermi-GBM Discovery of GRB 221009A: An Extraordinarily Bright GRB from Onset to Afterglow}",
      journal = {\apjl},
         year = 2023,
        month = aug,
       volume = {952},
       number = {2},
          eid = {L42},
        pages = {L42},
          doi = {10.3847/2041-8213/ace5b4},
archivePrefix = {arXiv},
       eprint = {2303.14172},
 primaryClass = {astro-ph.HE},
       adsurl = {https://ui.adsabs.harvard.edu/abs/2023ApJ...952L..42L}
}

@ARTICLE{Levan23,
       author = {{Levan}, A.~J. and {Lamb}, G.~P. and {Schneider}, B. and {Hjorth}, J. and {Zafar}, T. and {de Ugarte Postigo}, A. and {Sargent}, B. and {Mullally}, S.~E. and {Izzo}, L. and {D'Avanzo}, P. and {Burns}, E. and {Ag{\"u}{\'\i} Fern{\'a}ndez}, J.~F. and {Barclay}, T. and {Bernardini}, M.~G. and {Bhirombhakdi}, K. and {Bremer}, M. and {Brivio}, R. and {Campana}, S. and {Chrimes}, A.~A. and {D'Elia}, V. and {Della Valle}, M. and {De Pasquale}, M. and {Ferro}, M. and {Fong}, W. and {Fruchter}, A.~S. and {Fynbo}, J.~P.~U. and {Gaspari}, N. and {Gompertz}, B.~P. and {Hartmann}, D.~H. and {Hedges}, C.~L. and {Heintz}, K.~E. and {Hotokezaka}, K. and {Jakobsson}, P. and {Kann}, D.~A. and {Kennea}, J.~A. and {Laskar}, T. and {Le Floc'h}, E. and {Malesani}, D.~B. and {Melandri}, A. and {Metzger}, B.~D. and {Oates}, S.~R. and {Pian}, E. and {Piranomonte}, S. and {Pugliese}, G. and {Racusin}, J.~L. and {Rastinejad}, J.~C. and {Ravasio}, M.~E. and {Rossi}, A. and {Saccardi}, A. and {Salvaterra}, R. and {Sbarufatti}, B. and {Starling}, R.~L.~C. and {Tanvir}, N.~R. and {Th{\"o}ne}, C.~C. and {van der Horst}, A.~J. and {Vergani}, S.~D. and {Watson}, D. and {Wiersema}, K. and {Wijers}, R.~A.~M.~J. and {Xu}, Dong},
        title = "{The First JWST Spectrum of a GRB Afterglow: No Bright Supernova in Observations of the Brightest GRB of all Time, GRB 221009A}",
      journal = {\apjl},
         year = 2023,
        month = mar,
       volume = {946},
       number = {1},
          eid = {L28},
        pages = {L28},
          doi = {10.3847/2041-8213/acc2c1},
archivePrefix = {arXiv},
       eprint = {2302.07761},
 primaryClass = {astro-ph.HE},
       adsurl = {https://ui.adsabs.harvard.edu/abs/2023ApJ...946L..28L}
}

@ARTICLE{LHAASO23,
       author = {{LHAASO Collaboration} and {Cao}, Z. and {Aharonian}, F. and {An}, Q. and {Axikegu}, A. and {Bai}, L.~X. and {Bai}, Y.~X. and {Bao}, Y.~W. and {Bastieri}, D. and {Bi}, X.~J. and et al.},
        title = "{A tera-electron volt afterglow from a narrow jet in an extremely bright gamma-ray burst.}",
      journal = {Science},
         year = 2023,
        month = jun,
       volume = {380},
       number = {6652},
        pages = {1390-1396},
          doi = {10.1126/science.adg9328},
archivePrefix = {arXiv},
       eprint = {2306.06372},
 primaryClass = {astro-ph.HE},
       adsurl = {https://ui.adsabs.harvard.edu/abs/2023Sci...380.1390L}
}

@ARTICLE{Marti08,
       author = {{Mart{\'\i}-Vidal}, I. and {Marcaide}, J.~M.},
        title = "{Spurious source generation in mapping from noisy phase-self-calibrated data}",
      journal = {\aap},
         year = 2008,
        month = mar,
       volume = {480},
       number = {1},
        pages = {289-295},
          doi = {10.1051/0004-6361:20078690},
archivePrefix = {arXiv},
       eprint = {0801.1272},
 primaryClass = {astro-ph},
       adsurl = {https://ui.adsabs.harvard.edu/abs/2008A&A...480..289M}
}

@MISC{Maciel13,
       author = {Maciel, Walter J.},
       title = {General Overview of the Interstellar Medium},
       howpublished = {In: Astrophysics of the Interstellar Medium, Springer, New York, NY, pp. 1--15},
       year = {2013}
}

@ARTICLE{Meszaros98,
       author = {{M{\'e}sz{\'a}ros}, P. and {Rees}, M.~J. and {Wijers}, R.~A.~M.~J.},
        title = "{Viewing Angle and Environment Effects in Gamma-Ray Bursts: Sources of Afterglow Diversity}",
      journal = {\apj},
         year = 1998,
        month = may,
       volume = {499},
       number = {1},
        pages = {301-308},
          doi = {10.1086/305635},
archivePrefix = {arXiv},
       eprint = {astro-ph/9709273},
 primaryClass = {astro-ph},
       adsurl = {https://ui.adsabs.harvard.edu/abs/1998ApJ...499..301M}
}

@ARTICLE{Mirabel94,
       author = {{Mirabel}, I.~F. and {Rodr{\'\i}guez}, L.~F.},
        title = "{A superluminal source in the Galaxy}",
      journal = {\nat},
         year = 1994,
        month = sep,
       volume = {371},
       number = {6492},
        pages = {46-48},
          doi = {10.1038/371046a0},
       adsurl = {https://ui.adsabs.harvard.edu/abs/1994Natur.371...46M}
}

@ARTICLE{Mooley18,
       author = {{Mooley}, K.~P. and {Deller}, A.~T. and {Gottlieb}, O. and {Nakar}, E. and {Hallinan}, G. and {Bourke}, S. and {Frail}, D.~A. and {Horesh}, A. and {Corsi}, A. and {Hotokezaka}, K.},
        title = "{Superluminal motion of a relativistic jet in the neutron-star merger GW170817}",
      journal = {\nat},
         year = 2018,
        month = sep,
       volume = {561},
       number = {7723},
        pages = {355-359},
          doi = {10.1038/s41586-018-0486-3},
archivePrefix = {arXiv},
       eprint = {1806.09693},
 primaryClass = {astro-ph.HE},
       adsurl = {https://ui.adsabs.harvard.edu/abs/2018Natur.561..355M}
}

@ARTICLE{Mooley22,
       author = {{Mooley}, Kunal P. and {Anderson}, Jay and {Lu}, Wenbin},
        title = "{Optical superluminal motion measurement in the neutron-star merger GW170817}",
      journal = {\nat},
         year = 2022,
        month = oct,
       volume = {610},
       number = {7931},
        pages = {273-276},
          doi = {10.1038/s41586-022-05145-7},
archivePrefix = {arXiv},
       eprint = {2210.06568},
 primaryClass = {astro-ph.HE},
       adsurl = {https://ui.adsabs.harvard.edu/abs/2022Natur.610..273M}
}

@ARTICLE{OConnor23,
       author = {{O'Connor}, Brendan and {Troja}, Eleonora and {Ryan}, Geoffrey and {Beniamini}, Paz and {van Eerten}, Hendrik and {Granot}, Jonathan and {Dichiara}, Simone and {Ricci}, Roberto and {Lipunov}, Vladimir and {Gillanders}, James H. and {Gill}, Ramandeep and {Moss}, Michael and {Anand}, Shreya and {Andreoni}, Igor and {Becerra}, Rosa L. and {Buckley}, David A.~H. and {Butler}, Nathaniel R. and {Cenko}, Stephen B. and {Chasovnikov}, Aristarkh and {Durbak}, Joseph and {Francile}, Carlos and {Hammerstein}, Erica and {van der Horst}, Alexander J. and {Kasliwal}, Mansi M. and {Kouveliotou}, Chryssa and {Kutyrev}, Alexander S. and {Lee}, William H. and {Srinivasaragavan}, Gokul P. and {Topolev}, Vladislav and {Watson}, Alan M. and {Yang}, Yuhan and {Zhirkov}, Kirill},
        title = "{A structured jet explains the extreme GRB 221009A}",
      journal = {Sci. Adv.},
         year = 2023,
        month = jun,
       volume = {9},
       number = {23},
          eid = {eadi1405},
        pages = {eadi1405},
          doi = {10.1126/sciadv.adi1405},
archivePrefix = {arXiv},
       eprint = {2302.07906},
 primaryClass = {astro-ph.HE},
       adsurl = {https://ui.adsabs.harvard.edu/abs/2023SciA....9I1405O}
}

@ARTICLE{Panaitescu99,
       author = {{Panaitescu}, A. and {M{\'e}sz{\'a}ros}, P.},
        title = "{Dynamical Evolution, Light Curves, and Spectra of Spherical and Collimated Gamma-Ray Burst Remnants}",
      journal = {\apj},
         year = 1999,
        month = dec,
       volume = {526},
       number = {2},
        pages = {707-715},
          doi = {10.1086/308005},
archivePrefix = {arXiv},
       eprint = {astro-ph/9806016},
 primaryClass = {astro-ph},
       adsurl = {https://ui.adsabs.harvard.edu/abs/1999ApJ...526..707P}
}

@ARTICLE{Peer12,
       author = {{Pe'er}, Asaf},
        title = "{Dynamical Model of an Expanding Shell}",
      journal = {\apjl},
         year = 2012,
        month = jun,
       volume = {752},
       number = {1},
          eid = {L8},
        pages = {L8},
          doi = {10.1088/2041-8205/752/1/L8},
archivePrefix = {arXiv},
       eprint = {1203.5797},
 primaryClass = {astro-ph.HE},
       adsurl = {https://ui.adsabs.harvard.edu/abs/2012ApJ...752L...8P}
}

@ARTICLE{Pihlstrom07,
       author = {{Pihlstr{\"o}m}, Y.~M. and {Taylor}, G.~B. and {Granot}, J. and {Doeleman}, S.},
        title = "{Stirring the Embers: High-Sensitivity VLBI Observations of GRB 030329}",
      journal = {\apj},
         year = 2007,
        month = jul,
       volume = {664},
       number = {1},
        pages = {411-415},
          doi = {10.1086/518955},
archivePrefix = {arXiv},
       eprint = {0704.2085},
 primaryClass = {astro-ph},
       adsurl = {https://ui.adsabs.harvard.edu/abs/2007ApJ...664..411P}
}

@ARTICLE{Ravasio24,
       author = {{Ravasio}, Maria Edvige and {Salafia}, Om Sharan and {Oganesyan}, Gor and {Mei}, Alessio and {Ghirlanda}, Giancarlo and {Ascenzi}, Stefano and {Banerjee}, Biswajit and {Macera}, Samanta and {Branchesi}, Marica and {Jonker}, Peter G. and {Levan}, Andrew J. and {Malesani}, Daniele B. and {Mulrey}, Katharine B. and {Giuliani}, Andrea and {Celotti}, Annalisa and {Ghisellini}, Gabriele},
        title = "{A mega{\textendash}electron volt emission line in the spectrum of a gamma-ray burst}",
      journal = {Science},
         year = 2024,
        month = jul,
       volume = {385},
       number = {6707},
        pages = {452-455},
          doi = {10.1126/science.adj3638},
archivePrefix = {arXiv},
       eprint = {2303.16223},
 primaryClass = {astro-ph.HE},
       adsurl = {https://ui.adsabs.harvard.edu/abs/2024Sci...385..452R}
}

@ARTICLE{Ren23,
       author = {{Ren}, Jia and {Wang}, Yun and {Zhang}, Lu-Lu and {Dai}, Zi-Gao},
        title = "{The Possibility of Modeling the Very High Energy Afterglow of GRB 221009A in a Wind Environment}",
      journal = {\apj},
         year = 2023,
        month = apr,
       volume = {947},
       number = {2},
          eid = {53},
        pages = {53},
          doi = {10.3847/1538-4357/acc57d},
archivePrefix = {arXiv},
       eprint = {2210.10673},
 primaryClass = {astro-ph.HE},
       adsurl = {https://ui.adsabs.harvard.edu/abs/2023ApJ...947...53R}
}

@ARTICLE{Ren24,
       author = {{Ren}, Jia and {Wang}, Yun and {Dai}, Zi-Gao},
        title = "{Jet Structure and Burst Environment of GRB 221009A}",
      journal = {\apj},
         year = 2024,
        month = feb,
       volume = {962},
       number = {2},
          eid = {115},
        pages = {115},
          doi = {10.3847/1538-4357/ad1bcd},
archivePrefix = {arXiv},
       eprint = {2310.15886},
 primaryClass = {astro-ph.HE},
       adsurl = {https://ui.adsabs.harvard.edu/abs/2024ApJ...962..115R}
}

@ARTICLE{Rhoads99,
       author = {{Rhoads}, James E.},
        title = "{The Dynamics and Light Curves of Beamed Gamma-Ray Burst Afterglows}",
      journal = {\apj},
         year = 1999,
        month = nov,
       volume = {525},
       number = {2},
        pages = {737-749},
          doi = {10.1086/307907},
archivePrefix = {arXiv},
       eprint = {astro-ph/9903399},
 primaryClass = {astro-ph},
       adsurl = {https://ui.adsabs.harvard.edu/abs/1999ApJ...525..737R}
}

@ARTICLE{Rhodes24,
       author = {{Rhodes}, L. and {van der Horst}, A.~J. and {Bright}, J.~S. and {Leung}, J.~K. and {Anderson}, G.~E. and {Fender}, R. and {Ag{\"u}{\'\i} Fernandez}, J.~F. and {Bremer}, M. and {Chandra}, P. and {Dobie}, D. and {Farah}, W. and {Giarratana}, S. and {Gourdji}, K. and {Green}, D.~A. and {Lenc}, E. and {Micha{\l}owski}, M.~J. and {Murphy}, T. and {Nayana}, A.~J. and {Pollak}, A.~W. and {Rowlinson}, A. and {Schussler}, F. and {Siemion}, A. and {Starling}, R.~L.~C. and {Scott}, P. and {Th{\"o}ne}, C.~C. and {Titterington}, D. and {de Ugarte Postigo}, A.},
        title = "{Rocking the BOAT: the ups and downs of the long-term radio light curve for GRB 221009A}",
      journal = {\mnras},
         year = 2024,
        month = oct,
       volume = {533},
       number = {4},
        pages = {4435-4449},
          doi = {10.1093/mnras/stae2050},
archivePrefix = {arXiv},
       eprint = {2408.16637},
 primaryClass = {astro-ph.HE},
       adsurl = {https://ui.adsabs.harvard.edu/abs/2024MNRAS.533.4435R}
}

@ARTICLE{Rossi02,
       author = {{Rossi}, Elena and {Lazzati}, Davide and {Rees}, Martin J.},
        title = "{Afterglow light curves, viewing angle and the jet structure of {\ensuremath{\gamma}}-ray bursts}",
      journal = {\mnras},
         year = 2002,
        month = jun,
       volume = {332},
       number = {4},
        pages = {945-950},
          doi = {10.1046/j.1365-8711.2002.05363.x},
archivePrefix = {arXiv},
       eprint = {astro-ph/0112083},
 primaryClass = {astro-ph},
       adsurl = {https://ui.adsabs.harvard.edu/abs/2002MNRAS.332..945R}
}

@BOOKLET{Rybicki79,
       author = {{Rybicki}, George B. and {Lightman}, Alan P.},
       title = "{Radiative Processes in Astrophysics}",
       howpublished = {Wiley-Interscience Publication, New York},
       year = 1979,
       note = {1st edn}
}

@ARTICLE{Ryan20,
       author = {{Ryan}, Geoffrey and {van Eerten}, Hendrik and {Piro}, Luigi and {Troja}, Eleonora},
        title = "{Gamma-Ray Burst Afterglows in the Multimessenger Era: Numerical Models and Closure Relations}",
      journal = {\apj},
         year = 2020,
        month = jun,
       volume = {896},
       number = {2},
          eid = {166},
        pages = {166},
          doi = {10.3847/1538-4357/ab93cf},
archivePrefix = {arXiv},
       eprint = {1909.11691},
 primaryClass = {astro-ph.HE},
       adsurl = {https://ui.adsabs.harvard.edu/abs/2020ApJ...896..166R}
}

@ARTICLE{Sari98,
       author = {{Sari}, Re'em and {Piran}, Tsvi and {Narayan}, Ramesh},
        title = "{Spectra and Light Curves of Gamma-Ray Burst Afterglows}",
      journal = {\apjl},
         year = 1998,
        month = apr,
       volume = {497},
       number = {1},
        pages = {L17-L20},
          doi = {10.1086/311269},
archivePrefix = {arXiv},
       eprint = {astro-ph/9712005},
 primaryClass = {astro-ph},
       adsurl = {https://ui.adsabs.harvard.edu/abs/1998ApJ...497L..17S}
}

@ARTICLE{Sari99,
       author = {{Sari}, Re'em and {Piran}, Tsvi and {Halpern}, J.~P.},
        title = "{Jets in Gamma-Ray Bursts}",
      journal = {\apjl},
         year = 1999,
        month = jul,
       volume = {519},
       number = {1},
        pages = {L17-L20},
          doi = {10.1086/312109},
archivePrefix = {arXiv},
       eprint = {astro-ph/9903339},
 primaryClass = {astro-ph},
       adsurl = {https://ui.adsabs.harvard.edu/abs/1999ApJ...519L..17S}
}

@ARTICLE{Sari01,
       author = {{Sari}, Re'em and {Esin}, Ann A.},
        title = "{On the Synchrotron Self-Compton Emission from Relativistic Shocks and Its Implications for Gamma-Ray Burst Afterglows}",
      journal = {\apj},
         year = 2001,
        month = feb,
       volume = {548},
       number = {2},
        pages = {787-799},
          doi = {10.1086/319003},
archivePrefix = {arXiv},
       eprint = {astro-ph/0005253},
 primaryClass = {astro-ph},
       adsurl = {https://ui.adsabs.harvard.edu/abs/2001ApJ...548..787S}
}

@ARTICLE{Sato23,
       author = {{Sato}, Yuri and {Murase}, Kohta and {Ohira}, Yutaka and {Yamazaki}, Ryo},
        title = "{Two-component jet model for multiwavelength afterglow emission of the extremely energetic burst GRB 221009A}",
      journal = {\mnras},
         year = 2023,
        month = jun,
       volume = {522},
       number = {1},
        pages = {L56-L60},
          doi = {10.1093/mnrasl/slad038},
archivePrefix = {arXiv},
       eprint = {2212.09266},
 primaryClass = {astro-ph.HE},
       adsurl = {https://ui.adsabs.harvard.edu/abs/2023MNRAS.522L..56S}
}

@ARTICLE{Schlafly11,
       author = {{Schlafly}, Edward F. and {Finkbeiner}, Douglas P.},
        title = "{Measuring Reddening with Sloan Digital Sky Survey Stellar Spectra and Recalibrating SFD}",
      journal = {\apj},
         year = 2011,
        month = aug,
       volume = {737},
       number = {2},
          eid = {103},
        pages = {103},
          doi = {10.1088/0004-637X/737/2/103},
archivePrefix = {arXiv},
       eprint = {1012.4804},
 primaryClass = {astro-ph.GA},
       adsurl = {https://ui.adsabs.harvard.edu/abs/2011ApJ...737..103S}
}

@ARTICLE{Pradel06,
       author = {{Pradel}, N. and {Charlot}, P. and {Lestrade}, J.-F.},
        title = "{Astrometric accuracy of phase-referenced observations with the VLBA and EVN}",
      journal = {\aap},
         year = 2006,
        month = jun,
       volume = {452},
       number = {3},
        pages = {1099-1106},
          doi = {10.1051/0004-6361:20053021},
archivePrefix = {arXiv},
       eprint = {astro-ph/0603015},
 primaryClass = {astro-ph},
       adsurl = {https://ui.adsabs.harvard.edu/abs/2006A&A...452.1099P}
}

@INPROCEEDINGS{Shepherd97,
       author = {{Shepherd}, M.~C.},
        title = "{Difmap: an Interactive Program for Synthesis Imaging}",
    booktitle = {Astronomical Data Analysis Software and Systems VI},
         year = 1997,
       editor = {{Hunt}, Gareth and {Payne}, Harry},
       series = {Astronomical Society of the Pacific Conference Series},
       volume = {125},
        month = jan,
        pages = {77},
       adsurl = {https://ui.adsabs.harvard.edu/abs/1997ASPC..125...77S}
}

@ARTICLE{Sironi15,
       author = {{Sironi}, L. and {Keshet}, U. and {Lemoine}, M.},
        title = "{Relativistic Shocks: Particle Acceleration and Magnetization}",
      journal = {\ssr},
         year = 2015,
        month = oct,
       volume = {191},
       number = {1-4},
        pages = {519-544},
          doi = {10.1007/s11214-015-0181-8},
archivePrefix = {arXiv},
       eprint = {1506.02034},
 primaryClass = {astro-ph.HE},
       adsurl = {https://ui.adsabs.harvard.edu/abs/2015SSRv..191..519S}
}

@ARTICLE{Sironi11,
       author = {{Sironi}, Lorenzo and {Spitkovsky}, Anatoly},
        title = "{Particle Acceleration in Relativistic Magnetized Collisionless Electron-Ion Shocks}",
      journal = {\apj},
         year = 2011,
        month = jan,
       volume = {726},
       number = {2},
          eid = {75},
        pages = {75},
          doi = {10.1088/0004-637X/726/2/75},
archivePrefix = {arXiv},
       eprint = {1009.0024},
 primaryClass = {astro-ph.HE},
       adsurl = {https://ui.adsabs.harvard.edu/abs/2011ApJ...726...75S}
}

@ARTICLE{Taylor04,
       author = {{Taylor}, G.~B. and {Frail}, D.~A. and {Berger}, E. and {Kulkarni}, S.~R.},
        title = "{The Angular Size and Proper Motion of the Afterglow of GRB 030329}",
      journal = {\apjl},
         year = 2004,
        month = jul,
       volume = {609},
       number = {1},
        pages = {L1-L4},
          doi = {10.1086/422554},
archivePrefix = {arXiv},
       eprint = {astro-ph/0405300},
 primaryClass = {astro-ph},
       adsurl = {https://ui.adsabs.harvard.edu/abs/2004ApJ...609L...1T}
}

@ARTICLE{Taylor05,
       author = {{Taylor}, G.~B. and {Momjian}, E. and {Pihlstr{\"o}m}, Y. and {Ghosh}, T. and {Salter}, C.},
        title = "{Late-Time Observations of the Afterglow and Environment of GRB 030329}",
      journal = {\apj},
         year = 2005,
        month = apr,
       volume = {622},
       number = {2},
        pages = {986-990},
          doi = {10.1086/428346},
archivePrefix = {arXiv},
       eprint = {astro-ph/0412483},
 primaryClass = {astro-ph},
       adsurl = {https://ui.adsabs.harvard.edu/abs/2005ApJ...622..986T}
}

@ARTICLE{Tchekhovskoy08,
       author = {{Tchekhovskoy}, Alexander and {McKinney}, Jonathan C. and {Narayan}, Ramesh},
        title = "{Simulations of ultrarelativistic magnetodynamic jets from gamma-ray burst engines}",
      journal = {\mnras},
         year = 2008,
        month = aug,
       volume = {388},
       number = {2},
        pages = {551-572},
          doi = {10.1111/j.1365-2966.2008.13425.x},
archivePrefix = {arXiv},
       eprint = {0803.3807},
 primaryClass = {astro-ph},
       adsurl = {https://ui.adsabs.harvard.edu/abs/2008MNRAS.388..551T}
}

@ARTICLE{Troja17,
       author = {{Troja}, E. and {Piro}, L. and {van Eerten}, H. and {Wollaeger}, R.~T. and {Im}, M. and {Fox}, O.~D. and {Butler}, N.~R. and {Cenko}, S.~B. and {Sakamoto}, T. and {Fryer}, C.~L. and {Ricci}, R. and {Lien}, A. and {Ryan}, R.~E. and {Korobkin}, O. and {Lee}, S. -K. and {Burgess}, J.~M. and {Lee}, W.~H. and {Watson}, A.~M. and {Choi}, C. and {Covino}, S. and {D'Avanzo}, P. and {Fontes}, C.~J. and {Gonz{\'a}lez}, J. Becerra and {Khandrika}, H.~G. and {Kim}, J. and {Kim}, S. -L. and {Lee}, C. -U. and {Lee}, H.~M. and {Kutyrev}, A. and {Lim}, G. and {S{\'a}nchez-Ram{\'\i}rez}, R. and {Veilleux}, S. and {Wieringa}, M.~H. and {Yoon}, Y.},
        title = "{The X-ray counterpart to the gravitational-wave event GW170817}",
      journal = {\nat},
         year = 2017,
        month = nov,
       volume = {551},
       number = {7678},
        pages = {71-74},
          doi = {10.1038/nature24290},
archivePrefix = {arXiv},
       eprint = {1710.05433},
 primaryClass = {astro-ph.HE},
       adsurl = {https://ui.adsabs.harvard.edu/abs/2017Natur.551...71T}
}

@ARTICLE{Troja19,
       author = {{Troja}, E. and {van Eerten}, H. and {Ryan}, G. and {Ricci}, R. and {Burgess}, J.~M. and {Wieringa}, M.~H. and {Piro}, L. and {Cenko}, S.~B. and {Sakamoto}, T.},
        title = "{A year in the life of GW 170817: the rise and fall of a structured jet from a binary neutron star merger}",
      journal = {\mnras},
         year = 2019,
        month = oct,
       volume = {489},
       number = {2},
        pages = {1919-1926},
          doi = {10.1093/mnras/stz2248},
archivePrefix = {arXiv},
       eprint = {1808.06617},
 primaryClass = {astro-ph.HE},
       adsurl = {https://ui.adsabs.harvard.edu/abs/2019MNRAS.489.1919T}
}

@ARTICLE{Uhm11,
       author = {{Uhm}, Z. Lucas},
        title = "{A Semi-analytic Formulation for Relativistic Blast Waves with a Long-lived Reverse Shock}",
      journal = {\apj},
         year = 2011,
        month = jun,
       volume = {733},
       number = {2},
          eid = {86},
        pages = {86},
          doi = {10.1088/0004-637X/733/2/86},
archivePrefix = {arXiv},
       eprint = {1003.1115},
 primaryClass = {astro-ph.HE},
       adsurl = {https://ui.adsabs.harvard.edu/abs/2011ApJ...733...86U}
}

@ARTICLE{vanEerten12,
       author = {{van Eerten}, Hendrik J. and {MacFadyen}, Andrew I.},
        title = "{Observational Implications of Gamma-Ray Burst Afterglow Jet Simulations and Numerical Light Curve Calculations}",
      journal = {\apj},
         year = 2012,
        month = jun,
       volume = {751},
       number = {2},
          eid = {155},
        pages = {155},
          doi = {10.1088/0004-637X/751/2/155},
archivePrefix = {arXiv},
       eprint = {1105.2485},
 primaryClass = {astro-ph.HE},
       adsurl = {https://ui.adsabs.harvard.edu/abs/2012ApJ...751..155V}
}

@ARTICLE{vanEerten18,
       author = {{van Eerten}, Hendrik},
        title = "{Gamma-ray burst afterglow blast waves}",
      journal = {International Journal of Modern Physics D},
         year = 2018,
        month = jan,
       volume = {27},
       number = {13},
          eid = {1842002-314},
        pages = {1842002-314},
          doi = {10.1142/S0218271818420026},
archivePrefix = {arXiv},
       eprint = {1801.01848},
 primaryClass = {astro-ph.HE},
       adsurl = {https://ui.adsabs.harvard.edu/abs/2018IJMPD..2742002V}
}

@ARTICLE{Wang24,
       author = {{Wang}, Hao and {Dastidar}, Ranadeep G. and {Giannios}, Dimitrios and {Duffell}, Paul C.},
        title = "{jetsimpy: A Highly Efficient Hydrodynamic Code for Gamma-Ray Burst Afterglow}",
      journal = {\apjs},
         year = 2024,
        month = jul,
       volume = {273},
       number = {1},
          eid = {17},
        pages = {17},
          doi = {10.3847/1538-4365/ad4d9d},
archivePrefix = {arXiv},
       eprint = {2402.19359},
 primaryClass = {astro-ph.HE},
       adsurl = {https://ui.adsabs.harvard.edu/abs/2024ApJS..273...17W}
}

@ARTICLE{Williams23,
       author = {{Williams}, Maia A. and {Kennea}, Jamie A. and {Dichiara}, S. and {Kobayashi}, Kohei and {Iwakiri}, Wataru B. and {Beardmore}, Andrew P. and {Evans}, P.~A. and {Heinz}, Sebastian and {Lien}, Amy and {Oates}, S.~R. and {Negoro}, Hitoshi and {Cenko}, S. Bradley and {Buisson}, Douglas J.~K. and {Hartmann}, Dieter H. and {Jaisawal}, Gaurava K. and {Kuin}, N.~P.~M. and {Lesage}, Stephen and {Page}, Kim L. and {Parsotan}, Tyler and {Pasham}, Dheeraj R. and {Sbarufatti}, B. and {Siegel}, Michael H. and {Sugita}, Satoshi and {Younes}, George and {Ambrosi}, Elena and {Arzoumanian}, Zaven and {Bernardini}, M.~G. and {Campana}, S. and {Capalbi}, Milvia and {Caputo}, Regina and {D'A{\`\i}}, Antonino and {D'Avanzo}, P. and {D'Elia}, V. and {De Pasquale}, Massimiliano and {Eyles-Ferris}, R.~A.~J. and {Ferrara}, Elizabeth and {Gendreau}, Keith C. and {Gropp}, Jeffrey D. and {Kawai}, Nobuyuki and {Klingler}, Noel and {Laha}, Sibasish and {Melandri}, A. and {Mihara}, Tatehiro and {Moss}, Michael and {O'Brien}, Paul and {Osborne}, Julian P. and {Palmer}, David M. and {Perri}, Matteo and {Serino}, Motoko and {Sonbas}, E. and {Stamatikos}, Michael and {Starling}, Rhaana and {Tagliaferri}, G. and {Tohuvavohu}, Aaron and {Zane}, Silvia and {Ziaeepour}, Houri},
        title = "{GRB 221009A: Discovery of an Exceptionally Rare Nearby and Energetic Gamma-Ray Burst}",
      journal = {\apjl},
         year = 2023,
        month = mar,
       volume = {946},
       number = {1},
          eid = {L24},
        pages = {L24},
          doi = {10.3847/2041-8213/acbcd1},
archivePrefix = {arXiv},
       eprint = {2302.03642},
 primaryClass = {astro-ph.HE},
       adsurl = {https://ui.adsabs.harvard.edu/abs/2023ApJ...946L..24W}
}

@ARTICLE{Wygoda11,
       author = {{Wygoda}, N. and {Waxman}, E. and {Frail}, D.~A.},
        title = "{Relativistic Jet Dynamics and Calorimetry of Gamma-ray Bursts}",
      journal = {\apjl},
         year = 2011,
        month = sep,
       volume = {738},
       number = {2},
          eid = {L23},
        pages = {L23},
          doi = {10.1088/2041-8205/738/2/L23},
archivePrefix = {arXiv},
       eprint = {1102.5618},
 primaryClass = {astro-ph.HE},
       adsurl = {https://ui.adsabs.harvard.edu/abs/2011ApJ...738L..23W}
}

@ARTICLE{Yabe01,
       author = {{Yabe}, Takashi and {Xiao}, Feng and {Utsumi}, Takayuki},
        title = "{The Constrained Interpolation Profile Method for Multiphase Analysis}",
      journal = {Journal of Computational Physics},
         year = 2001,
        month = may,
       volume = {169},
       number = {2},
        pages = {556-593},
          doi = {10.1006/jcph.2000.6625},
       adsurl = {https://ui.adsabs.harvard.edu/abs/2001JCoPh.169..556Y}
}

@ARTICLE{ZhangB02,
       author = {{Zhang}, Bing and {M{\'e}sz{\'a}ros}, Peter},
        title = "{Gamma-Ray Burst Beaming: A Universal Configuration with a Standard Energy Reservoir?}",
      journal = {\apj},
         year = 2002,
        month = jun,
       volume = {571},
       number = {2},
        pages = {876-879},
          doi = {10.1086/339981},
archivePrefix = {arXiv},
       eprint = {astro-ph/0112118},
 primaryClass = {astro-ph},
       adsurl = {https://ui.adsabs.harvard.edu/abs/2002ApJ...571..876Z}
}

@ARTICLE{Zhang05,
       author = {{Zhang}, Bing and {Kobayashi}, Shiho},
        title = "{Gamma-Ray Burst Early Afterglows: Reverse Shock Emission from an Arbitrarily Magnetized Ejecta}",
      journal = {\apj},
         year = 2005,
        month = jul,
       volume = {628},
       number = {1},
        pages = {315-334},
          doi = {10.1086/429787},
archivePrefix = {arXiv},
       eprint = {astro-ph/0404140},
 primaryClass = {astro-ph},
       adsurl = {https://ui.adsabs.harvard.edu/abs/2005ApJ...628..315Z}
}

@ARTICLE{Zhang11,
       author = {{Zhang}, Bing and {Yan}, Huirong},
        title = "{The Internal-collision-induced Magnetic Reconnection and Turbulence (ICMART) Model of Gamma-ray Bursts}",
      journal = {\apj},
         year = 2011,
        month = jan,
       volume = {726},
       number = {2},
          eid = {90},
        pages = {90},
          doi = {10.1088/0004-637X/726/2/90},
archivePrefix = {arXiv},
       eprint = {1011.1197},
 primaryClass = {astro-ph.HE},
       adsurl = {https://ui.adsabs.harvard.edu/abs/2011ApJ...726...90Z}
}

@ARTICLE{ZhangB23,
       author = {{Zhang}, Bing and {Wang}, Xiang-Yu and {Zheng}, Jian-He},
        title = "{The BOAT GRB 221009A: A Poynting-flux-dominated narrow jet surrounded by a matter-dominated structured jet wing}",
      journal = {J. High Energy Astrophys.},
         year = 2024,
        month = mar,
       volume = {41},
        pages = {42-53},
          doi = {10.1016/j.jheap.2024.01.002},
archivePrefix = {arXiv},
       eprint = {2311.14180},
 primaryClass = {astro-ph.HE},
       adsurl = {https://ui.adsabs.harvard.edu/abs/2024JHEAp..41...42Z}
}

@ARTICLE{ZhangBT25,
       author = {{Zhang}, B. Theodore and {Murase}, Kohta and {Ioka}, Kunihito and {Zhang}, Bing},
        title = "{The origin of very-high-energy gamma-rays from GRB 221009A: Implications for reverse shock proton synchrotron emission}",
      journal = {Journal of High Energy Astrophysics},
         year = 2025,
        month = mar,
       volume = {45},
        pages = {392-408},
          doi = {10.1016/j.jheap.2025.01.007},
archivePrefix = {arXiv},
       eprint = {2311.13671},
 primaryClass = {astro-ph.HE},
       adsurl = {https://ui.adsabs.harvard.edu/abs/2025JHEAp..45..392Z}
}

@ARTICLE{ZhangW09,
       author = {{Zhang}, Weiqun and {MacFadyen}, Andrew},
        title = "{The Dynamics and Afterglow Radiation of Gamma-Ray Bursts. I. Constant Density Medium}",
      journal = {\apj},
         year = 2009,
        month = jun,
       volume = {698},
       number = {2},
        pages = {1261-1272},
          doi = {10.1088/0004-637X/698/2/1261},
archivePrefix = {arXiv},
       eprint = {0902.2396},
 primaryClass = {astro-ph.HE},
       adsurl = {https://ui.adsabs.harvard.edu/abs/2009ApJ...698.1261Z}
}

@ARTICLE{ZhangX25,
       author = {{Zhang}, X. and {Yu}, W. and {Carotenuto}, F. and {Fender}, R. and {Motta}, S. and {Bahramian}, A. and {Miller-Jones}, J.~C.~A. and {Russell}, T.~D. and {Corbel}, S. and {Woudt}, P.~A. and {Atri}, P. and {Knigge}, C. and {Sivakoff}, G.~R. and {Hughes}, A.~K. and {van den Eijnden}, J. and {Matthews}, J.~H. and {Baglio}, M.~C. and {Saikia}, P.},
        title = "{Jets from a stellar-mass black hole are as relativistic as those from supermassive black holes}",
      journal = {Nature Communications},
         year = 2026,
        month = may,
       volume = {17},
       number = {1},
          eid = {6477},
        pages = {6477},
          doi = {10.1038/s41467-026-72897-5},
archivePrefix = {arXiv},
       eprint = {2504.11945},
 primaryClass = {astro-ph.HE},
       adsurl = {https://ui.adsabs.harvard.edu/abs/2026NatCo..17.6477Z}
}

@ARTICLE{Zheng24,
       author = {{Zheng}, Jian-He and {Wang}, Xiang-Yu and {Liu}, Ruo-Yu and {Zhang}, Bing},
        title = "{A Narrow Uniform Core with a Wide Structured Wing: Modeling the TeV and Multiwavelength Afterglows of GRB 221009A}",
      journal = {\apj},
         year = 2024,
        month = may,
       volume = {966},
       number = {1},
          eid = {141},
        pages = {141},
          doi = {10.3847/1538-4357/ad3949},
archivePrefix = {arXiv},
       eprint = {2310.12856},
 primaryClass = {astro-ph.HE},
       adsurl = {https://ui.adsabs.harvard.edu/abs/2024ApJ...966..141Z}
}

@ARTICLE{Zou05,
       author = {{Zou}, Y.~C. and {Wu}, X.~F. and {Dai}, Z.~G.},
        title = "{Early afterglows in wind environments revisited}",
      journal = {\mnras},
         year = 2005,
        month = oct,
       volume = {363},
       number = {1},
        pages = {93-106},
          doi = {10.1111/j.1365-2966.2005.09411.x},
archivePrefix = {arXiv},
       eprint = {astro-ph/0508602},
 primaryClass = {astro-ph},
       adsurl = {https://ui.adsabs.harvard.edu/abs/2005MNRAS.363...93Z}
}

@ARTICLE{Planck20,
       author = {{Planck Collaboration} and {Aghanim}, N. and {Akrami}, Y. and {Ashdown}, M. and {Aumont}, J. and {Baccigalupi}, C. and {Ballardini}, M. and {Banday}, A.~J. and {Barreiro}, R.~B. and {Bartolo}, N. and {Basak}, S. and {Battye}, R. and {Benabed}, K. and {Bernard}, J. -P. and {Bersanelli}, M. and {Bielewicz}, P. and {Bock}, J.~J. and {Bond}, J.~R. and {Borrill}, J. and {Bouchet}, F.~R. and {Boulanger}, F. and {Bucher}, M. and {Burigana}, C. and {Butler}, R.~C. and {Calabrese}, E. and {Cardoso}, J. -F. and {Carron}, J. and {Challinor}, A. and {Chiang}, H.~C. and {Chluba}, J. and {Colombo}, L.~P.~L. and {Combet}, C. and {Contreras}, D. and {Crill}, B.~P. and {Cuttaia}, F. and {de Bernardis}, P. and {de Zotti}, G. and {Delabrouille}, J. and {Delouis}, J. -M. and {Di Valentino}, E. and {Diego}, J.~M. and {Dor{\'e}}, O. and {Douspis}, M. and {Ducout}, A. and {Dupac}, X. and {Dusini}, S. and {Efstathiou}, G. and {Elsner}, F. and {En{\ss}lin}, T.~A. and {Eriksen}, H.~K. and {Fantaye}, Y. and {Farhang}, M. and {Fergusson}, J. and {Fernandez-Cobos}, R. and {Finelli}, F. and {Forastieri}, F. and {Frailis}, M. and {Fraisse}, A.~A. and {Franceschi}, E. and {Frolov}, A. and {Galeotta}, S. and {Galli}, S. and {Ganga}, K. and {G{\'e}nova-Santos}, R.~T. and {Gerbino}, M. and {Ghosh}, T. and {Gonz{\'a}lez-Nuevo}, J. and {G{\'o}rski}, K.~M. and {Gratton}, S. and {Gruppuso}, A. and {Gudmundsson}, J.~E. and {Hamann}, J. and {Handley}, W. and {Hansen}, F.~K. and {Herranz}, D. and {Hildebrandt}, S.~R. and {Hivon}, E. and {Huang}, Z. and {Jaffe}, A.~H. and {Jones}, W.~C. and {Karakci}, A. and {Keih{\"a}nen}, E. and {Keskitalo}, R. and {Kiiveri}, K. and {Kim}, J. and {Kisner}, T.~S. and {Knox}, L. and {Krachmalnicoff}, N. and {Kunz}, M. and {Kurki-Suonio}, H. and {Lagache}, G. and {Lamarre}, J. -M. and {Lasenby}, A. and {Lattanzi}, M. and {Lawrence}, C.~R. and {Le Jeune}, M. and {Lemos}, P. and {Lesgourgues}, J. and {Levrier}, F. and {Lewis}, A. and {Liguori}, M. and {Lilje}, P.~B. and {Lilley}, M. and {Lindholm}, V. and {L{\'o}pez-Caniego}, M. and {Lubin}, P.~M. and {Ma}, Y. -Z. and {Mac{\'\i}as-P{\'e}rez}, J.~F. and {Maggio}, G. and {Maino}, D. and {Mandolesi}, N. and {Mangilli}, A. and {Marcos-Caballero}, A. and {Maris}, M. and {Martin}, P.~G. and {Martinelli}, M. and {Mart{\'\i}nez-Gonz{\'a}lez}, E. and {Matarrese}, S. and {Mauri}, N. and {McEwen}, J.~D. and {Meinhold}, P.~R. and {Melchiorri}, A. and {Mennella}, A. and {Migliaccio}, M. and {Millea}, M. and {Mitra}, S. and {Miville-Desch{\^e}nes}, M. -A. and {Molinari}, D. and {Montier}, L. and {Morgante}, G. and {Moss}, A. and {Natoli}, P. and {N{\o}rgaard-Nielsen}, H.~U. and {Pagano}, L. and {Paoletti}, D. and {Partridge}, B. and {Patanchon}, G. and {Peiris}, H.~V. and {Perrotta}, F. and {Pettorino}, V. and {Piacentini}, F. and {Polastri}, L. and {Polenta}, G. and {Puget}, J. -L. and {Rachen}, J.~P. and {Reinecke}, M. and {Remazeilles}, M. and {Renzi}, A. and {Rocha}, G. and {Rosset}, C. and {Roudier}, G. and {Rubi{\~n}o-Mart{\'\i}n}, J.~A. and {Ruiz-Granados}, B. and {Salvati}, L. and {Sandri}, M. and {Savelainen}, M. and {Scott}, D. and {Shellard}, E.~P.~S. and {Sirignano}, C. and {Sirri}, G. and {Spencer}, L.~D. and {Sunyaev}, R. and {Suur-Uski}, A. -S. and {Tauber}, J.~A. and {Tavagnacco}, D. and {Tenti}, M. and {Toffolatti}, L. and {Tomasi}, M. and {Trombetti}, T. and {Valenziano}, L. and {Valiviita}, J. and {Van Tent}, B. and {Vibert}, L. and {Vielva}, P. and {Villa}, F. and {Vittorio}, N. and {Wandelt}, B.~D. and {Wehus}, I.~K. and {White}, M. and {White}, S.~D.~M. and {Zacchei}, A. and {Zonca}, A.},
        title = "{Planck 2018 results. VI. Cosmological parameters}",
      journal = {Astronomy and Astrophysics},
         year = 2020,
        month = sep,
       volume = {641},
          eid = {A6},
        pages = {A6},
          doi = {10.1051/0004-6361/201833910},
archivePrefix = {arXiv},
       eprint = {1807.06209},
 primaryClass = {astro-ph.CO},
       adsurl = {https://ui.adsabs.harvard.edu/abs/2020A&A...641A...6P}
}

\begin{appendix}
\nolinenumbers
\onecolumn
\section{Materials and Methods}\label{sec:supp_methods}
\subsection{VLBA observations}\label{sec:vlba_obs}

The VLBA campaign used four observing programs (\SUPTAB{tab:obslog}).
The pilot VLBA observation of GRB 221009A was performed at 15 GHz on October 14, 2022 (project code: TG015), approximately 5 days after the burst, to detect the GRB and determine its precise coordinates. We performed 15 GHz observations at 9, 18 and 26 d after the burst (project code: BA161). We also analyzed the four-epoch archive data observed at 15 GHz from 44 to 262 d (BA160) and seven 8.3-GHz epochs from 12 to 299 d (BL307). \SUPTAB{tab:obslog} summarizes the observation logs for the data used in this study.

TG015 was correlated before the accurate GRB position was known. The target lay about 2--3~arcmin from the correlator and pointing position. A phase rotation moved the visibilities to the measured emission peak. This operation changes only the visibility phase and therefore does not change the fitted angular size. The pointing offset attenuated the flux density through the primary beam. A simulation with a uniform 2.5-arcmin offset applied to every antenna attenuated the injected source by a factor of 0.14 while leaving the fitted FWHM unchanged within numerical precision. Applying this factor to the observed flux density of 0.917 mJy recovers an intrinsic value of $0.917/0.14 \approx 6.6$ mJy, consistent with the contemporaneous Arcminute Microkelvin Imager Large Array (AMI-LA) measurement of 7.05 mJy at 15.5 GHz and the Australia Telescope Compact Array (ATCA) extrapolation at this epoch \citep{Rhodes24}. In a complementary experiment, 20,000 trials with independent random offsets drawn uniformly within a 2-arcmin radius gave a median size shift of $+0.001$~mas, and the central 68\% of shifts lay between $-0.040$ and $+0.038$~mas. The angular size is therefore insensitive to realistic station-dependent pointing variations. TG015 is retained as a broad early angular-scale constraint, while its flux density, which reflects the pointing-induced attenuation, is treated separately.

To achieve robust detection and accurate measurement of this relatively faint source ($\sim$ 1-3 mJy at 15 GHz), we implemented an optimized phase-referencing observation strategy. We selected the brightest nearby radio source in the International Celestial Reference Frame (ICRF), ICRF J190536.4+194308 (hereafter J1905+1943, located $\sim 1.7^{\circ}$ from the target), as the phase-reference calibrator, and used rapid switching between the calibrator and the target with a cycle time of 1.5 minutes (30 s - 60 s for calibrator--target) to overcome the phase variations at 15 GHz. 
Another nearby bright source ICRF J192313.1+201039 (J1923+2010), was observed as a phase-referencing check source, with several isolated scans interspersed throughout the observation.
This cycle time was carefully chosen based on empirical atmospheric coherence time estimates to optimize phase solution accuracy while maintaining sufficient on-source time.
The data were recorded at a high rate of 4 Gbps with 512 MHz total bandwidth split into four 128 MHz intermediate frequencies, each with 256 channels. 
Raw data were transferred to the DiFX correlator for initial processing, including 2-second integration times to minimize time smearing effects.

To improve temporal sampling and augment the dataset in our analysis, we incorporated archival VLBI observations. 
Project BA160 has four 15 GHz VLBA epochs (b, c, c1, d) spanning 44–262 days post-burst \citep{Giarratana24}.
Project BL307 provides multi-epoch 8.4-GHz VLBA monitoring from 21 to 299 days post-burst (\SUPTAB{tab:obslog}). The BA160 phase-referencing cycle alternated 30s on J1905+1943, 30s on another phase calibrator ICRF J192559.6+210626 (J1925+2106), and 80~s on the GRB. The BL307 X-band cycle used 60~s on J1905+1943 and 180~s on the GRB. Simultaneous 22-GHz BL307 data were reduced but did not provide a reliable detection or angular-size constraint, so they are not used in this work.

\begin{table}[h]
\scriptsize
\caption{VLBI observing setups and antennas contributing to the fitted visibility data.}
\label{tab:obslog}
\centering
\resizebox{\textwidth}{!}{%
\begin{tabular}{lccccll}
\hline\hline
Program & $t$ & $\nu$ & BW & $N_{\rm ant}$ & Phase calibrator & Antennas used \\
 & (d) & (GHz) & (MHz) & & & \\
\hline
TG015    &   5.491 & 15.168 & 512 &  9 & J1905+1943 & BR, FD, HN, LA, MK, NL, OV, PT, SC \\
BA161a   &   9.443 & 15.168 & 512 &  8 & J1905+1943 & BR, FD, HN, LA, MK, NL, OV, PT \\
BA161b   &  18.453 & 15.168 & 512 &  9 & J1905+1943 & BR, FD, HN, LA, MK, NL, OV, PT, SC \\
BA161c   &  26.432 & 15.168 & 512 &  9 & J1905+1943 & BR, FD, HN, LA, MK, NL, OV, PT, SC \\
BA160b   &  44.384 & 15.168 & 512 &  7 & J1905+1943, J1925+2106 & FD, HN, MK, NL, OV, PT, SC \\
BA160c   & 114.186 & 15.200 & 512 &  8 & J1905+1943, J1925+2106 & BR, FD, LA, MK, NL, OV, PT, SC \\
BA160c1  & 204.975 & 15.168 & 512 & 10 & J1905+1943, J1925+2106 & BR, FD, HN, KP, LA, MK, NL, OV, PT, SC \\
BA160d   & 261.733 & 15.168 & 512 & 10 & J1905+1943, J1925+2106 & BR, FD, HN, KP, LA, MK, NL, OV, PT, SC \\
BL307a   &  12.434 &  8.368 & 512 &  9 & J1905+1943 & BR, FD, HN, LA, MK, NL, OV, PT, SC \\
BL307b   &  21.421 &  8.368 & 512 &  8 & J1905+1943 & BR, FD, HN, LA, MK, NL, OV, PT \\
BL307c   &  36.356 &  8.368 & 512 &  8 & J1905+1943 & FD, HN, LA, MK, NL, OV, PT, SC \\
BL307d   &  64.256 &  8.368 & 512 & 10 & J1905+1943 & BR, EB, FD, HN, LA, MK, NL, OV, PT, SC \\
BL307e   & 109.187 &  8.368 & 512 &  7 & J1905+1943 & BR, FD, LA, MK, NL, OV, PT \\
BL307f   & 201.923 &  8.368 & 512 & 10 & J1905+1943 & BR, FD, HN, KP, LA, MK, NL, OV, PT, SC \\
BL307g   & 298.672 &  8.368 & 512 & 10 & J1905+1943 & BR, FD, HN, KP, LA, MK, NL, OV, PT, SC \\
\hline
\end{tabular}}

{\footnotesize The antenna lists refer to the visibility products used for the reported fits. BR is Brewster, EB is Effelsberg, FD is Fort Davis, HN is Hancock, KP is Kitt Peak, LA is Los Alamos, MK is Mauna Kea, NL is North Liberty, OV is Owens Valley, PT is Pie Town, and SC is Saint Croix.}
\end{table}

\subsection{VLBA data reduction}\label{sec:vlba_reduction}

Calibration and imaging of the correlated data were executed at the China SKA Regional Centre prototype \citep{An19}, using a Python-based VLBI pipeline \citep{An22} interfacing with the Astronomical Image Processing System (\textsc{AIPS}) software \citep{Greisen03}. The process involved quality evaluation using VLBASUMM and POSSUM, with the Fort Davis antenna serving as the reference. Subsequent calibration steps corrected for Earth orientation, atmospheric factors, gain amplitudes, opacities, and a detailed assessment of calibration consistency across epochs. Fringe fitting applied to the calibrator 3C 345 corrected delays and phases of the visibilities, with these corrections applied across all data. Global phase errors were determined by fringe fitting the phase calibrator J1905+1943 and corrected. Finally, bandpass functions derived from 3C 345 were applied to all data.
To ensure reliable calibration across different epochs, we maintained consistent calibration procedures and carefully evaluated potential systematic effects. 

\begin{figure}
    \centering
    \includegraphics[width=0.32\textwidth]{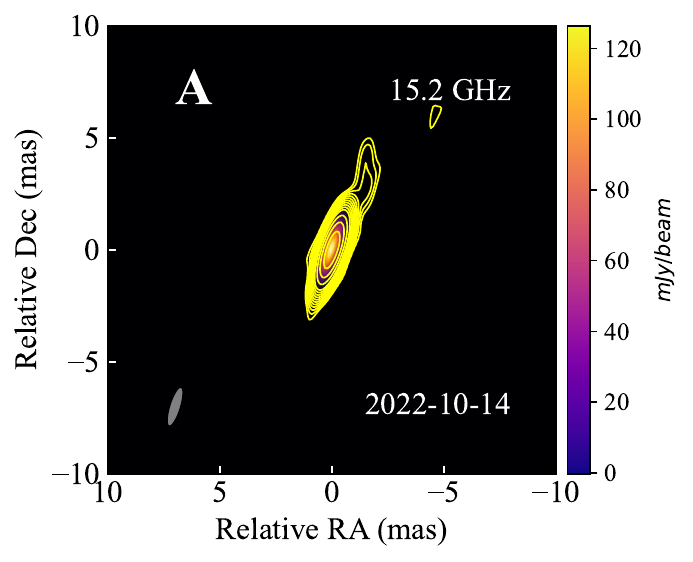}
    \includegraphics[width=0.32\textwidth]{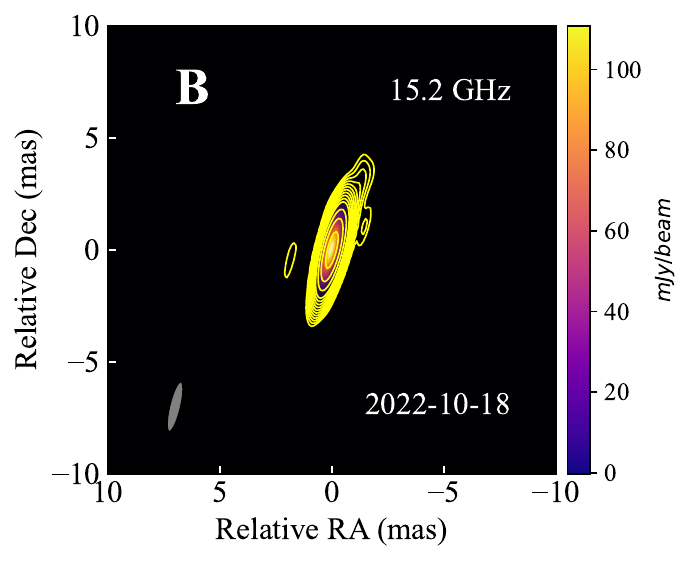}
    \includegraphics[width=0.32\textwidth]{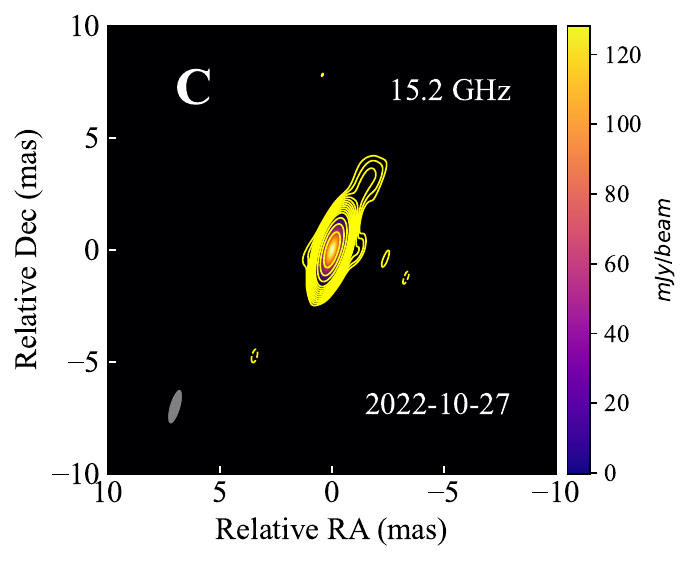} 
    \\
    \includegraphics[width=0.32\textwidth]{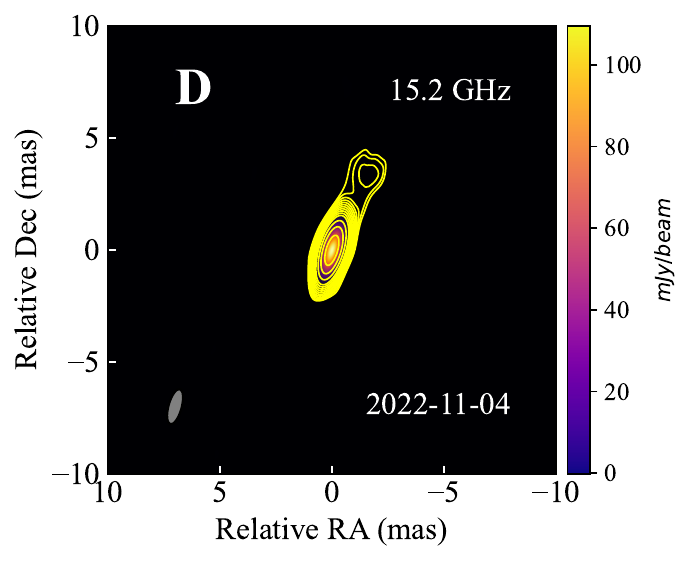}
    \includegraphics[width=0.32\textwidth]{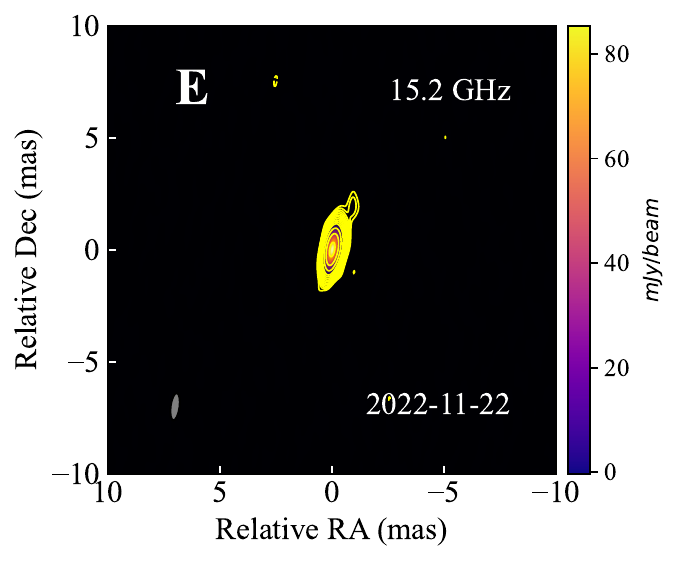}
    \includegraphics[width=0.32\textwidth]{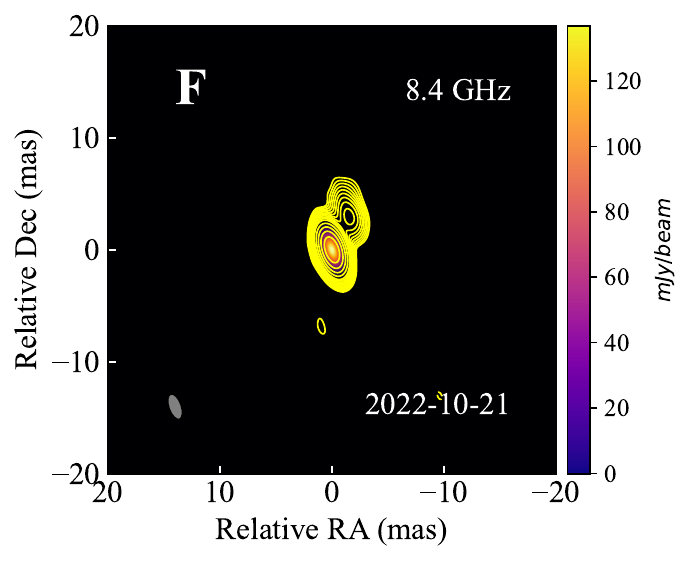}
    \\
    \includegraphics[width=0.32\textwidth]{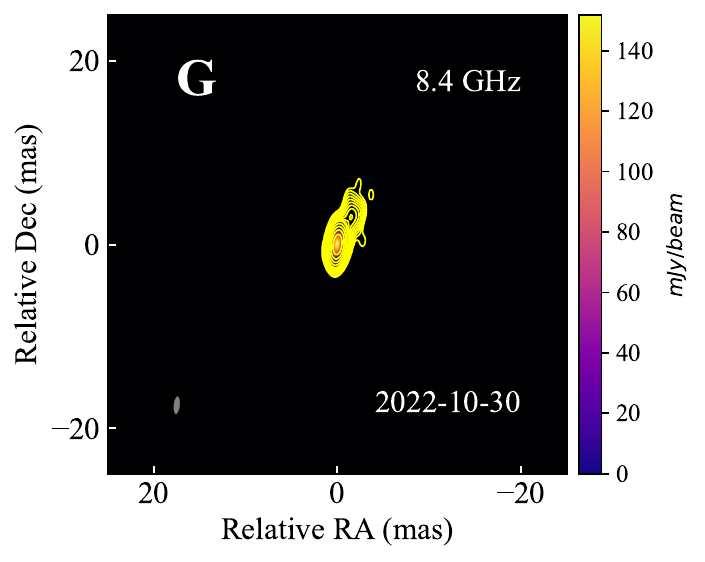}
    \includegraphics[width=0.32\textwidth]{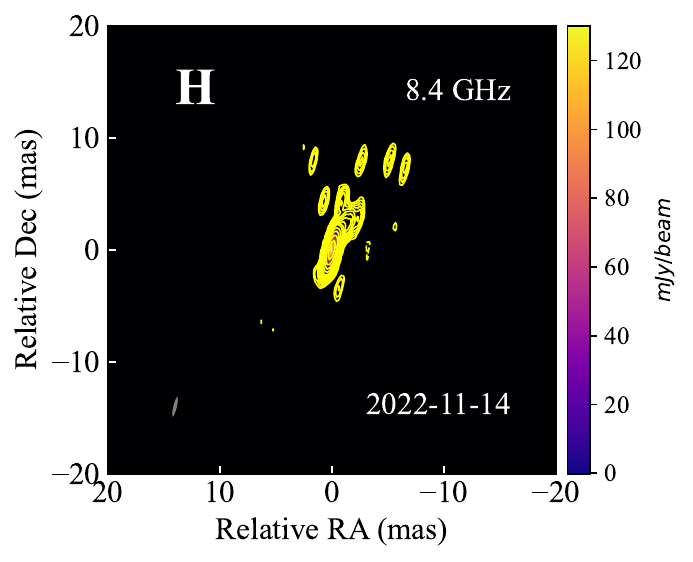}
    \includegraphics[width=0.32\textwidth]{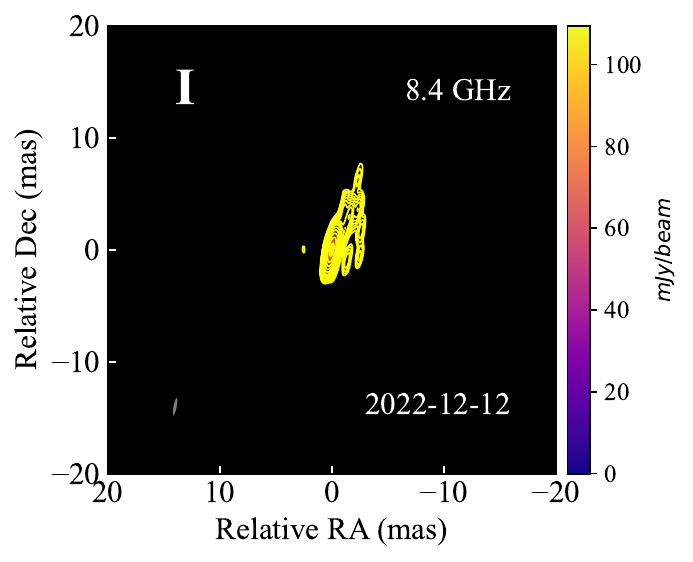}
    \\
    \includegraphics[width=0.32\textwidth]{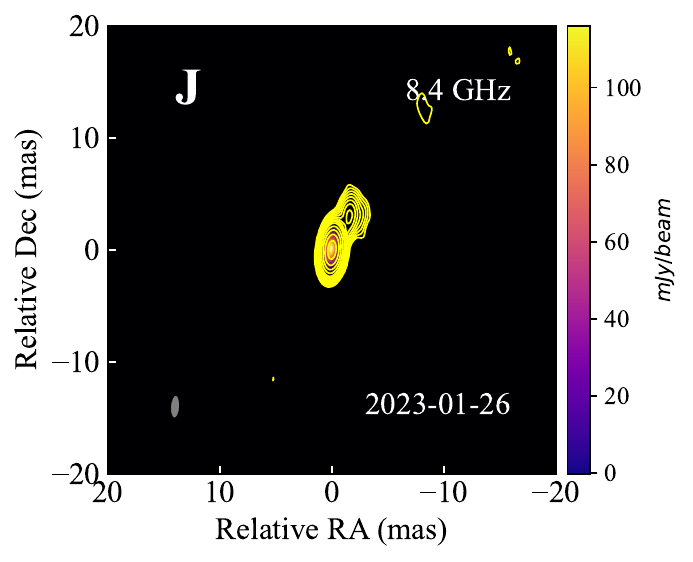}
    \includegraphics[width=0.32\textwidth]{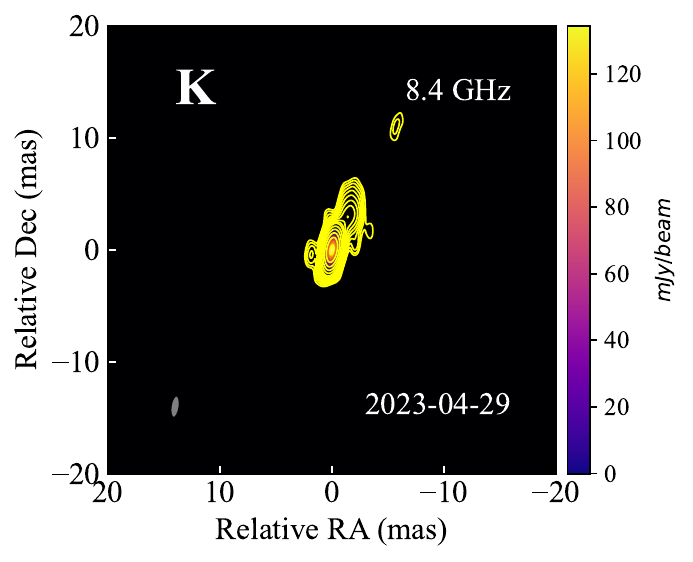}
    \includegraphics[width=0.32\textwidth]{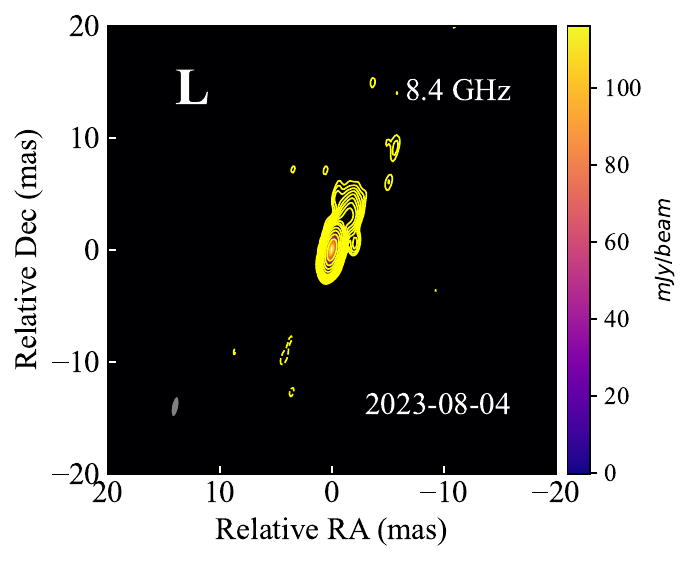}
    \caption{
    \textbf{Naturally weighted VLBI images of the phase calibrator (J1905+1943) at 15.2 GHz and 8.4 GHz.} 
    Each panel corresponds to the epoch indicated in \FIG{fig:Image}. 
    The restoring beam is displayed as a grey ellipse in the bottom‑left corner of each panel. All images are shifted to align their centers with the brightness peak.
    The lowest contour level in each map is set to 3 times the post‑CLEAN r.m.s. noise, with subsequent contours increasing by a factor of 2.}
    \label{fig:Image_ref}
\end{figure}

The calibrated J1905+1943 data were exported to Difmap \citep{Shepherd97}. CLEAN and phase-only self-calibration were repeated until the residual image noise ceased to improve. Phase and amplitude self-calibration then refined the calibrator model. A 30$^{\circ}$ elevation threshold was applied only while constructing this model, to avoid transferring poor phase solutions derived from low-elevation atmospheric turbulence on the calibrator. No elevation cut was applied to the target visibilities. This strategy preserves the long-baseline coverage essential for resolving the source. For the target, data flagging was strictly limited to scans with demonstrable instrumental failures or loss of phase coherence. The resulting model was returned to AIPS for a second fringe fit. Normalized antenna and intermediate-frequency gain corrections were applied with CLCOR.
J1905+1943 had a mean flux density of 126.7~mJy and an epoch-to-epoch spread of 11.5~mJy, with calibrated images and visibilities shown in \SUPFIG{fig:Image_ref}.
Two aspects of the amplitude calibration are treated separately in what follows. First, an absolute 5\% common-gain scale uncertainty was added in quadrature to the reported flux-density errors; a common attenuation of this kind only rescales the zero-baseline value $S_\nu$ and does not shift $\theta_{\rm G}$. Second, station-dependent gain perturbations at the same 5\% level can, in principle, distort the visibility curvature and therefore bias $\theta_{\rm G}$. This differential effect is quantified by dedicated injection experiments and is propagated into the angular-size analysis as the term $\sigma_{\rm amp}$ shown in Appendix~\ref{sec:vlbi_error_structure} and \SUPTAB{tab:vlbi_uncertainty_components}.

For the remaining epochs, the same calibration procedure as for BA161 was applied to the BA160 and BL307 data, again using J1905+1943 as the phase calibrator.
We performed additional quality control in a small subset of epochs where the data were demonstrably compromised (e.g., limited time coverage for a given station, persistently unstable phase transfer, discrepant amplitudes, or known antenna \& IF issues). In these cases, we excluded only the affected scans or IFs \& stations based on calibration diagnostics, rather than applying any generic elevation threshold to the GRB target visibilities. 
During the BL307 quality checks, the phase-calibrator solutions occasionally showed abrupt phase and amplitude excursions within individual scans, indicative of calibration failures. 
Such excursions can arise from several practical causes, including incomplete antenna slewing (most relevant for stations with sparse participation, such as Effelsberg in these observations), enhanced atmospheric/ionospheric phase variability, which is more likely at low elevations. 
We therefore applied standard scan-level editing: we used the QUACK task in AIPS to flag the initial 30 seconds of each problematic scan, and rejected scans with failed or clearly unstable phase transfer.

The BA160 visibilities were processed with the same pipeline, and the resulting angular sizes agree with the independently published values \citep{Giarratana24}, confirming the reproducibility of the reduction. The target was clearly detected in all BA160 (15 GHz) epochs.
With all calibrations applied, the target data were exported to Difmap for imaging. Naturally weighted CLEAN images were made to verify detection and locate the emission peak (\FIG{fig:Image}). 
Given the moderate SNR of the target, we conservatively refrained from applying self-calibration to the GRB 221009A visibilities. This precaution eliminates the risk of absorbing phase errors into the source structure model \citep{Marti08}, ensuring that the derived sizes are driven purely by the measured visibilities rather than model assumptions.

\subsection{Source size derivation and evolution}
\label{sec:vlbi_size_analysis}
\subsubsection{Visibility model and measured angular size}
\label{sec:vlbi_error_structure}

We measure angular size and flux density directly from the calibrated visibilities. This approach avoids the non-linear steps of image-plane Gaussian fitting, CLEAN deconvolution, and convolution with a restoring beam, which depend on imaging weights and can bias the recovered size when the source is comparable to or smaller than the beam. The image-plane fitting also bypasses the correlated noise between adjacent image pixels, which invalidates standard $\chi^2$ error estimates. In the visibility domain, the Fourier samples are statistically independent, their errors are set by the measured thermal noise, and the model is compared directly with the observed complex visibilities without intermediate image reconstruction.

\begin{figure}[htbp]
\centering
\includegraphics[width=0.8\textwidth]{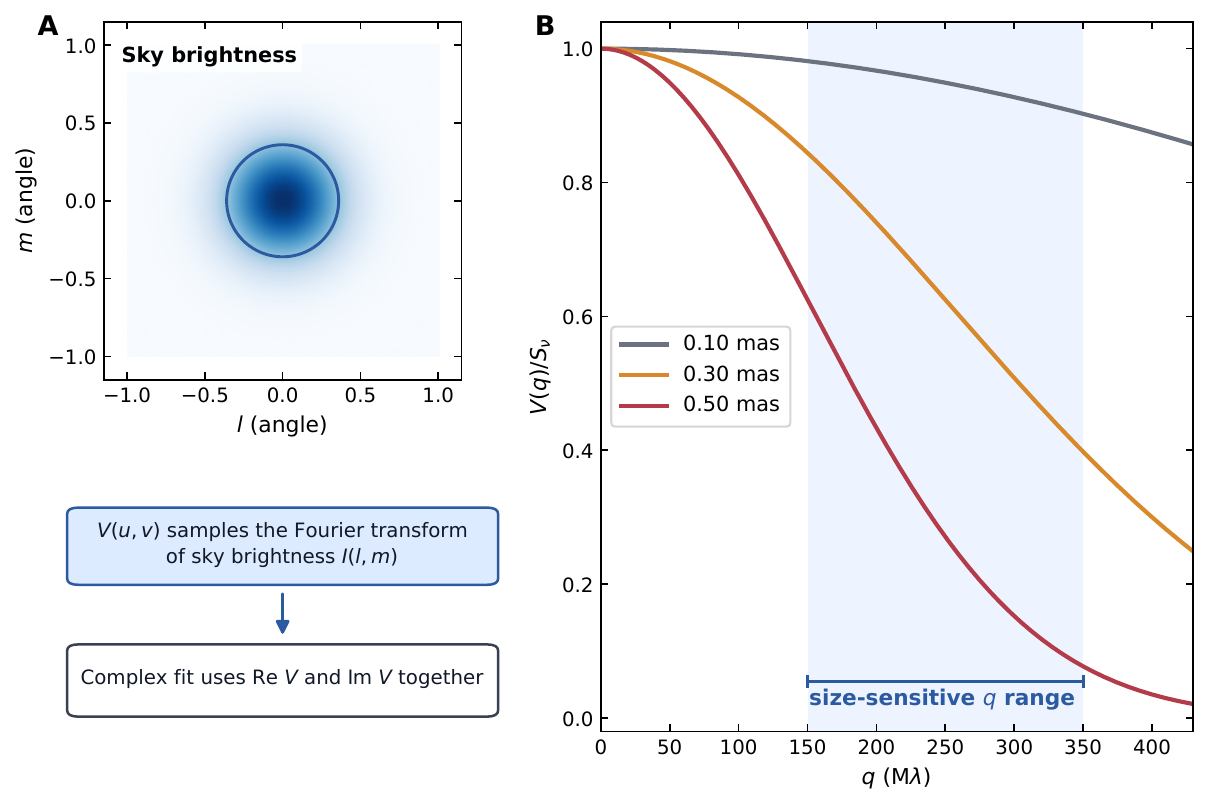}
\caption{\textbf{Circular-Gaussian size measurement.} (\textbf{A}) Schematic of the visibility-based method. (\textbf{B}) Theoretical visibility profiles as a function of projected baseline $q$. A larger source produces a steeper decline. The shaded band marks $q \approx 150$--$350$~M$\lambda$, corresponding to the Fisher-information peak $q_{\rm F}$ for the angular scales measured in GRB~221009A (from $\theta_{\rm G} \approx 0.3$~mas to $\approx 1.0$~mas; Eq.~\ref{eq:vlbi_qF}).}
\label{fig:vlbi_visibility_principle}
\end{figure}

Within the visibility domain, we fit the real and imaginary components of each Stokes-I visibility jointly. Complex visibility $V_k^{\rm obs}$ is the Fourier-domain sample of the sky brightness distribution measured at baseline $k$ at projected coordinates $(u_k,v_k)$, in units of wavelengths. For visibility $k$, the circular-Gaussian model is
\begin{equation}
 V_k^{\rm mod}=S_\nu
 \exp\!\left[-\frac{\pi^2\theta_{\rm G}^2q_k^2}{4\ln2}\right]
 \exp\!\left[-2\pi i(u_k\Delta\alpha+v_k\Delta\delta)\right],
 \label{eq:vlbi_visibility_model}
\end{equation}
where $q_k=(u_k^2+v_k^2)^{1/2}$ is the projected baseline length (in wavelengths, typically M$\lambda$). The fitted quantities are the flux density $S_\nu$, the circular-Gaussian FWHM $\theta_{\rm G}$, and the position offsets $\Delta\alpha$ and $\Delta\delta$ of the emission centroid from the phase-reference position. 
The zero-baseline intercept sets $S_\nu$, the curvature of visibility decline with $q$ sets $\theta_{\rm G}$ (\SUPFIG{fig:vlbi_visibility_principle}),
and the phase slope sets the position. The fit minimizes
\begin{equation}
 \chi^2=\sum_k W_k\left[({\rm Re}\,r_k)^2+({\rm Im}\,r_k)^2\right],
 \qquad r_k=V_k^{\rm obs}-V_k^{\rm mod},
 \label{eq:vlbi_chi2}
\end{equation}
where $W_k=1/\sigma_k^2$ is the inverse-variance weight assigned to visibility $k$ from its measured thermal noise, and $r_k$ is the residual between the observed and model visibilities.
Here, the circular Gaussian is adopted as the primary model. An elliptical Gaussian introduces an axial ratio and position angle that are poorly constrained in weak epochs. Disk, ring, and annulus models produce Bessel-function oscillations distinguishable from a Gaussian only when the data sample the first null and post-null lobe with sufficient coherent signal; when they do not, all such models return the same low-$q$ curvature and therefore the same Gaussian-equivalent scale.
The fitted results for all epochs are summarized in Table~\ref{tab:ObsSummary}.

The angular size is constrained by the curvature of the visibility envelope, with different baseline lengths contributing unequally. The Fisher information on $\theta_{\rm G}$ from Eq.~(\ref{eq:vlbi_visibility_model}) scales as ${\cal I}(\theta_{\rm G})\propto q^4\exp\!\left[-\frac{\pi^2\theta_{\rm G}^2q^2}{2\ln2}\right]$, which is suppressed at short baselines (by $q^4$) and at very long baselines (where coherent flux is exponentially weak). The information peaks at
\begin{equation}
 q_{\rm F}=\frac{2\sqrt{\ln2}}{\pi\,\theta_{\rm G}}
 \simeq365~{\rm M}\lambda\left(\frac{\theta_{\rm G}}{0.30~{\rm mas}}\right)^{-1},
 \label{eq:vlbi_qF}
\end{equation}
which migrates to shorter baselines as the source expands (\SUPFIG{fig:vlbi_visibility_principle}).
For the early 15-GHz epochs, $\theta_{\rm G}\simeq0.27$--0.42~mas and $q_{\rm F}\simeq260$--410~M$\lambda$, indicating that the longest VLBA baselines with Mauna Kea (MK) and St.\ Croix (SC) carry most of the size information. For the late 8.3-GHz epochs $\theta_{\rm G}\simeq0.5$--1.6~mas, $q_{\rm F}\simeq70$--220~M$\lambda$, so the size information moves to medium baselines and BL307g and BA160c1/d become weak constraints because their coherent flux at those baselines is close to the noise. 

\begin{figure}[htbp]
\centering
\includegraphics[width=0.92\textwidth]{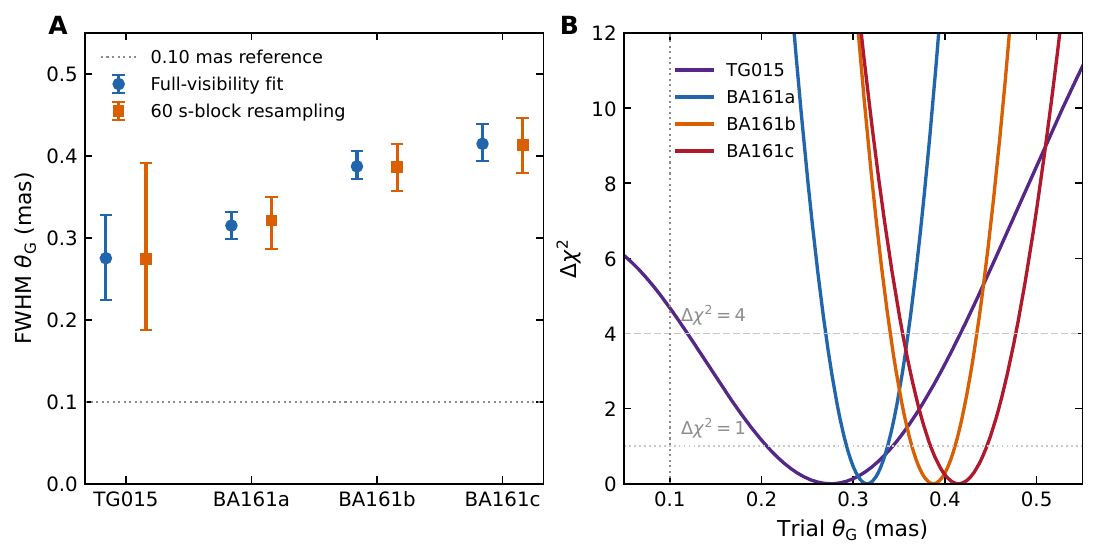}
\caption{\textbf{Statistical validation for the four early 15-GHz epochs.} (\textbf{A}) Circular-Gaussian FWHM $\theta_{\rm G}$ from the full complex-visibility fit (blue circles) and the median of a 60-second block bootstrap that preserves time-correlated visibility group (orange squares). Error bars mark 68\% intervals. The dotted line at 0.10~mas indicates a compact-source reference, excluded at high confidence for all epochs. (\textbf{B}) Profile-likelihood scans of $\theta_{\rm G}$. For each epoch, $\theta_{\rm G}$ was fixed on a grid while the remaining parameters ($S_\nu$, $\Delta\alpha$, $\Delta\delta$) were reoptimized; the resulting $\Delta\chi^2=\chi^2(\theta_{\rm G})-\chi^2_{\rm min}$ was normalized by the reduced $\chi^2$ so that thermal weights reflect the measured residual scatter. Dotted and dashed horizontal lines mark the 68\% and 95\% confidence bounds for one free parameter ($\Delta\chi^2=1$ and $4$). For BA161a, the $\Delta\chi^2$ at 0.10~mas exceeds 60, ruling out this reference size at very high confidence.}
\label{fig:vlbi_statistics}
\end{figure}

\begin{figure}[htbp]
\centering
\includegraphics[width=0.78\textwidth]{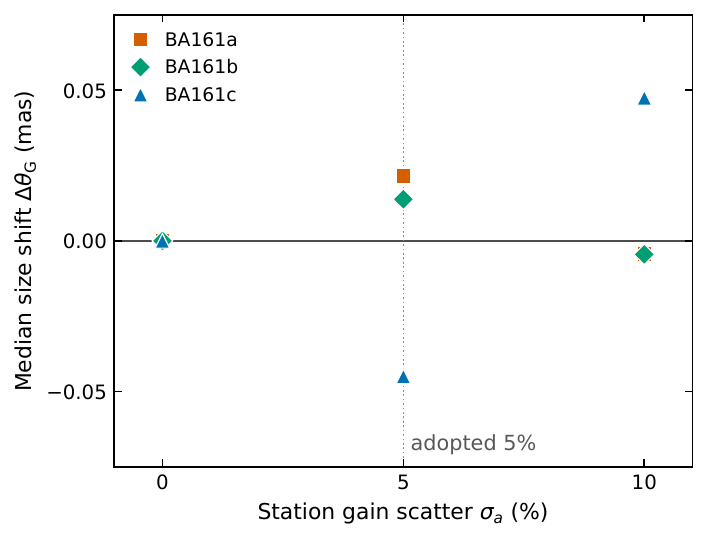}
\caption{\textbf{Station-dependent amplitude calibration test for the early 15-GHz epochs.}
Antenna gains were perturbed by a Gaussian scatter of standard deviation $\sigma_a$ and each epoch was refitted. Points show the median shift in recovered
$\theta_{\rm G}$ relative to the unperturbed fit, for BA161a (orange), BA161b (green) and BA161c (blue). The vertical dotted line marks the 5\% absolute amplitude-scale
uncertainty assumed for the phase calibrator J1905+1943. All three epochs are stable against gain perturbations up to $\sigma_a=10\%$.}
\label{fig:vlbi_amp_calibration}
\end{figure}

Three terms contribute to the continuous uncertainty on each measured angular size $\theta_{\rm G}$ (\SUPTAB{tab:vlbi_uncertainty_components}).
The statistical term $\sigma_{\rm stat}$ is obtained from the curvature of the $\chi^2$ surface at the best fit and verified by block-bootstrap resampling (\SUPFIG{fig:vlbi_statistics}). 
The amplitude-calibration term $\sigma_{\rm amp}$ is evaluated by drawing independent 5\% per-station gain perturbations and propagating them through the full visibility fit. The rms scatter of recovered $\theta_{\rm G}$ over many realizations defines $\sigma_{\rm amp}$ for each epoch. All early 15-GHz epochs are stable against gain perturbations up to 10\%, with $\sigma_{\rm amp} \le 0.05$~mas for BA161a and BA161b (\SUPFIG{fig:vlbi_amp_calibration}). BA161c shows a slightly larger response ($\sigma_{\rm amp} \simeq 0.11$~mas) because a greater fraction of its coherent flux lies on the long baselines that carry the size information, making it more sensitive to gain errors projected onto the visibility curvature.
The residual-coherence term $\sigma_{\rm coh}$ accounts for the size bias introduced by uncorrected atmospheric phase fluctuations, which attenuate the coherent visibility by approximately $\exp(-\sigma_{\phi}^2/2)$ and steepen the long-baseline visibility profile in a way that mimics a larger source. We bound this effect using an independent bright check source (J1923+2010) observed between GRB scans.
After scaling the measured phase rms to the GRB sightline with a Kolmogorov structure function, $\sigma_{\phi} \propto \psi^{5/6}$, where $\psi$ is the angular separation from the phase calibrator \citep{Asaki07,Pradel06}, the residual phases were applied to source injections with the observed $uv$ coverage. 
The scaled phases broadened a 0.10-mas injected source in every epoch, yet the broadened sizes remained well below the measured $\theta_{\rm G}$ values (\SUPTAB{tab:ba161_phase_coherence}), supporting that the resolved angular scale is a genuine source property rather than an artifact of residual phase decorrelation. The resulting size shifts of $\le 0.07$~mas define $\sigma_{\rm coh}$ as the quadrature contribution that conservatively covers this bias (\SUPTAB{tab:ba161_phase_coherence}).
These three terms are added in quadrature to give the continuous error $\sigma_{\rm cont}$ for each epoch (\SUPTAB{tab:vlbi_uncertainty_components}), which reflects the error under a fixed choice of data selection and brightness model.

\begin{figure}[htbp]
\centering
\includegraphics[width=0.98\textwidth]{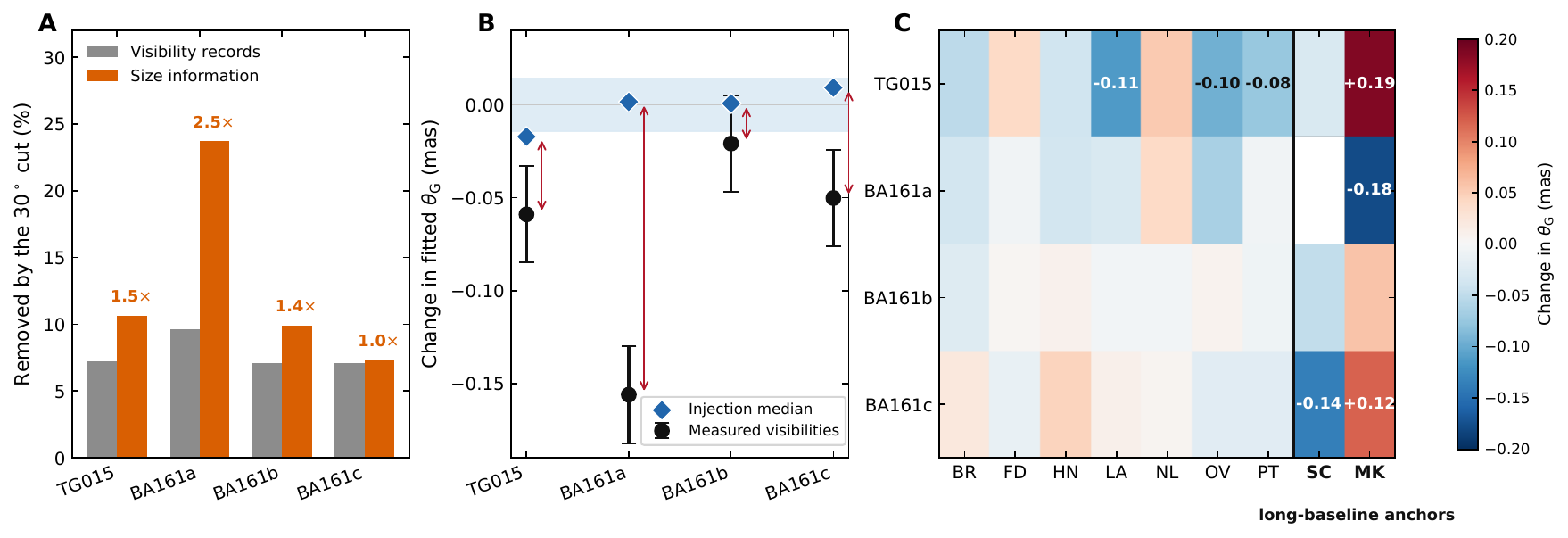}
\caption{\textbf{Visibility-selection tests for the four early 15-GHz epochs.} (\textbf{A}) Fraction of visibility records (grey) and Fisher information on size (orange) removed by a 30$^{\circ}$ elevation cut. Labels give the information-to-record removal ratio. (\textbf{B}) Change in $\theta_{\rm G}$ for the same cut: real data (black) and matched source injections (blue). The shadow band marks the 68\% scatter of the injection response. Red arrows mark the excess of the data over the injection expectation, showing that the selected long-baseline visibilities carry genuine source-size information. (\textbf{C}) Change in $\theta_{\rm G}$ after removing each station. SC and MK (bold) provide the longest baselines and produce the largest shifts. Small shifts from other stations confirm that no single antenna dominates the fitted size.}
\label{fig:vlbi_calibration_selection}
\end{figure}

The fitted size also depends on the analysis route. We assess this sensitivity through a set of qualified discrete branches: fits repeated after either (i) a localized, independently justified data removal (supported by calibrator amplitude, phase, or closure anomalies; \SUPFIG{fig:vlbi_calibration_selection}), or (ii) the adoption of an alternative single-component radial profile (uniform disk, thin ring, or annulus) that converges to a single identifiable scale. For each epoch, let $\theta_{\rm G}$ be the primary value, and let $\theta_{\rm min}$ and $\theta_{\rm max}$ be the extreme centres among all qualified branches. The signed branch deviations are $\Delta^{+}_{\rm branch} = \theta_{\rm max} - \theta_{\rm G}$ and $\Delta^{-}_{\rm branch} = \theta_{\rm G} - \theta_{\rm min}$ (\SUPTAB{tab:vlbi_uncertainty_components}). 
The conservative error envelope adopted for physical analysis is then $\sigma_{\rm cons}^{\pm} = \max [\sigma_{\rm cont}, \Delta^{\pm}_{\rm branch}]$, which does not assume a probability distribution over branches, but provides a worst-case coverage that acknowledges the systematic sensitivity to analysis choices. Epochs without qualified alternatives retain $\sigma_{\rm cont}$ as their total uncertainty. 

\begin{figure}[htbp]
\centering
\includegraphics[width=0.75\textwidth]{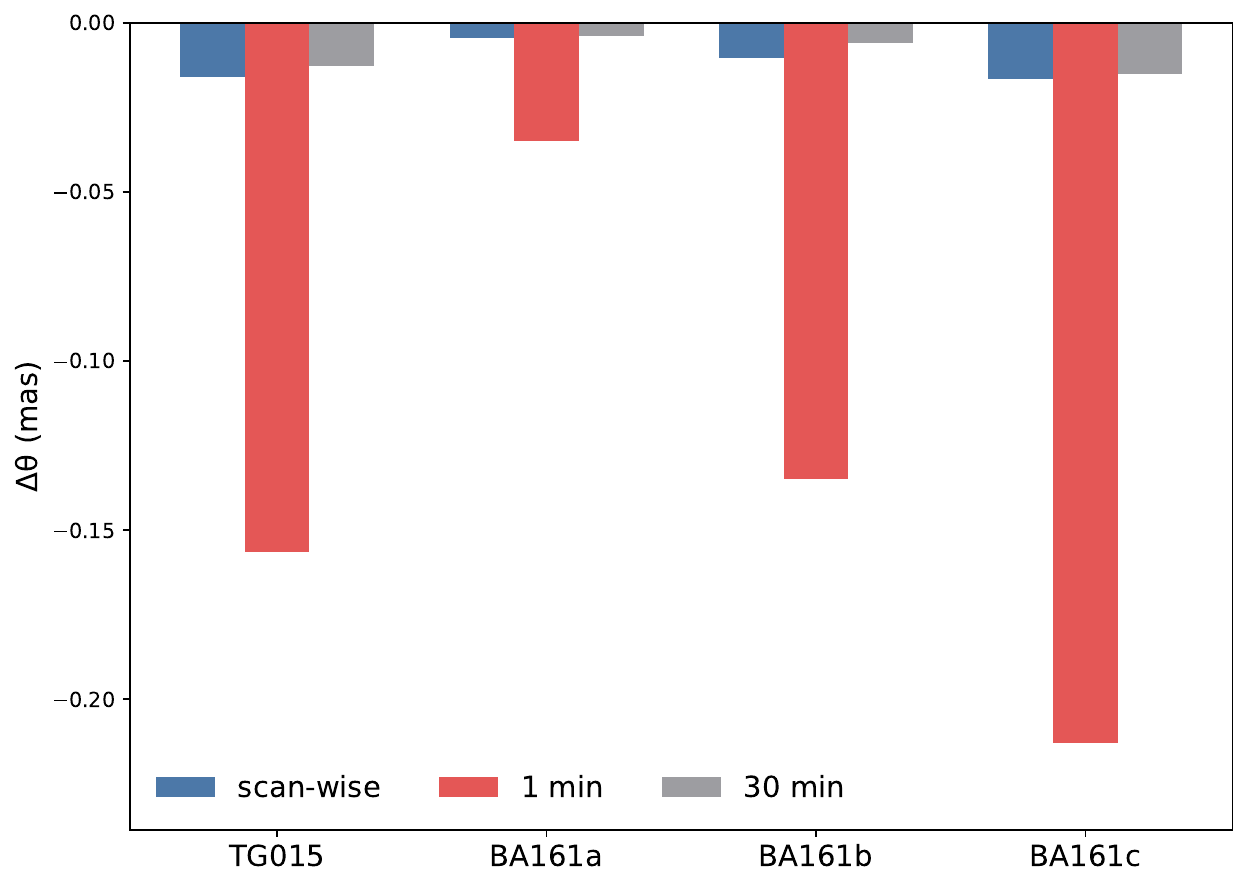}
\caption{\textbf{Target self-calibration bias for weak sources.} Mean shift $\Delta\theta = \theta_{\rm selfcal} - \theta_{\rm direct}$ from simulations using the actual $uv$ sampling and coherent flux of each early 15-GHz epoch. Colors denote the solution interval: scan-wise (blue), 1~min (red), and 30~min (grey). One-minute solutions compress the recovered size by 12--59\% in every realization, while scan-length and 30-min solutions produce negligible shifts of a few $\times 10^{-3}$~mas. The bias arises from the loss of long-baseline curvature information, and phase-reference calibration is therefore retained.}
\label{fig:vlbi_selfcal_bias}
\end{figure}

We retain phase-reference calibration throughout, rather than applying target self-calibration. For weak resolved sources such as GRB 221009A (median per-record S/N $\sim 1.5$; 30-s coherent S/N $\in [1.96, 5.44]$ for the early epochs), the long-baseline visibility phase is noise-dominated. Phase-only self-calibration would rotate this noise toward the starting-model phase, artificially flattening the long-baseline visibility envelope and biasing the fitted $\theta_{\rm G}$ downward. Simulations using the actual $uv$ sampling of each early epoch confirm this bias: one-minute solution intervals compress the recovered size by 12-59\% (\SUPFIG{fig:vlbi_selfcal_bias}). Scan-length and 30-minute solutions produce negligible mean shifts, confirming that the bias is driven by short-interval phase adjustments on noise-dominated baselines. 

VLBI flux densities for the BA161 epochs are compared with nearly coeval measurements from the AMI-LA (15.5~GHz) \citep{Rhodes24} and ATCA (16.7~GHz) connected-element arrays. BA161a and BA161c show no significant compact-flux deficit (VLBI/AMI $\ge$ 0.85; VLBI/ATCA $\ge$ 0.86). BA161b yields a lower VLBI/AMI ratio of 0.62 but a VLBI/ATCA ratio of 0.83. No epoch shows a persistent deficit against both arrays, indicating that broad, baseline-independent coherence loss is small and consistent with the 5\% absolute amplitude-scale uncertainty adopted for $\sigma_{\rm amp}$. Baseline-dependent phase decoherence, which affects the fitted $\theta_{\rm G}$ through the visibility curvature rather than the total flux, is accounted for separately by $\sigma_{\rm coh}$ as described above.

\begin{table}[htbp]
\centering
\caption{\textbf{Continuous uncertainty components and qualified discrete.} The three continuous terms (statistical, amplitude calibration, and residual coherence) are added in quadrature to obtain the continuous uncertainty. The final column reports the signed branch deviation from the primary results $\theta_{\rm G}$ among qualified alternative fits for that epoch. A dash means that no qualified alternative fit is available. BA160c is the independently published measurement.}
\label{tab:vlbi_uncertainty_components}
\resizebox{\textwidth}{!}{%
\begin{tabular}{lcccccccc}
\hline\hline
Epoch & $t$ & $\nu$ & $\theta_{\rm G}$ & $\sigma_{\rm stat}$ & $\sigma_{\rm amp}$ & $\sigma_{\rm coh}$ & $\sigma_{\rm cont}$ & $\Delta^{-}_{\rm branch}/\Delta^{+}_{\rm branch}$ \\
 & (d) & (GHz) & (mas) & (mas) & (mas) & (mas) & (mas) & (mas) \\
\hline
TG015   & 5.491   & 15.168 & 0.274 & 0.052 & 0.045 & 0.023 & 0.072 & 0.050/0.178 \\
BA161a  & 9.443   & 15.168 & 0.316 & 0.018 & 0.044 & 0.026 & 0.054 & 0.028/0.099 \\
BL307a  & 12.434  & 8.368  & 0.260 & 0.030 & 0.039 & 0.022 & 0.053 & 0.012/0.000 \\
BA161b  & 18.453  & 15.168 & 0.385 & 0.021 & 0.042 & 0.032 & 0.057 & 0.011/0.017 \\
BL307b  & 21.421  & 8.368  & 0.268 & 0.029 & 0.040 & 0.022 & 0.054 & 0.010/0.000 \\
BA161c  & 26.432  & 15.168 & 0.415 & 0.029 & 0.107 & 0.035 & 0.116 & 0.022/0.029 \\
BL307c  & 36.356  & 8.368  & 0.285 & 0.051 & 0.043 & 0.024 & 0.070 & 0.013/0.000 \\
BA160b  & 44.384  & 15.168 & 0.296 & 0.050 & 0.044 & 0.025 & 0.071 & -- \\
BL307d  & 64.256  & 8.368  & 0.326 & 0.096 & 0.049 & 0.027 & 0.111 & 0.004/0.000 \\
BL307e  & 109.187 & 8.368  & 0.491 & 0.130 & 0.073 & 0.041 & 0.155 & 0.033/0.000 \\
BA160c  & 114.186 & 15.200 & 0.450 & 0.190 & -- & -- & 0.190 & -- \\
BL307f  & 201.923 & 8.368  & 0.764 & 0.187 & 0.114 & 0.064 & 0.228 & 0.037/0.000 \\
BA160c1 & 204.975 & 15.168 & 1.584 & 0.821 & 0.237 & 0.132 & 0.865 & -- \\
BA160d  & 261.733 & 15.168 & 1.332 & 0.793 & 0.199 & 0.111 & 0.825 & -- \\
BL307g  & 298.672 & 8.368  & 0.899 & 0.396 & 0.135 & 0.075 & 0.425 & 0.056/0.000 \\
\hline
\end{tabular}}
\end{table}

\newpage

\begin{table}[htbp]
\centering
\caption{\textbf{Residual-phase check for the BA161 epochs.} The scaled values account for the smaller separation between J1905+1943 and the GRB. Injection results refer to an assumed small source with a size of 0.10-mas. The phase response is the estimated change in fitted size when the scaled phase response is evaluated near the measured angular scale.}
\label{tab:ba161_phase_coherence}
\centering
\begin{tabular}{l c c c c c c c}
\hline\hline
Epoch & \multicolumn{2}{c}{Raw phase rms (deg)} & \multicolumn{2}{c}{Scaled phase rms (deg)} & \multicolumn{2}{c}{Injection recovery (mas)} & Measured \\
\cline{2-3} \cline{4-5} \cline{6-7}
 & MK & SC & MK & SC & median & largest & $\theta_{\rm G}$ (mas) \\
\hline
BA161a & 62.9 & --   & 30.6 & --   & 0.179 & 0.249 & 0.316 \\
BA161b & 83.3 & 82.5 & 40.5 & 40.1 & 0.189 & 0.323 & 0.385 \\
BA161c & 96.4 & 90.2 & 46.9 & 43.9 & 0.208 & 0.335 & 0.415 \\
\hline
\multicolumn{8}{l}{Phase response (mas): BA161a = 0.036,\quad BA161b = 0.069,\quad BA161c = 0.062}\\
\hline
\end{tabular}
\end{table}

\clearpage

\subsubsection{Source size evolution}\label{sec:size_evolution}

To model the size evolution, we fit the measured angular sizes
$\theta_{\mathrm{G},i}$ at each epoch $i$ with both a single power-law (SPL, $\propto t_{\rm obs}^{\alpha}$) and a broken power-law (BPL, $\propto t_{\rm obs}^{\alpha_1}$ before a transition time $t_{\rm tr}$ and $\propto t_{\rm obs}^{\alpha_2}$ afterward). The uncertainties assigned to each data point are the conservative error envelopes $\sigma_{\rm cons}^{\pm}$ discussed above. 
To account for the resulting asymmetry in the error bars, we utilized a linear variance log-likelihood function \citep{Barlow26},
\begin{equation}
\ln (\mathcal{L}) = - \frac{1}{2} \chi^2 (\vec{x}) = - \frac{1}{2} \sum_{i=1}^{N_{\rm ep}} \left[ \frac{ (\theta_{\mathrm{G}}(t_{\mathrm{obs},i};\vec{x}) - \theta_{\mathrm{G},i})^2}{V_i +  (\theta_{\mathrm{G}}(t_{\mathrm{obs},i};\vec{x})- \theta_{\mathrm{G},i}) V_i^{\prime}} \right],  
\end{equation}
where $\theta_{\mathrm{G}} (t_{\mathrm{obs},i};\vec{x})$ represents the model size for the epoch at the observer time $t_{\mathrm{obs},i}$ (with zero point of Fermi-GBM trigger time $T_0$) and $\vec{x}$ denotes the set of free parameters in the model, $V_i = \sigma_{i,+} \sigma_{i,-}$ represents an effective symmetric variance, and $V_i^{\prime} = f_i (\sigma_{i,+}-\sigma_{i,-})$ introduces a correction for asymmetry.
A correction factor $f_i$ is applied to mitigate unphysical negative denominators, $f_i=1$ for deviations within $3\sigma_{i,\pm}$, and $f_i=0$ otherwise.

For GRB 221009A, an SPL fit to all VLBA size measurements at 15 and 8 GHz gives $\theta_{\mathrm{G}} \propto t_{\rm obs}^{0.18}$, with a reduced chi-square $\chi^2_{\rm red} = 0.87$. This expansion rate is shallower than the $t_{\rm obs}^{0.69}$ derived from late-time VLBI size measurements alone \citep{Giarratana24}.
A BPL fit yields a slow initial phase $\theta_{\mathrm{G}} \propto t_{\rm obs}^{0.11}$ followed by a faster expansion with $\theta_{\mathrm{G}} \propto t_{\rm obs}^{0.91}$ after a transition time $t_{\rm tr} \simeq 106$~day, with $\chi^2_{\rm red} = 0.63$.
The BPL fit provides only marginal improvement in $\chi^2_{\rm red}$, mainly because the late-epoch measurements carry large systematic uncertainties; hence, the data do not statistically require a break.
However, the transition time $t_{\rm tr}$ coincides with the onset of a prominent bump appearing in the 15 GHz lightcurve (\FIG{fig:SizeFit}), which is naturally expected from a change in the jet dynamics. This temporal coincidence lends independent support to the BPL description, despite the limited statistical significance.
For comparison, GRB 030329 is well described by an SPL with
$\theta_{\mathrm{G}} \propto t_{\rm obs}^{0.43}$ and $\chi^2_{\rm red} = 0.41$.

\section{Supplementary text}\label{sec:supp_text}

\subsection{Apparent superluminal motion}\label{sec:superluminal}
Our first size measurement at the epoch BA161a yields an apparent diameter of $D_{\perp,\mathrm{1st}} = \theta_{\mathrm{G,1st}} \times D_{\rm A} \simeq 2.7 \times 10^{18}$~cm around 9 days post-burst.   
Hereafter, we adopt an angular diameter distance $D_{\rm A} = 562.6\ \mathrm{Mpc}$ corresponding to $z = 0.151$ in a flat $\Lambda$CDM framework with Planck cosmological parameters \citep{Planck20}. 
This gives an average apparent expansion velocity of 
\begin{equation}
\beta_{\rm app} = \frac{(1+z) D_{\perp,\mathrm{1st}}}{2 t_{\rm obs} c} \simeq 63 \pm 11,
\end{equation} 
the highest value ever probed via VLBI observations, surpassing values of $\sim 5$ for GRB 030329 at 24 days \citep{Taylor04} and $\sim 8$ for GRB 170817A between 8 days and 75 days post-burst \citep{Ghirlanda19,Mooley18,Mooley22}.
For a conical jet geometry for GRBs, we have
$D_{\perp} = 2 R \times \sin(\min[1/\Gamma, \theta_{\rm j}])$,
where $R$ is the emitting distance from the explosion center, $\theta_{\rm j}$ is the jet half-opening angle, and $1/\Gamma$ accounts for the region visible due to the Doppler beaming in a relativistic outflow with bulk Lorentz factor $\Gamma$.
At approximately 9 days post-burst, the apparent diameter $D_{\perp,\mathrm{1st}}$ of GRB 221009A is about five times larger than that of GRB 030329 (\FIG{fig:size_evol}).
Given that the expansion of $R$ scales with burst energy (see detailed dependence below), this discrepancy directly reflects the difference in energy between the two bursts, about $10^{55}$ erg for GRB 221009A versus about $10^{52}$ erg for GRB 030329. Unlike the proper motion of discrete knots commonly reported for Active Galactic Nuclei (AGN) jets, $\beta_{\rm app}$ measures the growth rate of the Gaussian-equivalent radial scale. Both quantities can exceed $c$ due to light-travel effects, but $\beta_{\rm app}$ traces the outer boundary of the beaming-visible region, reflecting the emission radius in our case.

\subsection{Physical analysis of size evolution}\label{sec:size_analysis}

The overall VLBA observations indicate a slow-evolving stage
of $\theta_{\mathrm{G}} \propto t_{\rm obs}^{0.18}$ and a tentative fast-evolving stage for the period after $\sim 106$~days.
In the following, we show that the VLBI information provides insight into the complex jet dynamics of GRB 221009A.

\subsubsection{The top-hat jet scenario}\label{sec:tophat}
We start with a top-hat jet (\SUPFIG{fig:Schematic}A) characterized by an isotropic kinetic energy $E_{\rm K, iso}$, an initial bulk Lorentz factor $\Gamma_0$, and a half-opening angle $\theta_{\rm j,0}$, propagating into a circumburst medium with a power-law density profile $\propto R^{-s}$.
The jet begins to decelerate significantly when the blastwave sweeps up a certain amount of material. The deceleration timescale is 
\begin{equation}
t_{\rm dec} \simeq
\begin{cases}
10^2 \left(\frac{E_{\rm K,iso}}{10^{55}~\mathrm{erg}} \right)^{1/3}
\left(\frac{\Gamma_0}{300} \right)^{-8/3} 
\left(\frac{n_{\mathrm {ISM}}}{0.1~\mathrm{cm}^{-3}} \right)^{-1/3} ~\mathrm{s} \quad\quad &\mathrm{for~~~~~~ISM}, \\ 
40 \left(\frac{E_{\rm K,iso}}{10^{55}~\mathrm{erg}} \right)
\left(\frac{\Gamma_0}{300} \right)^{-4} 
\left(\frac{A_*}{0.1} \right)^{-1} ~\mathrm{s}  &\mathrm{for~~~~~~Wind}, \\
\end{cases}
\label{eq:tdec}
\end{equation}
where $n_{\mathrm {ISM}}$ is the interstellar medium (ISM) number density, and $A_*$ is the dimensionless wind parameter.
Our observations ($t_{\rm obs} > 10^5$~s) lie in the deceleration phase of the jet, with $\Gamma_0 \ge 30$ applicable to both ISM ($s = 0$) and wind-like ($s = 2$) environments.
During this stage, the jet dynamics follows the Blandford-McKee (BM) solution \citep{Blandford76}, i.e., $R \propto t^{1/(4-s)}_{\rm obs}$ and $\Gamma \propto t^{(s-3)/(8-2s)}_{\rm obs}$.
Assuming that $\theta_{\rm j} = \theta_{\rm j,0}$ is maintained during its propagation, the projected angular size is $\theta_{\mathrm{G}} = 2 R \sin (\min[\Gamma^{-1},\theta_{\mathrm{j},0}])/D_{\rm A}$, leading to
\begin{equation}
\theta_{\mathrm{G}} \propto
\begin{cases}
t_{\rm obs}^{5/8}  \quad\quad\quad &\mathrm{for~~~~~~} t_{\rm obs} < t_{\rm j}, \\
t_{\rm obs}^{1/4}  \quad\quad\quad &\mathrm{for~~~~~~} t_{\rm obs} \ge t_{\rm j}, 
\end{cases}
\end{equation}
for $s = 0$, and
\begin{equation}
\theta_{\mathrm{G}} \propto
\begin{cases}
t_{\rm obs}^{3/4}  \quad\quad\quad &\mathrm{for~~~~~~} t_{\rm obs} < t_{\rm j}, \\
t_{\rm obs}^{1/2}  \quad\quad\quad &\mathrm{for~~~~~~} t_{\rm obs} \ge t_{\rm j}, 
\end{cases}
\end{equation}
for $s = 2$. Here, $t_{\rm j}$ denotes the observational jet break time, defined by $\Gamma(t_{\rm j}) = \theta_{\rm j,0}^{-1}$. 
Jet breaks constrained by TeV \citep{LHAASO23} or X-ray/optical \citep{OConnor23} lightcurves suggest $t_{\rm j} \lesssim 10^5$~s.
The size measurements of the early epochs show that the angular size exceeds 0.2 mas around 9~days (1 $\sigma$; \FIG{fig:size_evol}), yielding a constraint on the jet angle of
\begin{equation}
\theta_{\rm j,0} \ge 0.15 \left( \frac{E_{\rm K,iso}}{10^{55}~\mathrm{erg}} \right)^{-1/4} \left( \frac{n_{\rm ISM}}{1~\mathrm{cm}^{-3}} \right)^{1/4} \left(\frac{\theta_{\mathrm{G}}}{0.2~\mathrm{mas}} \right) \left( \frac{t_{\rm obs}}{9~\mathrm{days}} \right)^{-1/4}~\mathrm{rad},
\label{eq:ll_ISM}
\end{equation}
for the ISM case, and
\begin{equation}
\theta_{\rm j,0} \ge 0.01 \left( \frac{E_{\rm K,iso}}{10^{55}~\mathrm{erg}} \right)^{-1/2} \left( \frac{A_*}{0.1} \right)^{1/2} \left(\frac{\theta_{\mathrm{G}}}{0.2~\mathrm{mas}} \right) \left( \frac{t_{\rm obs}}{9~\mathrm{days}} \right)^{-1/2}~\mathrm{rad},
\label{eq:ll_wind}
\end{equation}
for the wind case. Given the narrow jet nature ($\theta_{\rm j,0} < 0.1$) supported by other evidence~\citep{Burns23,Williams23,OConnor23}, these lower limits imply that the jet must have undergone an early fast-expansion phase in a wind-like environment to reach the observed angular size - unless the ISM density is extremely low, $n_{\rm ISM} \le 10^{-4}~\mathrm{cm}^{-3}$.
The temporal evolution of the X-ray and optical afterglow within $10^6$~s also favors a wind environment~\citep{OConnor23,Ren23}.
Since $E_{\rm K,iso}$ is unlikely to significantly exceed the prompt emission energy of $10^{55}$ erg for such a bright burst, and a moderately dense stellar wind ($A_{*} \ge 0.1$) is commonly adopted in afterglow modeling~\citep{Gill23,Ren24,Zheng24}, Eq.~(\ref{eq:ll_wind}) provides a robust lower limit of $\theta_{\rm j,0} \ge 0.01$ for GRB 221009A.
The higher $E_{\rm K,iso}$ and wind-like environment lead to significantly larger angular sizes compared to GRB 030329 (\FIG{fig:size_evol}).

In contrast, for epochs between $10^6$~s and $10^7$~s, the theoretical scaling $\theta_{\mathrm{G}} \propto t_{\rm obs}^{1/4}$ for the ISM case aligns better with the observed slow-evolution stage than the wind scaling $\theta_{\mathrm{G}} \propto t_{\rm obs}^{1/2}$ (\SUPFIG{fig:ScalingLaws}).
This suggests a change to an ISM-like environment at large distances, beyond the progenitor's stellar wind region (also see the afterglow modeling below).

\subsubsection{The structured jet scenario}\label{sec:structured}
In contrast to a top-hat jet, a realistic jet may possess a structured profile (\SUPFIG{fig:Schematic}B), shaped either by the central engine launching process or by jet–envelope interactions \citep{Tchekhovskoy08,Lazzati12,Geng16b,Gottlieb20}.
As commonly adopted in the literature, we assume that the isotropic-equivalent kinetic energy and the initial bulk Lorentz factor are uniform within a core of half-opening angle $\theta_{\rm c}$, and decay as power laws in the sheath ($\theta_{\rm c} < \theta \le \theta_{\rm edge}$): $d E_{\rm K,iso}/d \theta \propto \theta^{-k_{E}}$ and $\Gamma_0(\theta) \propto \theta^{-k_{\Gamma}}$ with two corresponding indices of $k_{E}$ and $k_{\Gamma}$.
In the period of interest ($t_{\rm obs} > 10^5$~s), the bulk Lorentz factor at different latitudes enters the BM self-similar regime (\EQ{eq:tdec}), so the initial $\Gamma_0(\theta)$ profile does not significantly affect the following analysis.
If the jet remains mildly relativistic at large latitudes, the observed source size is determined by the gradual entry of different latitudes into the field of view as deceleration proceeds. Using the condition $\Gamma(\theta,t_{\rm obs}) = \theta^{-1}$ and the BM scaling for $\Gamma$, the angular size evolution can be expressed as
\begin{equation}
\theta_{\mathrm{G}} \propto
\begin{cases}
t_{\rm obs}^{(5-k_{E})/(8-k_{E})},    \quad\quad \mathrm{for} \quad\quad s = 0, \\
t_{\rm obs}^{(3-k_{E})/(4-k_{E})},    \quad\quad \mathrm{for} \quad\quad s = 2. \\
\end{cases}
\label{eq:struc_size}
\end{equation}
While the temporal index can mathematically take any real value, physically plausible values of $k_E$ fall into three intervals separated by critical values.
Taking $s = 0$ as an example, for an interval of $k_E \in [8, \infty)$, it corresponds to an image of a bright core and an outer dim ring whose inner boundary moves inward. The faint ring remains undetectable due to the steep energy decline ($E_{\rm K,iso} \propto \theta^{-k_E}$) and does not influence the size evolution. For $k_E \in [4, 8]$, the sheath expands more slowly than the core, so the overall size evolution is dominated by the fast-growing core, asymptotically resembling a top-hat jet (\SUPFIG{fig:ScalingLaws}).
For $k_E \in [0, 4]$, the behavior spans from an extended spherical regime (effectively a top-hat jet with opening angle $\theta_{\rm edge}$, as $k_{E} \to 0$) to a jet-like regime (approaching a top-hat with angle $\theta_{\rm c}$, as $k_{E} \to 4$).
Our measured size evolution in the ISM environment requires $k_E = 4.34^{+0.33}_{-0.34}$, which is very close to the top-hat–like limit $k_{E} \simeq 4$. For the wind medium, the data would imply $k_E = 2.78^{+0.11}_{-0.11}$, which exceeds the critical value $k_{E} = 2$. When $k_{E} > 2$, the temporal slope exceeds the top-hat value ($\propto t_{\rm obs}^{1/2}$), which conflicts with the observed shallow rise.
Therefore, in the structured-jet scenario, the source size evolution also supports an ISM-like environment.

\subsubsection{The spreading jet scenario}\label{sec:spreading}
\begin{figure}
  \centering
  \includegraphics[width=\textwidth]{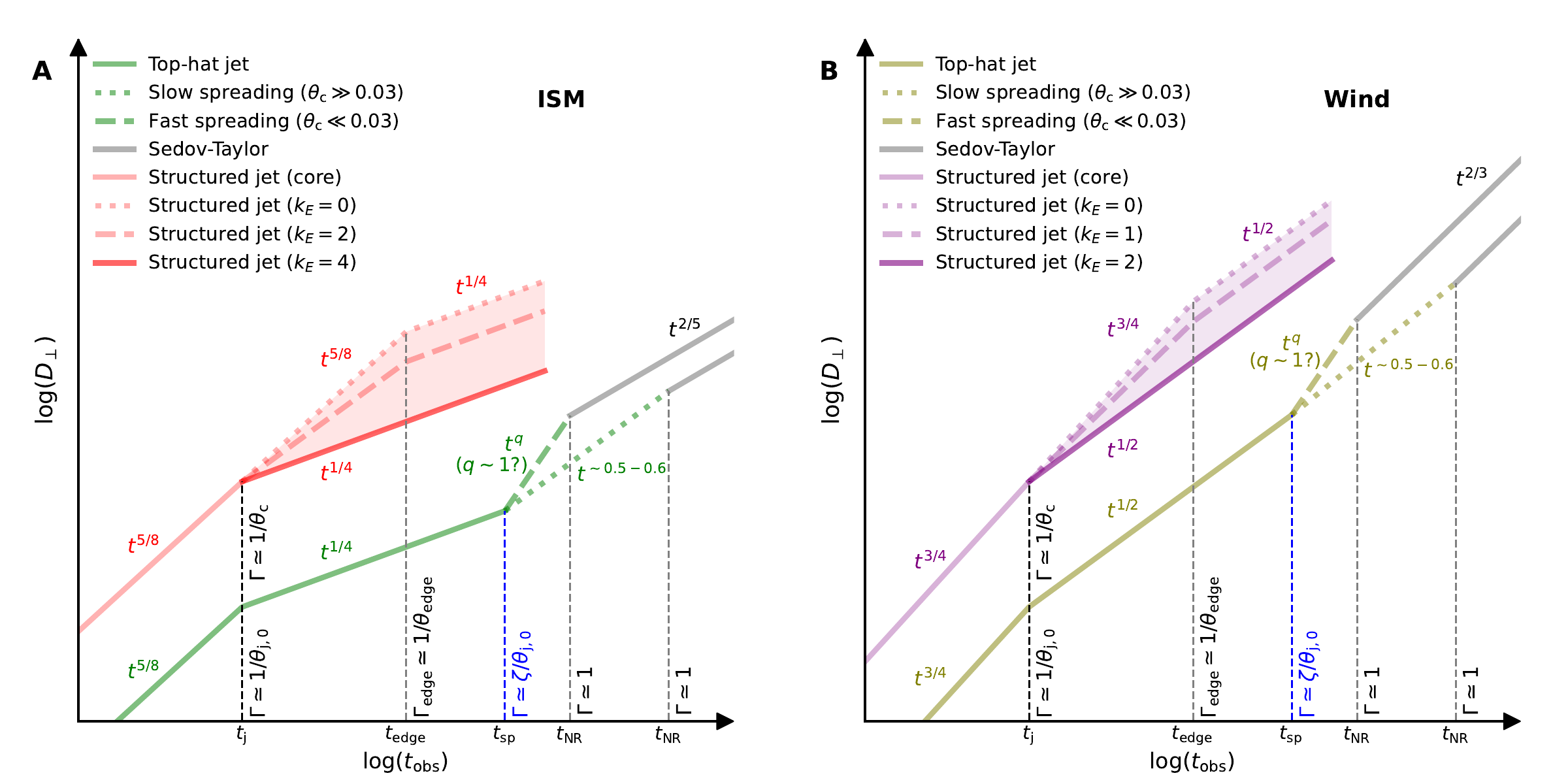}
  \caption{
  {\bf Summary of scaling laws for the angular diameter ($D_{\perp}$) evolution in different scenarios.} 
  (\textbf{A}) The analytical evolution of $D_{\perp}$ as a function of observer time in an ISM. The green-grey and red lines represent the case of a top-hat jet and a structured jet, respectively.
  For the top-hat jet case, the size evolution scalings in the spherical regime ($t_{\rm obs} < t_{\rm j}$), the non-spreading jet regime ($t_{\rm j} < t_{\rm obs} < t_{\rm sp}$), the lateral-spreading jet regime ($t_{\rm sp} < t_{\rm obs} < t_{\rm NR}$), and the non-relativistic Sedov-Taylor regime ($t_{\rm obs} > t_{\rm NR}$) are illustrated, with relevant temporal indices marked. The lateral spreading could be divided into the fast spreading and the slow spreading cases \citep{Granot12}. For the conventional structured jet case, three representative cases with critical $k_E$ (see \textbf{Supplementary text}) are shown in dotted, dashed, and solid lines, respectively.
  (\textbf{B}) The case of a wind environment. The exact temporal index of $D_{\perp}$ in the fast-spreading regime may vary according to the jet properties.
  } 
 \label{fig:ScalingLaws}
\end{figure}

The relativistic jet of GRB 221009A may undergo intrinsic evolution during its propagation (\SUPFIG{fig:Schematic}C). 
For relativistic outflows, the onset of lateral spreading is governed by causal contact. 
Initially, the jet follows the BM solution, maintaining a highly collimated beam due to the absence of transverse causal connection within the characteristic timescale $t_{\rm j}$.
Consider a sound wave originating at the jet edge when the shock Lorentz factor is $\Gamma_{\rm start}$, full causal connection across the shock front is established when the Lorentz factor decreases to $\Gamma_{\rm sp} \simeq [(3-s) \theta_{\rm j,0} + \Gamma_{\rm start}^{-1}]^{-1}$~\citep{Granot12,vanEerten18}. 
Thereafter, sideways expansion becomes significant for the outflow, marking a critical transition in the jet's dynamical evolution. 
The spreading behavior exhibits strong dependence on $\theta_{\rm j,0}$,
which could be slow spreading \citep{Panaitescu99,Rhoads99,Granot05} or fast spreading \citep{ZhangW09,Wygoda11,Granot12,Govreen24} (\SUPFIG{fig:ScalingLaws}).
For typical bursts with $\theta_{\rm j,0} = 0.1$ and $s = 0$, $\Gamma_{\rm sp} \le 3$ means a marginally relativistic expansion.
However, in the case of GRB 221009A's exceptionally narrow jet ($\theta_{\rm j,0} \le 0.03$), $\Gamma_{\rm sp}
\simeq 10$ indicates significantly faster spreading dynamics, which could explain its tentative faster expansion rate at the late stage.

We assume that significant lateral spreading occurs when the jet Lorentz factor approaches $\zeta/\theta_{\rm j,0}$ at $t_{\rm sp}$ in the observer frame \citep{Rhoads99,Sari99,Granot12}, where $\zeta$ is a dimensionless parameter less than unity and its accurate value is still uncertain in the literature \citep{ZhangW09,Wygoda11,Granot12,vanEerten12,Lazzati18,Govreen24,Wang24}.
If we attribute the observed slow-evolving stage to a top-hat jet, the fast spreading could in principle account for the fast-evolving stage of the size. 
The BM evolving law of $\Gamma$ gives
\begin{equation}
\zeta \simeq \left( \frac{t_{\rm sp}}{t_{\rm j}}\right)^{(s-3)/(8-2s)}.
\end{equation}
Our VLBI observations, which give $t_{\rm sp} \ge 106$~days and $t_{\rm j} \le 5$~days, provide the first direct observational constraints on this parameter. These yield upper limits of $\zeta \le 0.3$ and $\zeta \le 0.5$ for the ISM and the wind case, respectively.

\subsubsection{Emergence of a new emission component from large latitudes}
\begin{figure}
  \centering
  \includegraphics[width=0.67\textwidth]{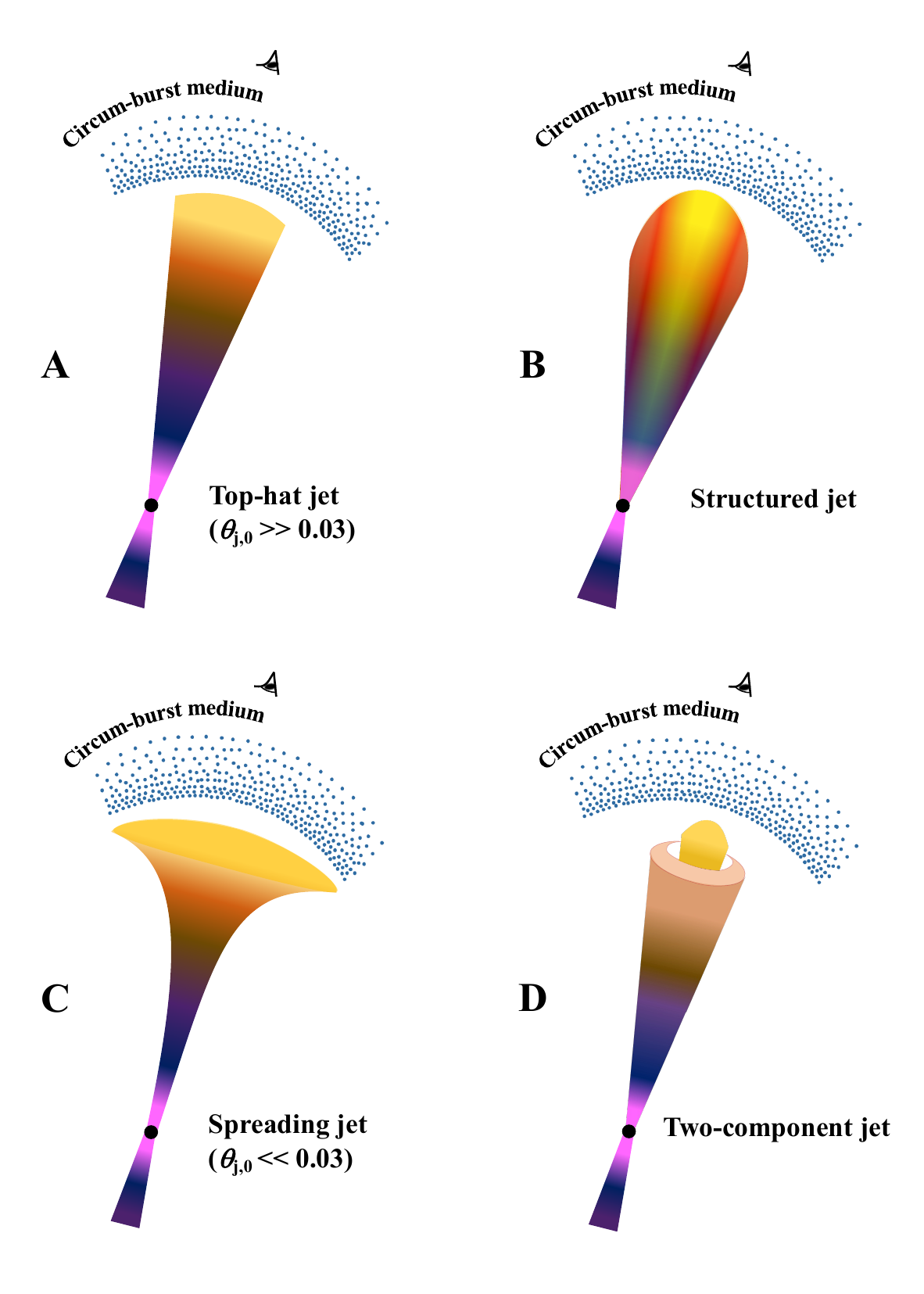}
  \caption{
  {\bf A schematic illustration of different scenarios of GRB jets in the relativistic phase.} 
  In each panel, the eye indicates the on-axis observer. 
  (\textbf{A}) A classical top-hat jet with a uniform distribution of energy and velocity across the jet surface. (\textbf{B}) A prevailing structured jet with the conventional profile. (\textbf{C}) A core-dominated jet spreading at the late stage. (\textbf{D}) A two-component jet with an inner narrow outflow and an outer wide one separated by a hollow region. 
  } 
 \label{fig:Schematic}
\end{figure}

Another possibility is that the slow-evolving stage is attributed to a top-hat jet or a structured jet with large $k_E$, while the tentative fast-evolving stage is caused by the emergence of another emission component appearing abruptly at larger latitudes at the late stage (\SUPFIG{fig:Schematic}D).
The absence of a significant shift of the centroid in the radio image indicates this component should be concentric with the inner component, e.g., another concentric wider jet with an inner edge angle of $\theta_2$.
At the jet break time of the wide component, $t_{\rm j,2}$, the bulk Lorentz factor of the wide jet is
\begin{equation}
\Gamma_{\rm j,2} \simeq \frac{D_{\perp}}{4 c t_{\rm j,2}} \simeq 3 \left( \frac{D_{\perp}}{7 \times 10^{18}~\mathrm{cm}} \right) \left( \frac{t_{\rm j,2}}{200~\mathrm{days}} \right)^{-1},
\end{equation}
as inferred from \FIG{fig:size_evol}. The corresponding edge angle is $\theta_2 \simeq \Gamma_{\rm j,2}^{-1} \simeq 0.3$~rad. 
On the other hand, the equivalent isotropic energy of the wide jet should satisfy a ratio of
\begin{equation}
\frac{E_{\rm K,iso,2}}{E_{\rm K,iso,1}} \simeq \left( \frac{\theta_{\rm j,0}}{\theta_2} \right)^{2 (4-s)} \left( \frac{t_{\rm j,1}}{t_{\rm j,2}} \right)^{s-3}.    
\end{equation}
For $t_{\rm j,2}/t_{\rm j,1} \ge 10^2$ constrained from observations, the jet energy ratio is 
$E_{\rm K,iso,2} \theta_2^2 /E_{\rm K,iso,1} \theta_{\rm j,0}^2 \gtrsim 1$ if a conservative value of $\theta_2/\theta_{\rm j,0} \simeq 10$ is adopted, both for $s = 0$ or $s = 2$.
The required large jet energy from the outer wide jet has not been predicted from previous numerical simulations \citep{Lazzati12}. Such a component, if exists, may require an additional physical mechanism to launch. 
The size evolution of the wide component after $t_{\rm j,2}$ is again similar to that of a top-hat jet, the long-term monitoring could verify this scenario.

\subsection{Modeling of multi-wavelength emission}
Based on the analyses of the size observations above, we move to interpret the complex temporal and spectral behaviors of GRB 221009A and assess how our observations advance our understanding of the origin of the multi-wavelength emission. 

\subsubsection{External shock scenario}
In the standard external shock scenario \citep{Sari98}, the interstellar medium is shocked by a blastwave with a bulk Lorentz factor of $\Gamma$, and accelerated electrons emit multi-wavelength photons through synchrotron radiation.
Assuming the distribution of the accelerated electrons is a power-law with an index of $p$ and a minimum Lorentz factor of $\gamma_\mathrm{m}$,
the synchrotron radiation of these electrons will shine in the X-ray, optical, and radio bands over time,
as electrons cool in the comoving frame magnetic field of $B^{\prime}$ (the superscript of $\prime$ denotes the quantities in the comoving frame hereafter).
The cooling Lorentz factor $\gamma_{c} = \frac{6 \pi m_\mathrm{e} c (1+z)}{\sigma_T \Gamma B^{\prime 2} (1+Y) t_{\rm obs}}$ is thus commonly introduced to describe the evolving electron spectrum, where $m_\mathrm{e}$ is the electron mass, $c$ is the speed of light, $\sigma_T$ is the Thomson cross-section,
and $Y$ is the Compton parameter defined as the ratio of the inverse Compton power to the synchrotron power.
These two characteristic frequencies, $\nu_\mathrm{m} \propto \Gamma B^{\prime} \gamma_\mathrm{m}^2$ and $\nu_\mathrm{c} \propto \Gamma B^{\prime} \gamma_\mathrm{c}^2$,
together with the spectral peak flux $F_{\rm max}^{\rm Syn} \propto \Gamma N_\mathrm{e} B^{\prime}$ by a total of $N_\mathrm{e}$ electrons,
could well define the observed synchrotron spectra shape \citep{Sari98,Sari99}.
In addition, synchrotron photons would Compton scatter on electrons, producing an inverse Compton (IC) component at higher energies.
IC scattering by self-emitted synchrotron photons is referred to as the synchrotron self-Compton (SSC) process,
and the corresponding spectra could also be approximated as a broken power law with characteristic frequencies
of $\nu_\mathrm{m}^{\mathrm{IC}} = 2 \gamma_\mathrm{m}^2 \nu_\mathrm{m}$ and $\nu_\mathrm{c}^{\mathrm{IC}} = 2 \gamma_\mathrm{c}^2 \nu_\mathrm{c}$,
and a spectral peak flux of $F_{\mathrm{max}}^{\mathrm{IC}} \simeq x_0 \sigma_T n_{\mathrm {ISM}} R F_{\mathrm{max}}^{\mathrm{Syn}}$,
where $x_0 \sim 0.5$ is a correction factor, $n_{\mathrm{ISM}}$ is the number density of interstellar medium \citep{Sari01}.

In this external shock scenario, the shock initially coasts with a constant Lorentz factor $\Gamma \simeq \Gamma_0 = \mathrm{const}$. 
During this stage, the TeV emission flux follows the relation \citep{Sari01} $F_{\nu_{\rm TeV}} = F_{\mathrm{max}}^{\mathrm {IC}} \frac{p-1}{p+1} (\frac{\nu_\mathrm{c}}{\nu_\mathrm{m}}) (\frac{\nu_{\rm TeV}}{\nu_\mathrm{m}^{\mathrm {IC}}})^{-p/2} \propto t_{\rm obs}^2 \nu_{\rm TeV}^{-p/2}$ for $\nu_{\rm m}^{\rm IC} < \nu_{\rm c}^{\rm IC} < \nu_{\rm TeV}$, which successfully explains the observed temporal and spectral characteristics of the rising TeV emission \citep{LHAASO23}.
The TeV lightcurve peaks when significant shock deceleration begins, reaching a maximum flux of
\begin{equation}
F_{\nu_{\rm TeV}}^{\mathrm{Peak}} \propto n_{\rm ISM}
R(T_{\mathrm{Peak}}) F_{\mathrm{max}}^{\mathrm{Syn}} \left( \frac{\nu_\mathrm{c}}{\nu_\mathrm{m}} \right) \left( \frac{\nu_{\rm TeV}}{\nu_\mathrm{m}^{\mathrm {IC}}} \right)^{-p/2} \propto
\Gamma_0^{3p+2} n_{\rm ISM}^{(p+2)/4} \epsilon_{\mathrm{e}}^{2p-2} \epsilon_{B}^{(p-6)/4} T_{\mathrm{Peak}}^2 \nu_{\rm TeV}^{-p/2},
\end{equation}
at the peak time given by
\begin{equation}
T_{\mathrm{Peak}} = \left[\frac{3 (1+z)^3}{32 \pi m_{\rm p} c^5} \right]^{1/3} E_{\rm K,iso}^{1/3} n_{\rm ISM}^{-1/3} \Gamma_0^{-8/3},
\label{eq:Tpeak}
\end{equation}
where $E_{\mathrm{K,iso}}$ is the initial isotropic kinetic energy of the blastwave, $\epsilon_\mathrm{e}$ and $\epsilon_B$ are equipartition parameters for post-shock electrons and the magnetic field, respectively.
After the coasting stage, $\Gamma$ evolves as $\Gamma \propto t_{\mathrm {obs}}^{-3/8}$ according to the BM solution \citep{Blandford76}.
At the subsequent deceleration stage, the synchrotron radiation in the optical band is expressed as
$F_{\nu_{\rm opt}}^{\mathrm {Syn}} \propto \epsilon_\mathrm{e}^{p-1} \epsilon_B^{(p+1)/4} n_{\mathrm {ISM}}^{1/2} E_{\mathrm{K,iso}}^{(p+3)/4} \nu_{\rm opt}^{-(p-1)/2} t_{\mathrm {obs}}^{-3(p-1)/4}$
for $\nu_\mathrm{m} < \nu_{\rm opt} < \nu_\mathrm{c}$.
By combining Eqs.~(\ref{eq:Tpeak}) and (\ref{eq:Fopt}), we eliminate $\Gamma_0$ to derive the dependence of $\epsilon_{\rm e}$ on observables, including $T_{\rm Peak} \simeq 20$~s and $F^{\rm Peak}_{\nu_{\rm TeV}} \simeq 0.003$~$\mu$Jy at $\nu_{\rm TeV} = 1.62$~TeV.
It further gives the expected optical emission as
\begin{eqnarray}
F_{\nu_{\rm opt}}^{\rm Syn} (t_{\rm obs}) &\simeq&\label{eq:Fopt} 
2.0~x_0^{-1/2} (1+Y) \left(\frac{\epsilon_B}{6 \times 10^{-4}} \right)^{(p+8)/8}
\left(\frac{E_{\mathrm {K,iso}}}{1.5 \times 10^{55}~\mathrm{erg}} \right)^{(p+10)/16}
\left(\frac{n_{\mathrm {ISM}}}{0.4~\mathrm{cm}^{-3}} \right)^{(p+6)/16}
\\ \nonumber
& \times & 
\left(\frac{\nu_{\mathrm {TeV}}}{1.62~\mathrm{TeV}} \right)^{p/4}
\left(\frac{T_{\rm Peak}}{20~\mathrm{s}} \right)^{(9p-10)/16}
\left(\frac{F^{\rm Peak}_{\nu_{\rm TeV}}}{0.003~\mu \mathrm{Jy}} \right)^{1/2}
\left(\frac{\nu_{\mathrm {Opt}}}{r~\mathrm{band}} \right)^{-(p-1)/2} 
\left(\frac{t_{\mathrm {obs}}}{4000~\mathrm{s}} \right)^{-3(p-1)/4} 
~\mathrm{Jy},
\end{eqnarray}
for $p \in [2.1, 2.3]$ as constrained by spectral analysis, with microphysical parameters the same as those adopted by the LHAASO Collaboration \citep{LHAASO23}.
However, the earliest observed flux in the $r$ band at $\sim 4000$~s post-burst is only 0.1~Jy \citep{Kann23}, approximately 20 times fainter than the predicted value.
This discrepancy, along with a similar excess in the radio flux (see below), challenges the interpretation of the entire TeV emission as originating from a standard external shock in the ISM.

The above estimations do not account for precise jet dynamics or potential jet-break effects.
Additionally, it is difficult to capture the nonlinear effects like IC scattering in the Klein–Nishina regime \citep{Blumenthal70} and photon-photon annihilation \citep{Gould67} accurately under analytical approximation.
To address these limitations and demonstrate the inconsistencies in the early multi-wavelength data under the external shock scenario, we conducted detailed numerical calculations.
We adopted a generic energy-conservation-based model to describe the external shock dynamics \citep{Huang00,Peer12}.
The time-dependent electron spectrum accelerated by the blastwave is derived by solving the continuity equation in energy space, 
\begin{equation}
\frac{\partial}{\partial t^{\prime}}\left(\frac{d N_{\mathrm{e}}}{d \gamma_{\mathrm{e}}^{\prime}}\right)+\frac{\partial}{\partial \gamma_{\mathrm{e}}^{\prime}}\left[\dot{\gamma}_{\mathrm{e}, \mathrm {tot}}^{\prime}\left(\frac{d N_{\mathrm{e}}}{d \gamma_{\mathrm{e}}^{\prime}}\right)\right]=Q\left(\gamma_{\mathrm{e}}^{\prime}, t^{\prime}\right),
\end{equation}
using the constrained interpolation profile method \citep{Yabe01,Geng18},
where $\dot{\gamma}_{\mathrm{e}, \mathrm{tot}}^{\prime} = \dot{\gamma}_{\mathrm{e}, \mathrm{syn}}^{\prime} + \dot{\gamma}_{\mathrm{e}, \mathrm{adi}}^{\prime}$
is the total energy loss rate consisting of synchrotron radiation and adiabatic cooling,
$Q\left(\gamma_{\mathrm{e}}^{\prime}, t^{\prime} \right) \propto (\gamma/\gamma_{\rm m})^{-p}$ is the term for freshly shocked electrons.
From this electron distribution, we computed the synchrotron and SSC emissivity using standard formulae \citep{Rybicki79}.
In addition, the synchrotron self-absorption (low frequencies) and the photon-photon annihilation (high-energy) were included self-consistently.
The observed multi-wavelength emission was obtained by integrating over the equal-arrival-time surface (EATS),
and the five key parameters ($E_{\rm K, iso}$, $\Gamma_0$, $\epsilon_{\mathrm e}$, $\epsilon_B$, $n_{\rm ISM}$) were constrained via Bayesian inference.

\begin{figure}[h]
  \centering
  \includegraphics[width=\textwidth]{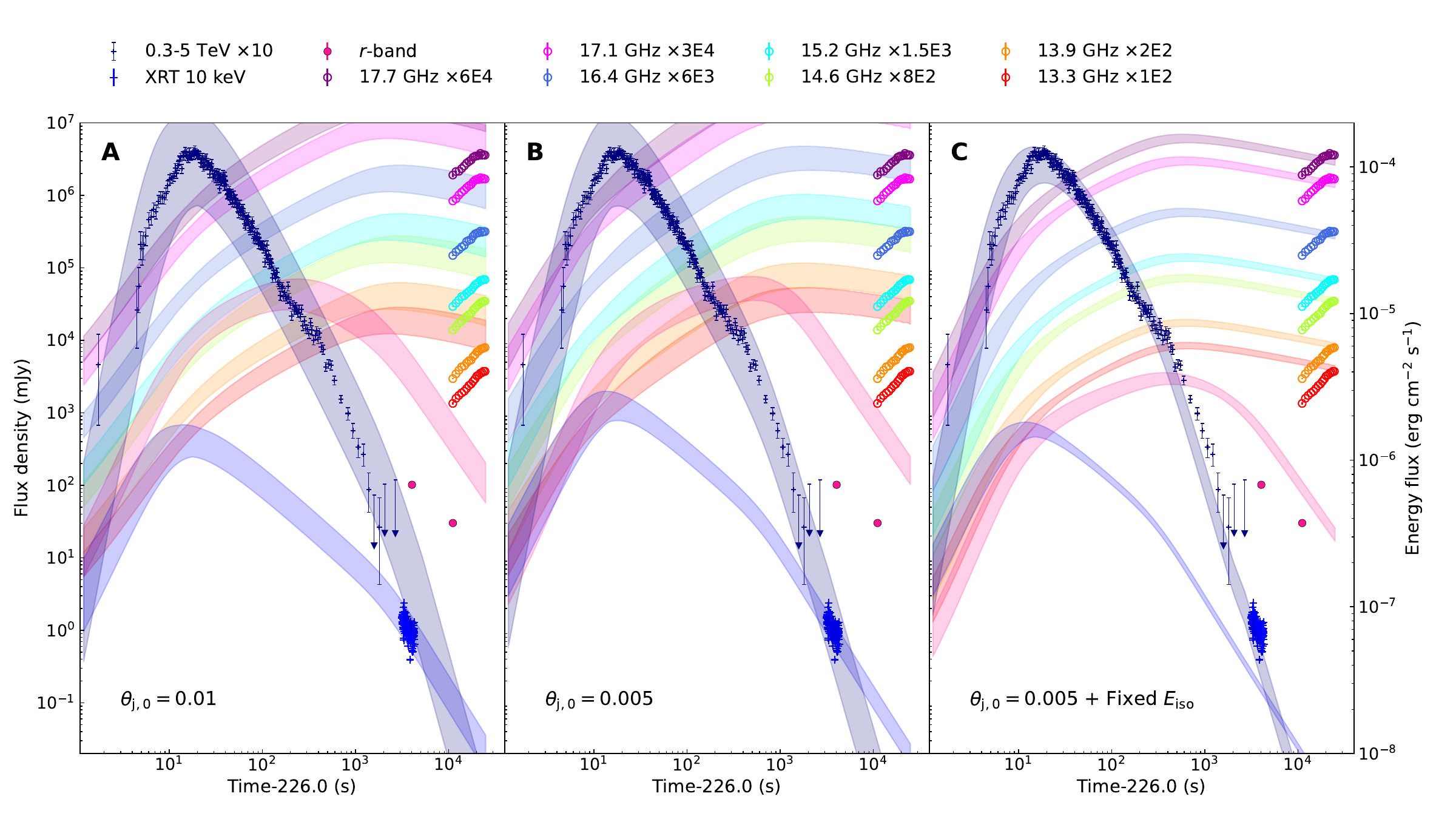}
  \caption{{\bf The fitting results with external shock model.} 
  Each panel shows the fitting results to the data using different strategies.
  The colored shaded areas correspond to the 3$\sigma$ confidence regions for lightcurves in different bands with posteriors constrained by the MCMC method (see \SUPTAB{tab:standard}).
  (\textbf{A}) Both the TeV lightcurve and spectra (in five time intervals) and the X-ray lightcurve are used in the posterior search. 
  A conical jet with an initial half-opening angle $\theta_{\rm j,0}$ of 0.01 rad is adopted.
  (\textbf{B}) Only the TeV information is used in the posterior search.
  A narrower jet with $\theta_{\rm j,0}$ = 0.005 rad is adopted.
  (\textbf{C}) Same as (\textbf{B}), but the parameter $E_{\rm K,iso}$ is fixed to a relatively small value ($9 \times 10^{54}$~erg) in order to alleviate the flux excess issue. 
  The corner plots of posteriors in (b) and (c) are shown as \SUPFIG{fig:cornerb} and \SUPFIG{fig:cornerc}.
  The flux predicted by the external shock model exceeds the observation data in the optical and radio bands significantly in these calculations.
  } 
 \label{fig:Standard}
\end{figure}

Given the sensitivity of low-frequency emission to jet geometry, we performed three sets of calculations.
First, we take an initial opening angle $\theta_{\rm j,0}$ of $0.01$~rad
and apply the standard model to fit the TeV lightcurve and spectra in five time intervals and the early X-ray lightcurve ($< 10^3$~s).
The fitting results (\SUPTAB{tab:standard}) yield a $3\sigma$ excess in optical/radio fluxes (\SUPFIG{fig:Standard}A). A narrower jet ($\theta_{\rm j,0} = 0.005$) is further considered, and only the TeV lightcurve and spectra are used in the fitting. Although the jet break at $\simeq 600$~s leads to rapid decline in the early optical ($\propto t_{\rm obs}^{-p}$) and radio ($\propto t_{\rm obs}^{-1/3}$) emission, the excess in the early optical and radio band still holds (see \SUPFIG{fig:Standard}B).
Even though $E_{\rm K,iso}$ is fixed to a relatively smaller value ($9 \times 10^{54}$~erg), this issue could not be eliminated in the third calculation (\SUPFIG{fig:Standard}C). Other attempts, like exploring the effect that only a portion (e.g., 10\%) of electrons are accelerated to emit, also failed. 
The VLBI-derived lower limit on $\theta_{\rm j,0}$ by \EQ{eq:ll_wind} exacerbates this tension.
Therefore, we conclude that if the TeV emission originates purely from an ISM-propagating external shock, the cooling of TeV-producing particles inevitably overpredicts optical/radio fluxes. This issue persists even in structured jets (core plus sheath), as the additional slower sheath preferentially enhances low-frequency emission.

\subsubsection{The two-shell collision scenario}\label{sec:twoshell}

\begin{figure}
  \centering
  \includegraphics[width=\textwidth]{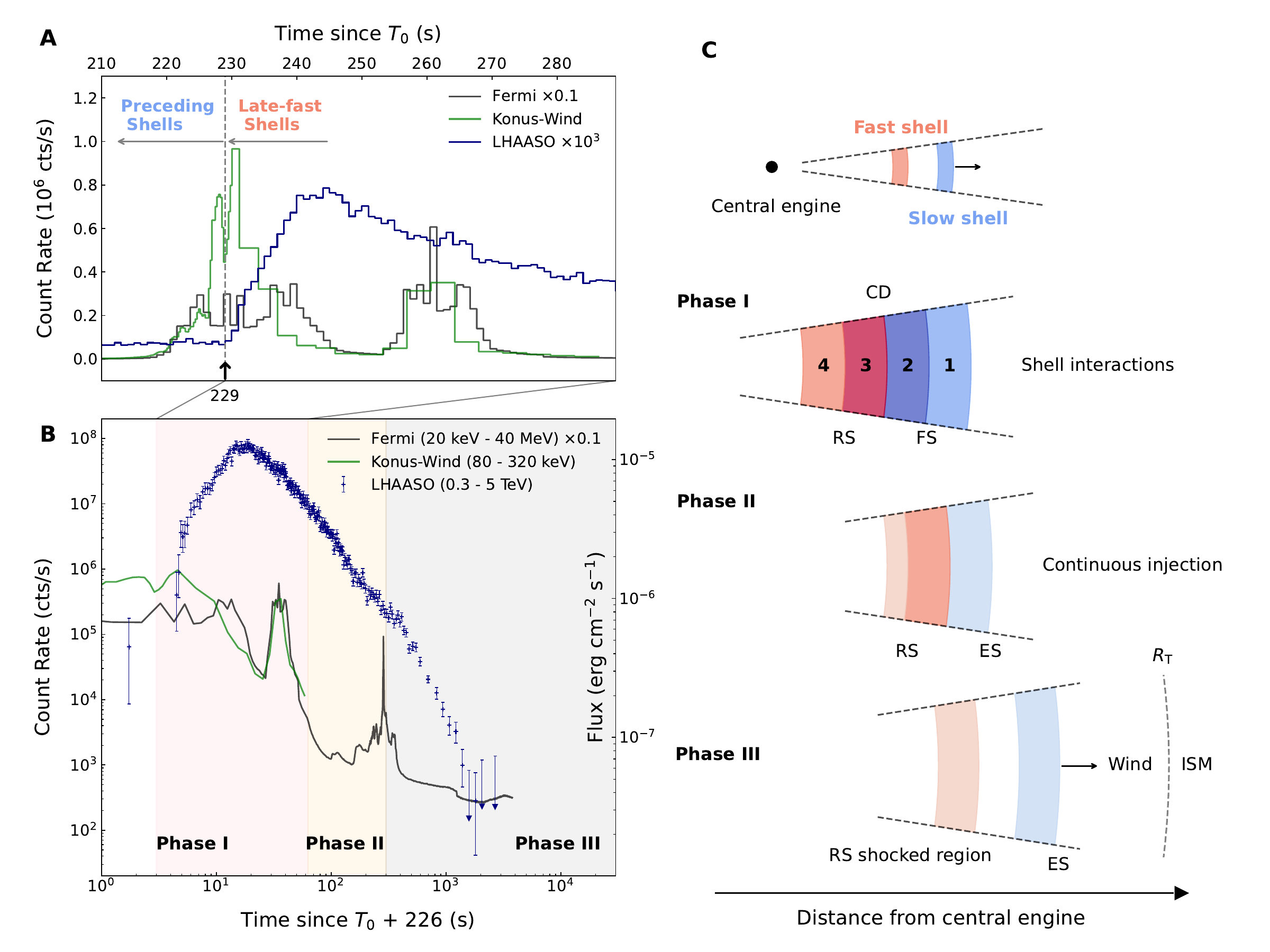}
  \caption{
  {\bf The schematic picture of timing analyses and the two-shell collision scenario.} 
  (\textbf{A}) The prompt gamma-ray lightcurve by Fermi-GBM and Konus-Wind indicates that some preceding shells were launched before the rise of TeV emission ($T_0 + 229$~s).
  (\textbf{B}) In the two-shell collision scenario, the TeV emission is decomposed into three phases, divided by the crossing time of the forward shock (FS) and the reverse shock (RS).
  (\textbf{C}) In Phase I, the collision between the two shells produces an FS propagating into the materials of the slow preceding shell, and an RS propagating into the fast rear shell, separated by the contact discontinuity (CD). This creates distinct shocked regions labeled with the number ``$i$'' ($i = 1-4$, see {\bf Supplementary text}). The short-lived FS contributes to the main peak in the TeV lightcurve. 
  After the breakout of the FS into the wind-type interstellar environment, the external shock (ES) is formed. 
  In Phase II, continuous outflow injections from the central engine result in a coasting stage for the ES.
  The ES generates the rest of the TeV lightcurve and the long-lasting GeV/X-ray/optical afterglows.
  The RS detaches from the ES gradually once the injection ceases in Phase III.
  The electrons shocked by the RS make brightening bumps in radio frequencies.
  At late stages, the radio emission is dominated again by the external shock, when the environment becomes a homogeneous interstellar medium (ISM).
  } 
\label{fig:twoshock}
\end{figure}

Following the detection of a weak precursor burst by the Fermi-GBM detector at $T_0$, the main prompt emission of GRB 221009A begins at $\sim T_0 + 185$~s, marked by a rapid rise in the count rate exceeding $10^4$~s$^{-1}$. As shown in \SUPFIG{fig:twoshock}, the prompt gamma-ray and TeV lightcurves observed by Fermi, Konus-Wind, and LHAASO exhibit a multi-episode structure lasting for $\sim 300$~s. Prior to $T_0 + 229$~s, several intense gamma-ray pulses were observed, with the onset of the TeV emission ($\sim T_0 + 230$~s) coinciding with a subsequent prominent gamma-ray pulse. 
Prompt pulses are attributed to internal dissipation within outflow elements ejected by the central engine, with each intense pulse forming an outgoing shell post-dissipation.
If earlier shells propagate at lower velocities, a subsequent faster and more energetic shell would collide with them before encountering the circumburst medium.
This situation is particularly relevant for GRB 221009A, given its prolonged duration and multi-episode structure.
We therefore adopt a two-shell collision scenario \citep{Kumar00}, wherein the initial TeV emission arises from shell interactions during the prompt phase. While the actual dynamics likely involve multiple shells, the ``two-shell'' approximation provides a tractable and representative framework. 
In this scenario, the TeV emission can be decomposed into three distinct phases: the shell collision phase (Phase I), the shell coasting phase with continuous outflow injection (Phase II), and the afterglow emission phase from the merged shell propagating into the circumburst environment (Phase III). 
Through detailed numerical modeling, we show that this two-shell collision scenario self-consistently explains the rich observations of GRB 221009A.

\noindent {\bf --- Modeling of the shock dynamics} \\
In Phase I, for a preceding shell with a bulk Lorentz factor of $\Gamma_1$, a later launched but faster shell (with a bulk Lorentz factor of $\Gamma_4$) will catch up at radius $R_0$, colliding with the slow shell. If the fast shell is not extremely magnetized, the collision would produce a forward shock (FS) propagating into materials of the preceding shell, and a reverse shock (RS) propagating into the fast shell.
After crossing the slow shell, the FS will break out into the circumburst material, becoming an external shock (ES), which produces long-lasting afterglows (see \SUPFIG{fig:twoshock} for illustration).

During the collision phase, two shocks separate the system into four regions: (1) the unshocked slow shell, (2) the shocked slow shell, (3) the shocked fast shell, and (4) the unshocked fast shell.
Hereafter, $\chi_i$ denotes the value of the quantity $\chi$ in Region ``$i$'' in its own rest frame (also see \SUPFIG{fig:twoshock}), and $\chi_{ij}$ denotes the value of the quantity $\chi$ in Region ``$i$'' measured in the rest frame of Region ``$j$''. 
Unlike the preceding shell that exhausts the magnetic energy in the early stage ($\sigma_1 \simeq 0$),
the later fast shell may keep the magnetic fields advected from the central engine,
which could be parameterized by the magnetization of 
\begin{equation}
\sigma_{4} = \frac{B_4^2}{4 \pi n_4 m_{\rm p} c^2},
\end{equation}
where $n_4$ is the particle density in the comoving frame of Region 4.
Let's introduce an equivalent ``luminosity'' of the total energy for the two shells measured in the lab frame, $L_1$ and $L_4$, the particle density is then
\begin{equation}
n_{i} = \frac{L_i}{4 \pi R^2 \beta_i \Gamma_i^2 m_{\rm p} c^3 \left[1 + \frac{4}{3} (\gamma_{{\rm p},i} - 1) \right] (1+\sigma_i)},
\end{equation}
where $i = 1,4$, $\beta_i = \sqrt{1-1/\Gamma_i^2}$, and $\gamma_{{\rm p},i}$ is the average Lorentz factor of protons in each upstream.
The particles in Region 1 are supposed to be shocked or heated already, i.e., $\gamma_{{\rm p},1} > 1$,
while the later outflow is cold ($\gamma_{{\rm p},4} = 1$).
The adiabatic cooling of hot protons is described by $\gamma_{{\rm p},1} \propto (R/R_0)^{-2/3}$.
For simplicity, both the luminosity of the two shells are assumed to be constant,
i.e., $L_4 = L_{\rm f}$ before RS crossing ($R < R_{\rm off,3}$) and $L_1 = L_{\rm s}$ before FS crossing ($R < R_{\rm off,2}$).
We also assume that Region 2 and Region 3 share a common bulk Lorentz factor ($\Gamma_2 = \Gamma_3$).
After applying the hydrodynamical/magnetohydrodynamical jump conditions \citep{Kumar00,Zhang05} to the FS/RS respectively
and the energy conservation law to the FS-RS system \citep{Beloborodov06,Uhm11,Geng16,Ai21},
the evolution of $\Gamma_2$ and relevant quantities within these regions could be solved numerically.
For an infinitesimal radius increasement of $dR$, 
the RS will sweep $dN_3 = 4 \pi R^2 n_4 \Gamma_4 dR (\beta_4 - \beta_{\rm RS})/ \beta_2$ particles into Region 3,
while the FS would sweep $dN_2 = 4 \pi R^2 n_1 \Gamma_1 dR (\beta_{\rm FS} -\beta_1) / \beta_2$ particles into Region 2, 
where $\beta_{\rm FS}$ and $\beta_{\rm RS}$ are the velocities of the FS and RS.
Given the initial total isotropic energy of each shell ($E_{\rm iso,f}$ and $E_{\rm iso,s}$), the energy conservation of $E_{\rm f} = \Gamma_4 m_{\rm p} c^2 (1+\sigma_4) \int_{R_0}^{R_{\rm off,3}} dN_3$ and $E_{\rm s} = \Gamma_1 [4/3 (\gamma_{{\rm p},1} - 1)+1] m_{\rm p} c^2 \int_{R_0}^{R_{\rm off,2}} dN_2$ determines the radius that the RS and the FS cross each shell, i.e., $R_{\rm off,3}$ and $R_{\rm off,2}$ respectively.

During Phase II, the FS crosses the entire preceding shell, and TeV emission from the ES emerges. 
Continuous outflows injected from the persistent central engine keep the ES in a coasting state.
After the main prompt emission ceases and the RS crossing, the system evolves into Phase III.
During this stage, the shocked Region 3 will gradually detach from Region 2 when the effective thermal Lorentz factor of its plasma is approaching 1 due to adiabatic expansion. The bulk Lorentz factor of Region 3 post-detachment exhibits a power-law decay as $\Gamma_3 \propto R^{-g}$ with $g \in [1/2, 3/2]$.
While hydrodynamic simulations are required to precisely determine $g$,
the temporal evolution of $\Gamma_3 (t_{\rm obs})$ remains relatively insensitive to its exact value \citep{Kobayashi00,Zou05}.
We adopt a median value of $g = 1$ for simplicity.

According to the particle-in-cell simulations for shocks \citep{Sironi11,Sironi15}, the distribution of the accelerated  electrons downstream could be generalized as
\begin{equation}
Q\left(\gamma_{\mathrm{e}}^{\prime}, t^{\prime}\right)=
\begin{cases}
Q_{\mathrm{inj}}(t^{\prime}) \left(\frac{\gamma_{\rm e}^{\prime}}{\gamma_{\rm th}^{\prime}}\right)^2 \exp \left(-\frac{\gamma_{\rm e}^{\prime}}{\gamma_{\rm th}^{\prime}}\right), \quad\quad\quad\quad\quad\gamma_{\mathrm{e}}^{\prime} \leq \gamma_{\mathrm{nth}}^{\prime}, \\
Q_{\mathrm{inj}}(t^{\prime}) \left(\frac{\gamma_{\rm nth}^{\prime}}{\gamma_{\rm th}^{\prime}}\right)^2 \exp \left(-\frac{\gamma_{\rm nth}^{\prime}}{\gamma_{\rm th}}\right)
\left( \frac{\gamma_{\rm e}^{\prime}}{\gamma_{\rm nth}^{\prime}} \right)^{-p},\gamma_{\mathrm{e}}^{\prime}>\gamma_{\mathrm{nth}}^{\prime},
\end{cases}
\end{equation}
where $Q_{\mathrm{inj}}$ is the normalized injection rate, $\gamma_{\rm th}^{\prime}$ and $\gamma_{\rm nth}^{\prime}$
are the characteristic thermal and non-thermal Lorentz factors in the comoving frame respectively \citep{Giannios09,Gao24}. By introducing the energy fraction parameter of the non-thermal part to the total energy $\delta$, $\gamma_{\rm th}^{\prime}$ and $\gamma_{\rm nth}^{\prime}$ could be solved from known $\epsilon_{\rm e}$ and shock jump conditions.
In our scenario, such a generalized electron distribution is adopted for all shocked regions, and the emission from these regions is subsequently obtained.

\noindent {\bf --- Size evolution modeling} \\
As mentioned above, our VLBI observations suggest that the jet is propagating in the ISM beyond $8 \times 10^5$~s post-burst. 
On the contrary, the shallow decay of early ($<10^5$~s) optical afterglow hints that the jet was propagating in the wind medium.
The BM solution of the jet in a pure wind medium gives that the number density it encounters is 
\begin{equation}
n(t_{\rm obs}) = 3 \times 10^{35} A_* / R(t_{\rm obs})^2 = 8 \times 10^{-4} \left(\frac{E_{\rm K,iso}}{10^{55}~ \mathrm{erg}} \right)^{-1} \left(\frac{A_*}{0.5}\right)^2 \left(\frac{t_{\rm obs}}{10^5~\mathrm{s}}\right)^{-1}~\mathrm{cm}^{-3}, \end{equation}
which becomes sufficiently smaller than the typical value of the ISM
$\sim 0.1~\mathrm{cm}^{-3}$~\citep{Maciel13} around $10^5$~s.
Meanwhile, the James Webb Space Telescope (JWST) observation reveals that the burst is sited at a dense star formation region \citep{Blanchard24}. 
Therefore, these lines of evidence indicate that the circumburst environment of GRB 221009A should transit from the wind type to the ISM, a common expectation in GRBs \citep{Eldridge06,Chrimes22}.
In our modeling, the jet encounters the ISM beyond the terminal radius ($R_{\mathrm{T}}$), and a smooth transition is adopted for simplicity, i.e.,
$n_{\rm ISM} = A/R_{\mathrm{T}}^2 = 3 \times 10^{35} A_* \mathrm{cm}^{-1} / R_{\mathrm{T}}^2$ for $R > R_{\mathrm{T}}$.

The fast size evolution indicates the significant lateral expansion of the jet at late times, which holds for jets with extremely narrow jet angles.
For the top-hat jet approximation considered here, the increasing rate of the half-opening angle is modeled as \citep{Granot12}
\begin{equation}
\frac{d \theta_{\rm j}}{d \ln R} = 
\begin{cases}
0,    \quad\quad\quad \mathrm{for} \quad\quad \Gamma \ge \Gamma_{\rm sp}, \\
\frac{1}{\Gamma^2 \theta_{\rm j}},    \quad\quad \mathrm{for} \quad\quad \Gamma < \Gamma_{\rm sp}, \\
\end{cases}
\label{eq:dthdR}
\end{equation}
which is incorporated in the equations mentioned above to determine the dynamics of a spreading jet consistently.
The theoretical image sizes are derived from the projection of the EATSs. Since the surface brightness profile in radio frequencies is limb-brightened \citep{Granot99}, we use 
\begin{equation}
\theta_{\mathrm{G}} (t_{\rm obs}) = 2 R_{\rm EATS}(t_{\rm obs},\theta_{\rm j}) \sin \theta_{\rm j} /D_{\mathrm{A}}
\end{equation}
to model source size, where $R_{\rm EATS}$ is determined from $t_{\rm obs} = (1+z) \int_{R_0}^{R_{\rm EATS}} (1-\beta \cos \theta) dr / \beta c$.
In the early wind-environment phase, the BM solution yields $R = [\frac{9}{8 \pi m_{\rm p} c^2}] \left(\frac{E_{\rm k,iso}}{A} \right) \Gamma^{-2} = \kappa \Gamma^{-2}$. 
Under the conditions $\Gamma \gg 1$ and $\theta \le 0.1$, the EATS relation can be approximated as $t_{\rm obs} \simeq (1+z) [\Gamma^{-4}/4+\theta^2 \Gamma^{-2}/2] \kappa / c = (1+z) [R_{\rm EATS}^2 + 2 \theta^2 \kappa R_{\rm EATS}]/(4 \kappa c)$, and its average apparent expansion speed is given by
\begin{equation}
    \beta_{\rm app} (t_{\rm obs},\theta_{\rm j}) = \frac{(1+z) R_{\rm EATS} \sin \theta_{\rm j}}{t_{\rm obs} c} \simeq \frac{(1+z) \kappa \theta_{\rm j} (\sqrt{\theta_{\rm j}^4 + f} - \theta_{\rm j}^2)}{t_{\rm obs} c},
\end{equation}
where $f = \frac{4 t_{\rm obs} c}{\kappa(1+z)}$.
Taking $E_{\rm k,iso} = 3 \times 10^{55}$~erg and $A_* = 0.1$, we then obtain
$\beta_{\rm app} (t_{\rm obs} = \mathrm{9~days},\theta_{\rm j,0} = 0.03~\mathrm{rad}) \simeq 60$ and $D_{\perp} (\mathrm{9~days}) \simeq 2.5 \times 10^{18}$~cm, consistent with the first-epoch observation BA161a (\FIG{fig:size_evol}). 
Following the change to the ISM regime, the apparent speed evolves as $\beta_{\rm app} \propto R_{\rm EATS} / t_{\rm obs} \propto R / t_{\rm obs} \propto t_{\rm obs}^{1/4}/t_{\rm obs} \propto t_{\rm obs}^{-3/4}$. Scaling from $\beta_{\rm app} \approx 60$ at 9 days, this relation gives $\beta_{\rm app} (t_{\rm obs} = 100~\mathrm{days}) \simeq 50 (109/9)^{-3/4} \simeq 9$, consistent with our derived value of $\beta_{\rm app} \approx 8$ at 109 days from the BL307 epoch, as well as with the values of $\sim 6$–$7$ between 114 and 118 days by VLBA and EVN \citep{Giarratana24}. 
The subsequent, more gradual decline in $\beta_{\rm app}$ after $\approx 100$ days could be attributed to the lateral spreading of the initially narrow jet, which becomes significant when causal contact is established across the shock front \citep{Granot12,vanEerten18} (see the discussion on the spreading jet scenario above).

\begin{figure}
  \centering
  \includegraphics[width=\textwidth]{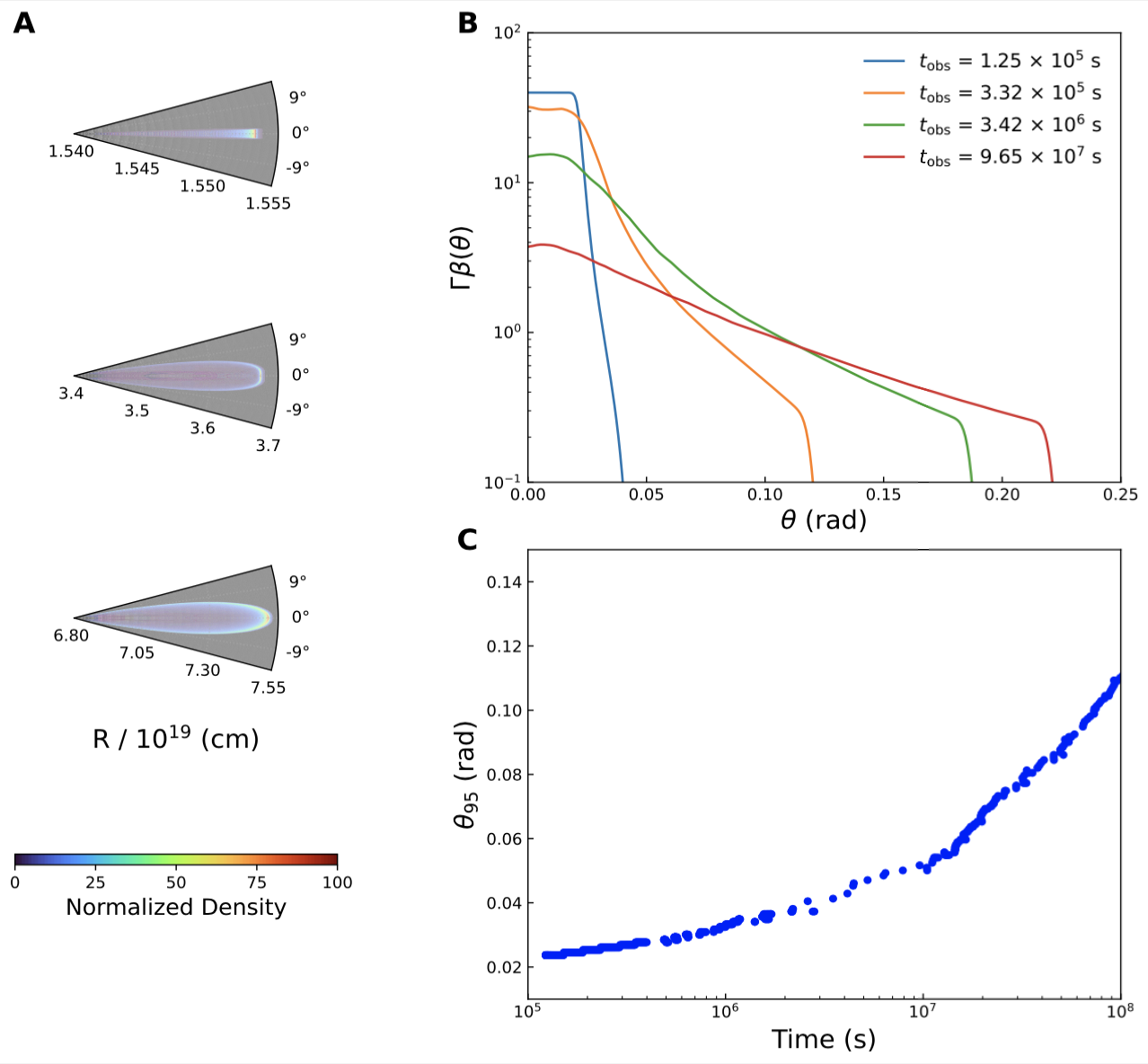}
  \caption{
  {\bf Hydrodynamic simulations of the jet and the structure evolution.}
  (\textbf{A}) Snapshots of the density distribution of the jet before spreading, during significant spreading, and near the end of the simulation.
  (\textbf{B}) The evolution of the $\Gamma \beta (\theta)$ profiles at different observational times. 
  The lab-frame time in the simulation is converted to observer-frame time by accounting for Doppler boosting along the jet axis ($\theta = 0$). Each profile corresponds to the jet structure at a fixed lab-frame time (light-travel effects for different latitudes are not included).
  (\textbf{C}) Time-dependent evolution of the effective core opening angle $\theta_{95}$ that encloses 95\% of the blastwave energy.
  } 
 \label{fig:simulation}
\end{figure}

To further test whether the lateral spreading holds for GRB 221009A, we conduct hydrodynamic simulations using the GAMMA code \citep{Ayache22}.
This code implements a finite-volume Godunov scheme with Riemann solvers and adaptive moving meshes, providing high-resolution treatment of relativistic outflows. 
Our simulation setup adopts an initial isotropic-equivalent kinetic energy of $E_{\rm K,iso} = 3 \times 10^{55}$ erg and a jet half-opening angle of $\theta_{\rm j,0} = 0.025$ rad, with the simulation beginning at the wind terminal radius $R_{\rm T}$. 
The simulation utilizes an initial polar grid of 500 ($\theta$) $\times$ 600 ($r$) zones with an adaptive cell split/merge function, achieving optimal resolution of jet dynamics while maintaining computational efficiency.
We characterize the jet evolution by defining an effective core opening angle that encloses 95\% of the blastwave energy. As shown in \SUPFIG{fig:simulation}, the jet core angle exhibits significant broadening from 0.025 to 0.12 rad during its evolution, in agreement with recent high-resolution numerical studies by other groups \citep{Govreen24}. The approach in Eq.~(\ref{eq:dthdR}) well reproduces this spreading behavior, and our fitting results align with the observed lightcurves and size evolution (\FIG{fig:SizeFit}).
The minor discrepancy in the spreading rate between our numerical approach and the exact hydrodynamic simulation arises as a limitation of the top-hat approximation and the assumptions underlying Eq.~(\ref{eq:dthdR}).
While comprehensive multi-wavelength radiation modeling based on hydrodynamic simulations and full parameter-space exploration is challenging and computationally expensive, such detailed investigations would be valuable in the future.

\begin{figure}
  \centering
  \includegraphics[width=0.6\textwidth]{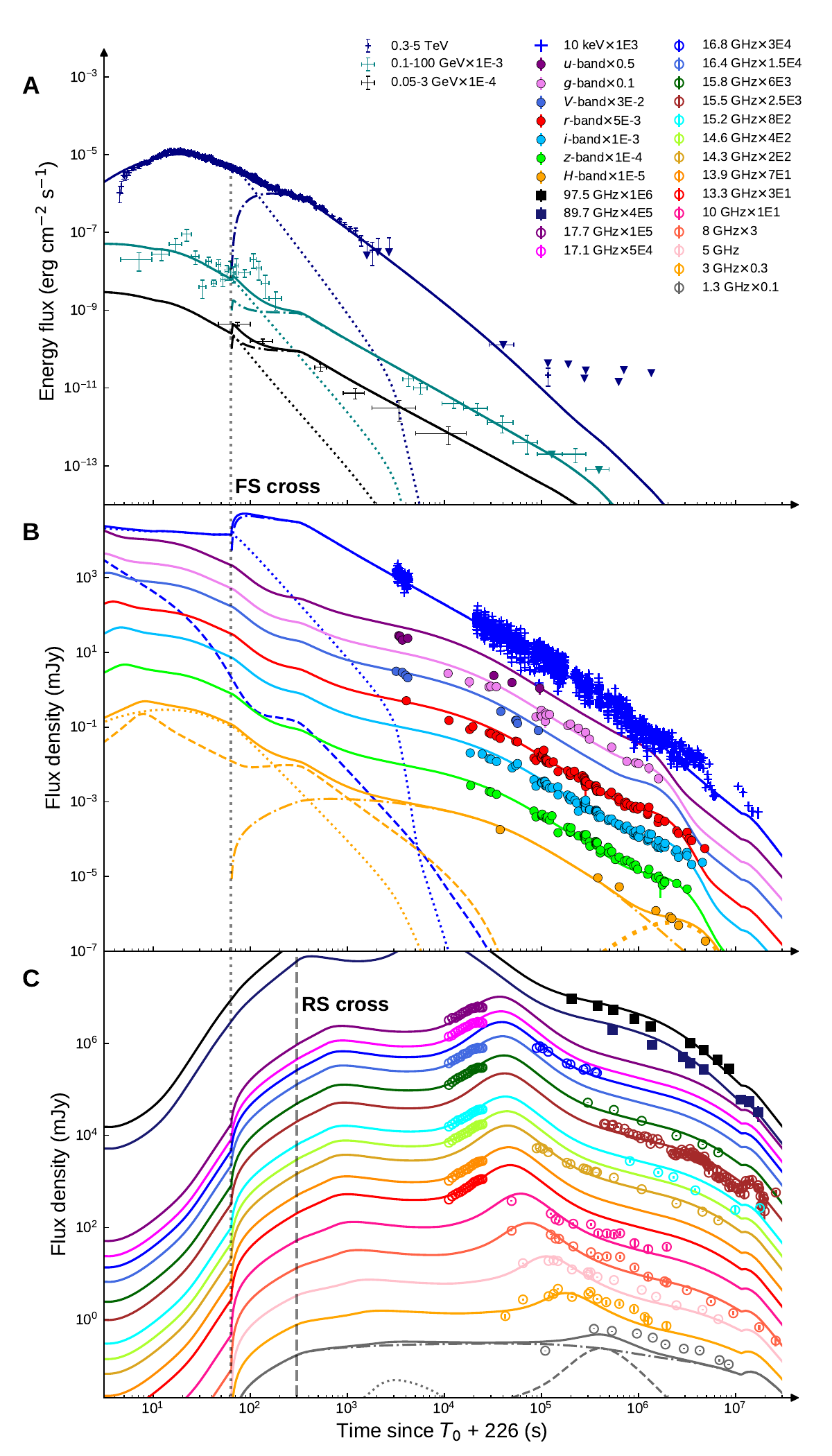}
  \caption{
  {\bf Fit to the multi-wavelength lightcurves of GRB 221009A.} 
  Panels (\textbf{A-C}) show our numerical fits to high-energy (TeV and GeV), X-ray/optical, and low-frequency emission, respectively. In each panel, the dotted and dashed lines represent forward shock (FS) and reverse shock (RS) emission, respectively, arising from the collision of the slow and fast outflows (\textbf{Supplementary text}).
  $T_0$ is the Fermi-GBM trigger time. 
  The dash-dotted lines represent the emission of the external shock propagating through the surrounding medium.
  Solid lines are the sum of all components.
  The grey vertical dotted and dashed lines mark the crossing time of FS and RS, respectively.
  All the data and upper limits (triangles) of different frequencies, including the TeV range \citep{LHAASO23,HAWC22,LST25}, the GeV range \citep{Tavani23,Axelsson25}, the optical bands \citep{Kann23,OConnor23,Levan23,Williams23}, and the radio bands \citep{Laskar23,Bright23,Rhodes24} are taken from the literature. The optical data have been corrected only for Galactic extinction of $E(B-V) = 1.32$ mag \citep{Schlafly11}, and the flux contribution from the supernova is accounted for as thick dotted lines.   
  } 
\label{fig:Fitting}
\end{figure}

The source size evolution has been argued to be frequency-dependent, due to the transition from being dominated by the RS emission to being dominated by the ES emission \citep{Giarratana24}. However, it is shown that the lightcurve at 15 GHz and 8 GHz is already dominated by the radiation of the ES beyond $\sim 3 \times 10^5$~s (see \FIG{fig:Fitting}). Meanwhile, the RS lags far behind the ES, and its projected size, if visible, should be much smaller than observed values, as also shown in \FIG{fig:SizeFit}. 
Thus it is reasonable to model the observed size at 15 GHz and 8 GHz using the ES directly.

\noindent {\bf --- Temporal and spectral modeling} \\
The timescales of Phase I and II in our scenario are inferred from the timing analyses of the prompt emission (\SUPFIG{fig:twoshock}). There are several observed critical times, the GRB trigger time by Fermi/GBM ($T_0$), the starting time of the first main gamma-ray episode ($T_1 \simeq T_0 + 185$~s), the zero point commonly adopted for TeV emission ($T_* = T_0 + 226$~s), the peak time of the second main gamma-ray episode ($T_{\rm 2,p} \simeq T_0 + 260$~s), and the peak time of the third main gamma-ray episode ($T_3 \simeq T_* + 300$~s).
The FS crossing occurs at $T_{\rm FS,cr} \simeq T_{\rm 2,p} + (T_{\rm 2,p} - T_1)(\Gamma_{\rm s}/\Gamma_{2})^2$.
Thus Phase I lasts for $T_{\rm FS,cr} - T_* \sim 60$~s since $T_*$ for $\Gamma_{\rm s} \sim \mathcal{O}(50)$ and $\Gamma_{2} \sim \mathcal{O}(100)$.
Meanwhile, Phase II should last for slightly longer than $300$~s since $T_*$ when the continuous strong outflow during the main prompt emission ceases.

In our scenario, the velocities of involving regions of the collision are roughly constant in Phase I. The early rising TeV flux is $F_{\rm TeV}^{\rm IC} \propto \nu_{\rm TeV}^{-p_2/2}$ for $\nu_{\rm m}^{\rm IC} < \nu_{\rm c}^{\rm IC} < \nu_{\rm TeV}$ \citep{Sari01}.
As the strong photon annihilation results in a softer observed spectrum in the range of $> 1$~TeV, the local spectral index of $-1.1$ in [0.2, 0.4] TeV given by LHAASO \citep{LHAASO23} hints at an intrinsic $p_2 \simeq 2.2$ for the forward shock region. The condition of $\nu_{\rm c}^{\rm IC} < 0.2~\mathrm{TeV}$ at $\sim T_* + 10$~s gives that $(L_{\rm s}/10^{52})^{-7/2} (\Gamma_{\rm s}/60)^{14} (\Gamma_2/200)^{26/3} (\epsilon_{B,2}/10^{-5})^{-7/2} < 1$. The peak flux of $\nu F_{\nu}^{\rm IC}$ spectrum at $\nu_{\rm c}^{\rm IC}$ should be larger than the observed value $3.7 \times 10^{-6}$~erg~cm$^2$~s$^{-1}$ at 0.2 TeV, indicating $(L_{\rm s}/10^{52})^{2p_2-3} (\Gamma_{\rm s}/60)^{16-10 p_2} (\Gamma_2/200)^{(40-16 p_2)/3} (\epsilon_{B,2}/10^{-5})^{2p_2-5}$  $(\epsilon_{\mathrm{e,2}}/0.1)^{2p_2-2} (\gamma_{\rm p,0}/30)^{2p_2-4} > 0.1$ with a moderate value of $\sim 10$ adopted for the Compton parameter. After the peak time of TeV emission, $\nu_{\rm c}^{\rm IC}$ is progressively crossing $\nu_{\rm TeV}$ and the TeV lightcurve is asymptotically determined by the evolution of the peak flux of the spectrum as $F_{\rm TeV}^{\rm IC} \propto F_{\nu,\mathrm{max}}^{\rm IC} \propto t_{\rm obs}^{-4/3}$, which is consistent with the temporal index of $\sim -1.1$ for the segment of [20-70]~s after $T_*$ given by LHAASO \citep{LHAASO23}.
After the FS crossing, the TeV emission in Phase II is the superposition of the high-latitude emission of the FS (evolves as $\propto t_{\rm obs}^{-2-p_2/2}$) and the onset emission of the ES, producing a sustaining decay and a potential transition to a shallow decay or a flare during $T_* + [320, 550]$~s as reported \citep{LHAASO23}. 
In Phase III, the steep decay of TeV emission could be attributed to the ES transition from a coasting phase to the deceleration phase.
Applying the BM solution in the wind, we have
$F_{\rm TeV}^{\rm IC} \propto F_{\nu,\mathrm{max}}^{\rm IC} \left(\nu_{\rm TeV} / \nu_{\rm m}^{\rm IC}\right)^{-(p_{\rm E}-1)/2} \propto t_{\rm obs}^{-p_{\rm E}} \nu_{\rm TeV}^{-(p_{\rm E}-1)/2}$ for the regime of $\nu_{\rm m}^{\rm IC} < \nu_{\rm TeV} < \nu_{\rm c}^{\rm IC}$ \citep{Sari01}.
Then the corresponding observed temporal index of $\sim -2.3$ and spectral index of $ \sim 1.3$ gives $p_{\rm E} \simeq 2.4$.
Meanwhile, the long-lasting GeV afterglow evolves as
$F_{\rm GeV} \propto t_{\rm obs}^{-(3p_{\rm E}-2)/4} \nu_{\rm GeV}^{-p_{\rm E}/2}$ for $\nu_{\rm m} < \nu_{\rm c} < \nu_{\rm GeV}$.
A value of $p_{\rm E} \simeq 2.4$ could also meet the observed 
temporal index of $-1.3$ and spectral index of $- [1.1, 1.5]$ as revealed by Fermi-LAT \citep{Axelsson25}.

The decaying index of the early X-ray lightcurve is $\simeq -1.5$, which also gives an electron spectral index of $p_{\rm E} \simeq 2.4$ with the closure relation for $\nu_{\rm m} < \nu_{\rm X} < \nu_{\rm c}$.
The transition of the X-ray/optical temporal indices from a shallow decay to a steep decay near $10^5$~s hints at the occurrence of the jet break, which implies that the initial half jet opening angle should satisfy
\begin{equation}
\theta_{\rm j,0} \le 0.03 \left( \frac{E_{\rm K,iso}}{10^{55}~\mathrm{erg}} \right)^{-1/4} \left( \frac{A_*}{0.5} \right)^{1/4} \left( \frac{t_{\rm j}}{10^5~\mathrm{s}} \right)^{1/4}~\mathrm{rad},    
\end{equation}
where $t_{\rm j}$ is the jet break time.
It further constrains the jet angle to be $\simeq$ 0.01-0.03~rad, jointly with the discussion of the VLBI observations and the condition for the significant jet lateral spreading discussed above.

\begin{figure}
  \centering
  \includegraphics[width=\textwidth]{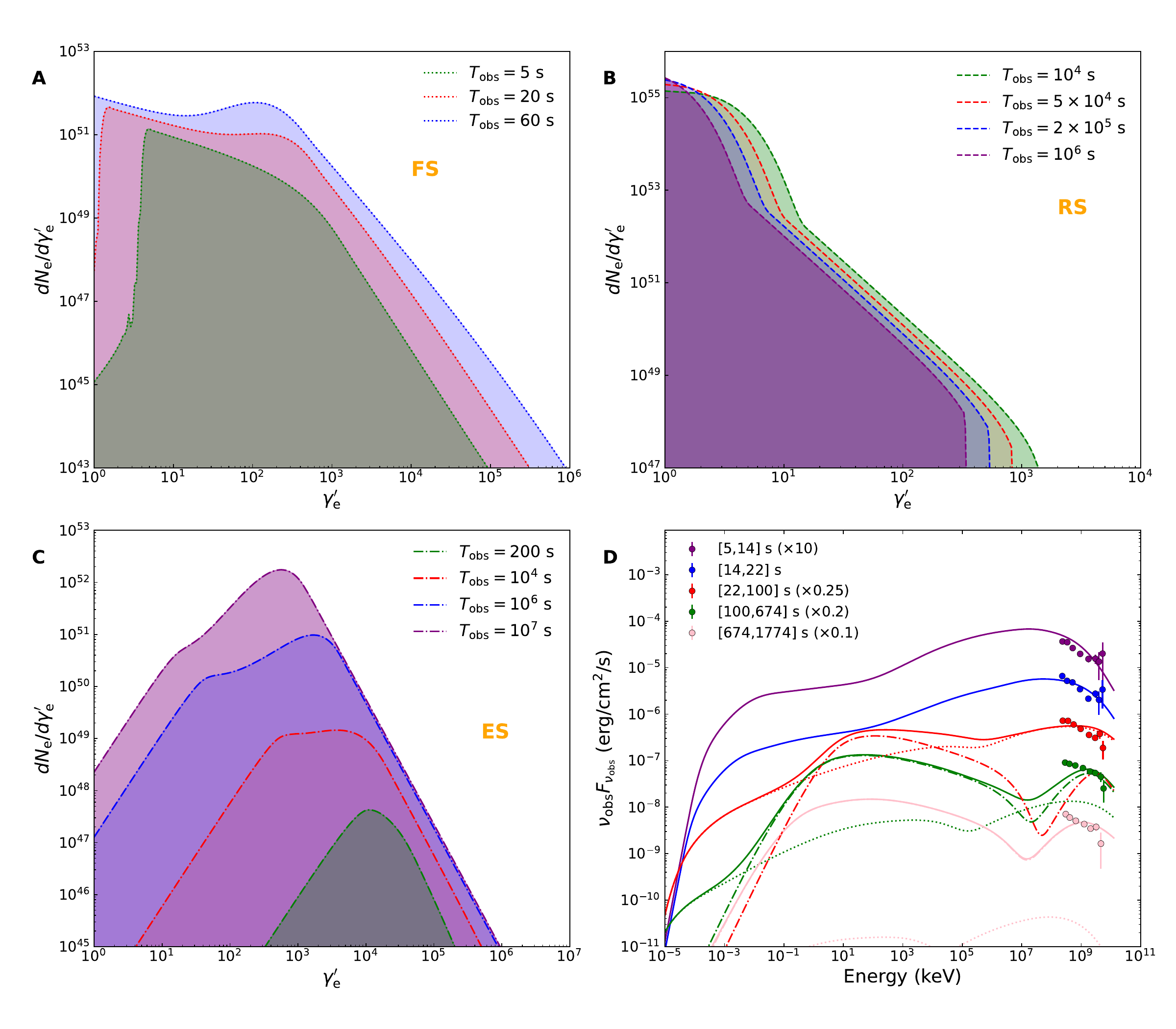}
  \caption{
  {\bf Snapshots of the electron and photon spectra.} 
  The distribution of electrons shocked by the FS (\textbf{A}), RS (\textbf{B}), and ES (\textbf{C}) at several representative moments in our calculation is shown.
  Panel \textbf{D} shows the spectra at five time intervals by LHAASO. The dotted and dashed-dotted lines are emission from the FS and the ES, respectively, and the solid lines are the sum of them.
  } 
 \label{fig:Spectra}
\end{figure}

The sophisticated behavior of radio afterglows is still challenging in the literature. In our scenario, the early radio bump is the optically thick rising emission of electrons accelerated by the RS. The radio spectral index transits from $5/2$ to $-(p_3-1)/2$ as the synchrotron self-absorption frequency passes through the observed band. The measured spectral index of $-0.2$ around $3 \times 10^5$~s thus implies $p_3 \simeq 1.4$, harder than an expected value~\citep{Bright23}. However, the emission at this period becomes increasingly affected by the emerging ES emission with a spectral index of $1/3$, introducing spectral complexity. As a result, $p_3$ is not tightly constrained by the data, and we adopt a typical value of 2.3 in our modeling.
At a later stage, the radio emission becomes dominated by the ES.
The radio spectral index at $\sim 2.5 \times 10^{6}$~s is $\beta \simeq -0.14$ which means that the characteristic synchrotron frequency $\nu_{\rm m}$ has just crossed 5~GHz band \citep{Rhodes24}, giving $\epsilon_{B,\mathrm{E}}^{1/2} \epsilon_{\mathrm{e,E}}^{2} \le 6 \times 10^{-5} E_{\rm K,iso,55}^{-1/2}$.
With a starting radius of $R_0 \simeq 10^{15}$~cm, $\theta_{\rm j,0} \simeq 0.025$~rad, and a set of parameters shown in \SUPTAB{tab:para-all} guided by the discussions above, the two-shock scenario could well interpret the multi-wavelength data (\FIG{fig:Fitting}).
The electron spectra of different regions at several characteristic times are shown in \SUPFIG{fig:Spectra}. Note that the SSC cooling of electrons in the FS has been self-consistently incorporated in the calculation of the electron spectra \citep{Geng18}.  
The predicted post-jet-break optical decay is slightly steeper than that observed in our modeling.
Nevertheless, the supernova signature only slightly fainter than that of SN 1998bw has been detected beyond $\sim 10^6$~s by both the Pan-STARRS2 Observatory \citep{Fulton23} and JWST \citep{Blanchard24}, and other photometric observations \citep{Kong24}. 
By accounting for the supernova emission using a set of parameters suggested in these works \citep{Fulton23}, the discrepancy in the late-time optical lightcurves can be relieved as shown in \FIG{fig:Fitting}.

Among the fitting parameters, the derived value of $\epsilon_{\rm e,2} \simeq 0.2$ is approximately one order of magnitude higher than $\epsilon_{\mathrm{e,E}}$. While $\epsilon_{\rm e,2}$ is primarily constrained by the inverse-Compton TeV flux ($F_{\rm TeV}^{\rm IC} \propto \epsilon_{\rm e,2}^{2 p_2 -2}$), some degeneracy remains with other model parameters.
This discrepancy can be understood in the context of particle-in-cell simulations, which suggest that the stronger turbulence and more efficient particle acceleration during shell collisions naturally lead to a higher $\epsilon_{\rm e,2}$ compared to the afterglow phase \citep{Sironi15}.
The realistic jet should be structured.
Our adoption of a uniform jet approximation serves to simplify the model by eliminating poorly constrained structural parameters.
A physically reasonable parameter set could in principle be identified to match the observations within the framework of structured jet models (e.g., a Gaussian jet).

\begin{figure}
  \centering
  \includegraphics[width=\textwidth]{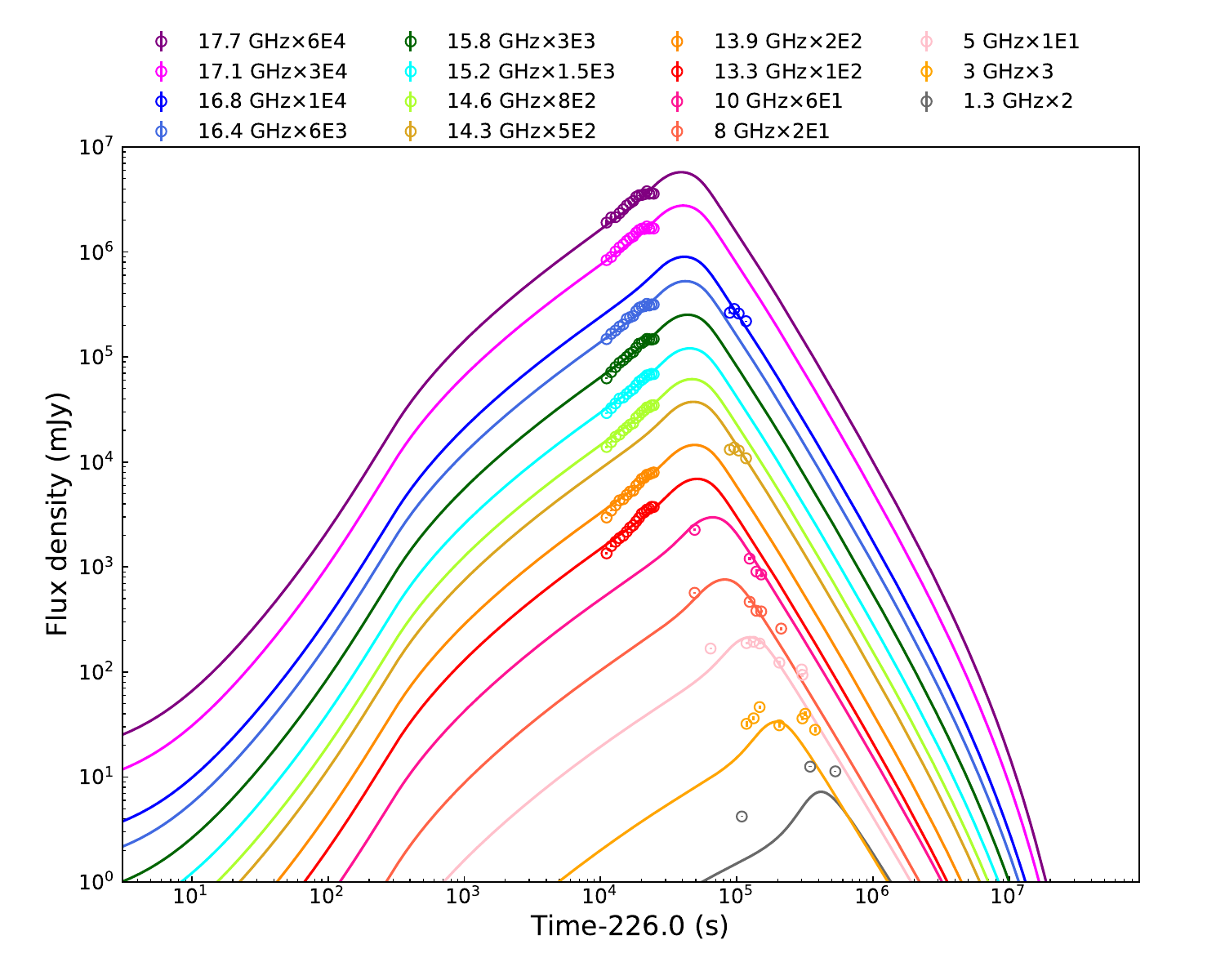}
  \caption{
  {\bf The fitting results of the reverse shock.} 
  Our best-fit results to the radio bumps are shown here.
  The posteriors of reverse shock parameters are listed in \SUPTAB{tab:RS}, and the corner plot is shown in \SUPFIG{fig:cornerRS}.  
  } 
 \label{fig:RS}
\end{figure}

\noindent {\bf --- The reverse shock} \\
Since the early radio bump is mainly the emission of the RS \citep{Bright23}, it provides a unique opportunity to probe the characteristics of the RS.
In our scenario, the dynamics of the bulk Lorentz factor of the RS could be approximated by a coasting phase during the collision and a
decelerating phase with a decaying law of $\propto R^{-g}$ afterwards \citep{Kobayashi00,Zou05}.  
The duration of the coasting phase corresponds to the RS crossing time,
which is $\sim$ 300 s as indicated by the end of the prompt emission \citep{Axelsson25}.
Therefore, as a simplified but generalized case, we adopt $\Gamma_{3} \simeq 200$ (an average value in the fitting of all multi-wavelength data), typical values of $g \simeq 1$ and $\delta_{3} \simeq 0.1$, and leave other crucial parameters of the fast shell (i.e., $L_{\rm f}$, $\Gamma_{\rm f}$, $\epsilon_{\mathrm{e,3}}$, $\sigma_{\rm f}$, and $p_3$) free, to perform Bayesian posterior searches.
The fitting result to the radio bump is shown in \SUPFIG{fig:RS} and \SUPTAB{tab:RS}.
This yields a derived magnetization of $3 \times 10^{-3}$ for the fast shell, consistent with our modeling for all data. 
Note that this magnetization applies to the outflow after experiencing energy release in the prompt phase. Therefore, the initial outflow was likely moderately magnetized before dissipation. By leveraging the exquisite radio bump, our analysis constrains the fast shell properties and demonstrates that the original jet had modest but non-negligible magnetization.

\clearpage

\clearpage

\newpage

\begin{figure}
  \centering
  \includegraphics[width=\textwidth]{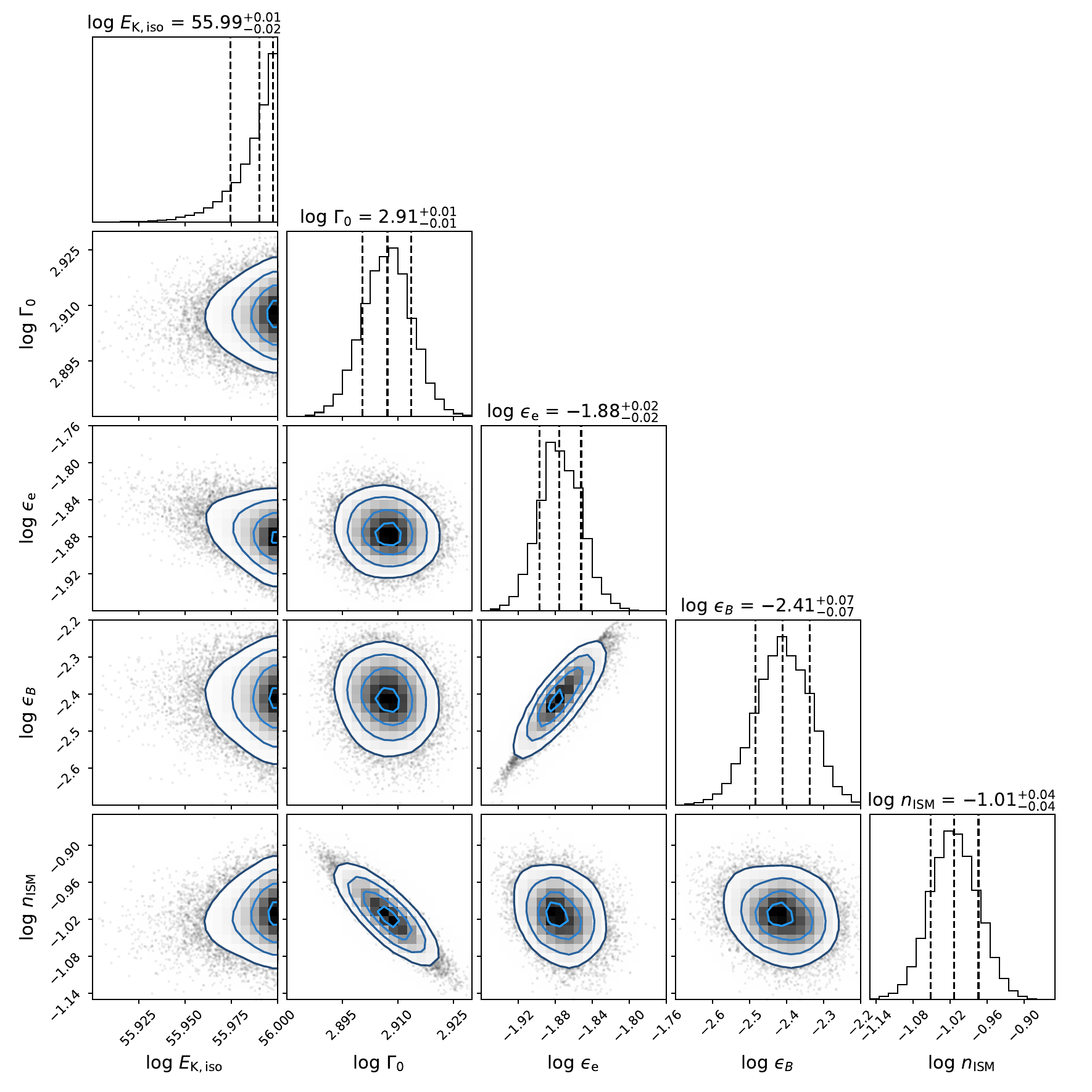}
  \caption{{\bf Corner plot of the fitting result of panel (B) in \SUPFIG{fig:Standard}.} 
  The corner plot shows all the one and two-dimensional projections of the posterior probability distributions of the initial isotropic kinetic energy ($E_{\rm K,iso}$), the initial Lorentz factor ($\Gamma_0$), the equipartition parameters for shocked electrons and magnetic field ($\epsilon_{\rm e}$ and $\epsilon_{B}$), and the circumburst number density ($n_{\rm ISM}$).
  The upper limit for priors of $E_{\rm K,iso}$ is set as $10^{56}$~erg by assuming that the radiation efficiency should not be less than 10\% for such a bright burst.
  The 1-dimensional histograms are marginal posterior distributions of these parameters. The vertical dashed lines indicate the 16th, 50th, and 84th percentiles of the samples, respectively, which are labeled on the top of each histogram.
  } 
 \label{fig:cornerb}
\end{figure}

\begin{figure}
  \centering
  \includegraphics[width=\textwidth]{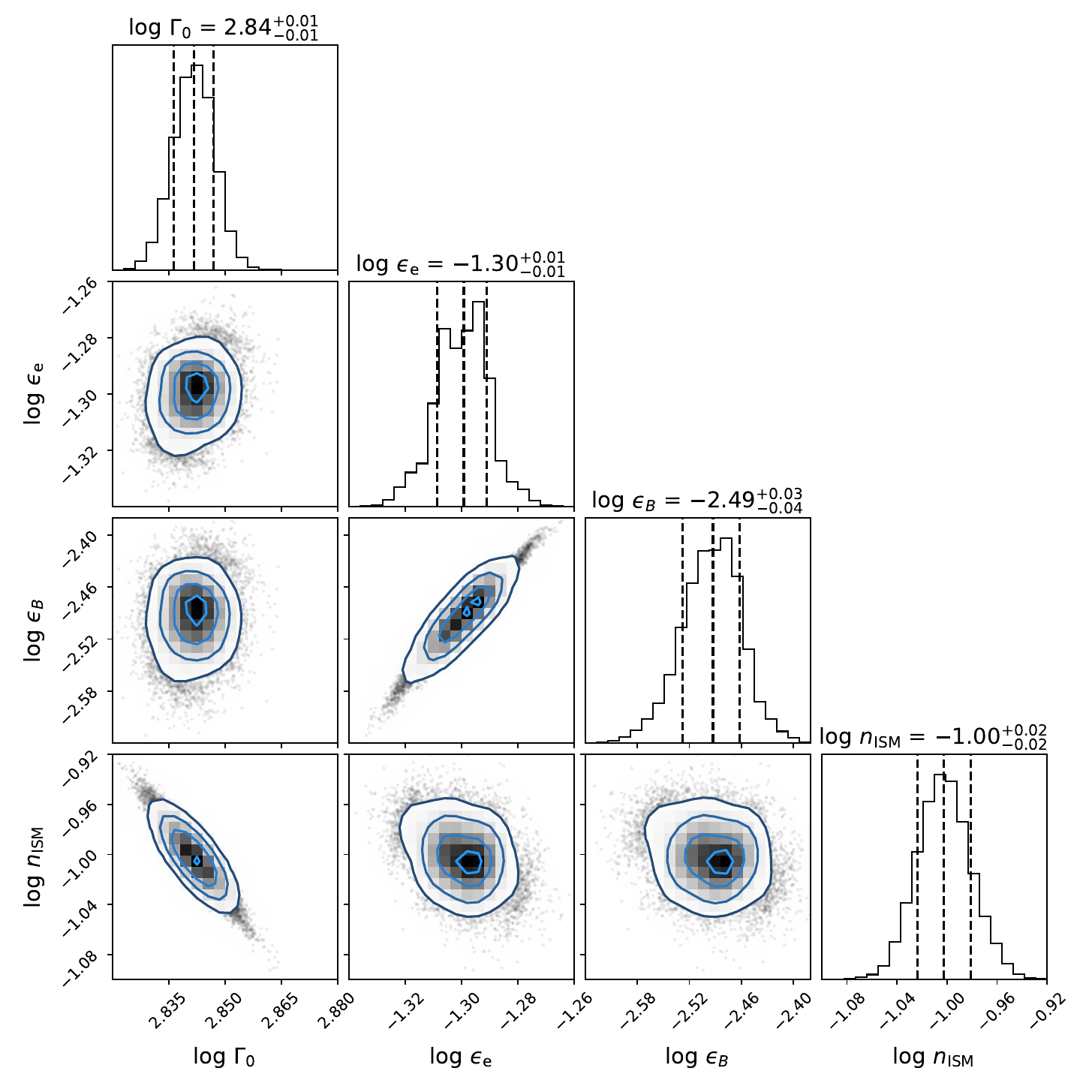}
  \caption{{\bf Corner plot of the fitting result of panel (C) in \SUPFIG{fig:Standard}.} 
  This figure is similar to \SUPFIG{fig:cornerb} but with $E_{\rm K,iso}$ fixed to be $9 \times 10^{54}$ erg \citep{LHAASO23}.
  } 
 \label{fig:cornerc}
\end{figure}

\newpage

\newpage

\newpage

\newpage

\begin{figure}
  \centering
  \includegraphics[width=\textwidth]{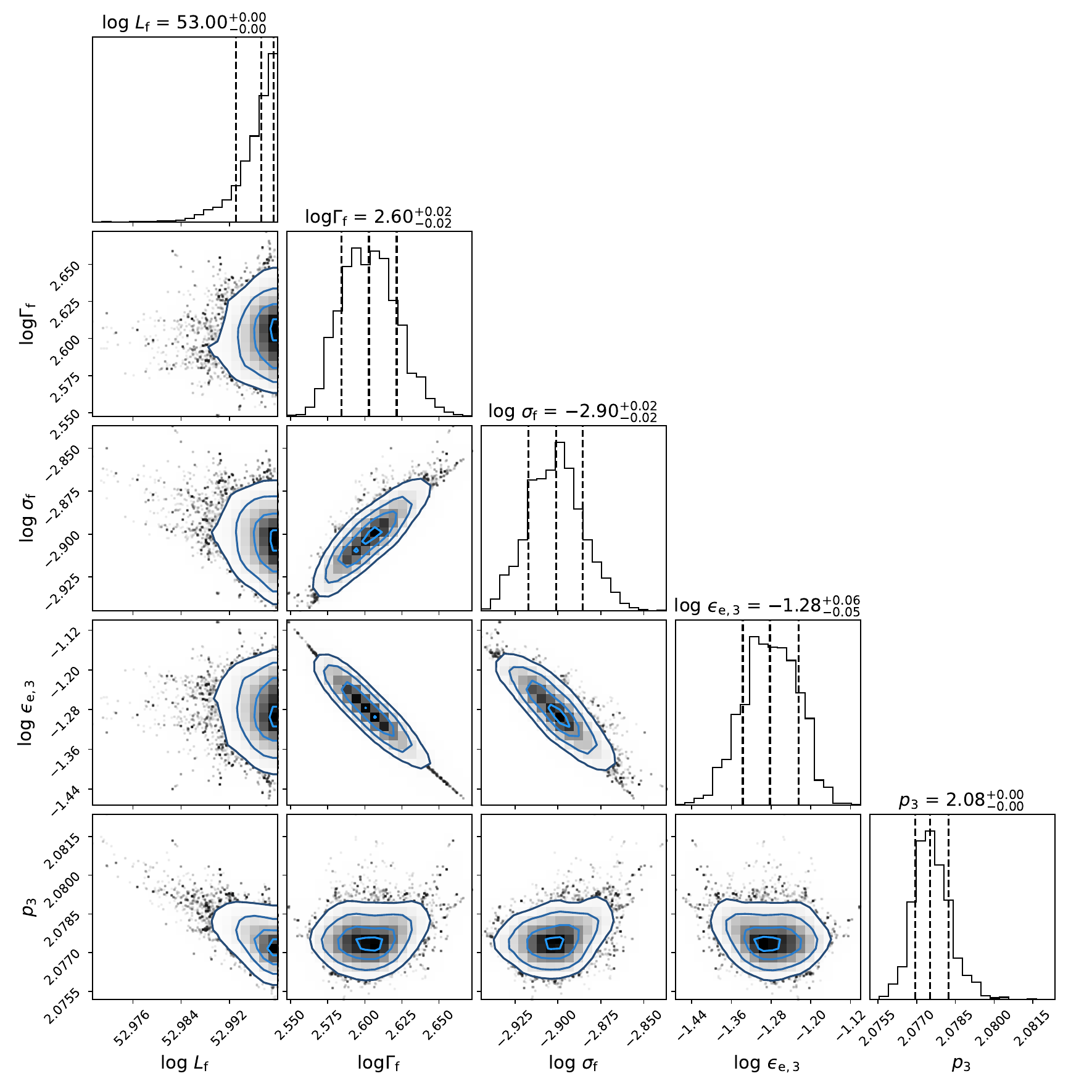}
  \caption{
  {\bf The corner plot of the fitting result in \SUPFIG{fig:RS}.} 
  Note that the marginal posterior distribution of $L_{\rm f}$ is close to the prior upper limit, but other parameters have converged well.} 
 \label{fig:cornerRS}
\end{figure}

\newpage

\begin{table}
\caption{Posteriors of the external shock parameters.}
\label{tab:standard}
\begin{center}
	\begin{tabular}{cccccc}
    \hline
    \hline
    Case & $\log E_{\rm K,iso}$  ($\mathrm{erg}$)  & $\log \Gamma_{\rm 0}$ & $\log \epsilon_{\mathrm{e}}$ & $\log \epsilon_{B}$ & $\log n_{\rm ISM}$ ($\mathrm{cm}^{-3}$)  \\
    \hline
    (A) & $55.98_{-0.03}^{+0.01}$ & $2.80_{-0.01}^{+0.01}$ & 
          $-2.14_{-0.01}^{+0.03}$ & $-3.33_{-0.03}^{+0.04}$ &
          $-0.14_{-0.04}^{+0.03}$ \\
    (B) & $55.99_{-0.02}^{+0.01}$ & $2.91_{-0.01}^{+0.01}$ & 
          $-1.88_{-0.02}^{+0.02}$ & $-2.41_{-0.07}^{+0.07}$ &
          $-1.01_{-0.04}^{+0.04}$ \\
    (C) & - & $2.84_{-0.01}^{+0.01}$ & 
          $-1.30_{-0.01}^{+0.01}$ & $-2.49_{-0.04}^{+0.03}$ &
          $-1.00_{-0.02}^{+0.02}$ \\
    \hline
    \hline
	\end{tabular}
\end{center}
\end{table}

\begin{table}
\caption{Parameters used in modeling the multi-wavelength emission of GRB 221009A.}
\label{tab:para-all}
\begin{center}
	\begin{tabular}{cccccccc}
    \hline
    \hline
 	\multicolumn{8}{c}{Properties of Two Shells} \\
 	\hline
    $L_{\mathrm{s}}$ & $L_{\mathrm{f}}$ & $E_{\mathrm{iso,s}}$ & $E_{\mathrm{iso,f}}$ & $\Gamma_{\mathrm{s}}$ & $\Gamma_{\mathrm{f}}$ & $\gamma_{\mathrm{p},0}$ & $\sigma_{\mathrm{f}}$ \\
    ($\mathrm{erg~s}^{-1}$)  &  ($\mathrm{erg~s}^{-1}$) &  ($\mathrm{erg}$)  &  ($\mathrm{erg}$)  & & & & \\ 
    $10^{52}$ & $10^{53}$ & $8.4 \times 10^{54}$ & $3.8 \times 10^{55}$ & $65$ & $450$ & $30$ & $10^{-3}$ \\
    \hline
    \hline
 	\multicolumn{8}{c}{Microphysical parameters of the FS} \\
 	\hline
    $\epsilon_{\mathrm{e},2}$ &     $\epsilon_{B,2}$  &   $p_2$   & $\delta_2$ & $\eta_2^{\dag}$ & & & \\
             $0.24$           &    $3 \times 10^{-5}$ &   $2.2$   &   $0.85$    &   $1.0$   & & \\
    \hline
    \hline
 	\multicolumn{8}{c}{Microphysical parameters of the RS} \\
 	\hline
    $\epsilon_{\mathrm{e},3}$ &     $\epsilon_{B,3}$  &   $p_3$   & $\delta_3$ & $\eta_3$ & & & \\
             $0.02$           &          --          &   $2.3$   &    $0.1$   &   $1.0$    & & & \\
    \hline
    \hline
 	\multicolumn{8}{c}{Parameters of the ES} \\
 	\hline
    $\epsilon_{\mathrm{e,E}}$ & $\epsilon_{B,\mathrm{E}}$ & $p_{\mathrm{E}}$ & $\delta_{\mathrm{E}}$ & $\eta_{\mathrm{E}}$ & $A_{*}$ & $R_{\mathrm{T}}$ \\ 
    & & & & & & ($\mathrm{cm}$) \\
    $0.02$          &        $2 \times 10^{-4}$          &    $2.4$   &      $0.75$    &     $0.1$  &   $0.1$ &  $1.9 \times 10^{19}$ \\
	\hline
	\end{tabular}
	\\	
	$\dag$. $\eta$ is the fraction of electrons that is assumed to be accelerated by the shock.   
\end{center}
\end{table}

\begin{table}
\caption{Posteriors of the reverse shock parameters.}
\label{tab:RS}
\begin{center}
	\begin{tabular}{ccccc}
    \hline
    \hline
    $\log L_{\rm f}$ ($\mathrm{erg~s}^{-1}$)  & $\log \Gamma_{\rm f}$ & $\log \sigma_{\rm f}$ & $\log \epsilon_{\mathrm{e},3}$ & $p_3$  \\
    $53.00_{-0.00}^{+0.00}$ & $2.60_{-0.02}^{+0.02}$ &
    $-2.90_{-0.02}^{+0.02}$ &
    $-1.28_{-0.05}^{+0.06}$ &  $2.08_{-0.00}^{+0.00}$ \\
    \hline
    \hline
	\end{tabular}
\end{center}
\end{table}

\clearpage
\clearpage
\end{appendix}

\end{document}